\documentclass[12pt]{article}
\usepackage{amssymb}
\usepackage{amsmath}
\usepackage{amstext}
\usepackage{graphicx,epsfig}
\usepackage{epsfig}
\usepackage{verbatim} 
\usepackage{fancybox}
\usepackage{color}
\usepackage{ulem}
\usepackage{enumitem}
\usepackage{subfigure}
\usepackage{bbm}
\usepackage{parskip}
\usepackage[utf8]{inputenc}

\usepackage{mathrsfs}
\usepackage{amsmath}
\usepackage{amssymb}
\usepackage{esint}

\newcommand{\Comment}[1]{{}}
\definecolor{MyDarkBlue}{rgb}{0.15,0.15,0.45}
\usepackage[linktocpage=true]{hyperref}
\hypersetup{
colorlinks=true,
citecolor=MyDarkBlue,
linkcolor=MyDarkBlue,
urlcolor=MyDarkBlue,
pdfauthor={Emel Altas},
pdftitle={Linearization Instability in Gravity Theories},
pdfsubject={hep-th}
}

\newcommand{\be}{\begin{equation}}
\newcommand{\ee}{\end{equation}}
\newcommand{\bea}{\begin{eqnarray}}
\newcommand{\eea}{\end{eqnarray}}
\newcommand{\beas}{\begin{eqnarray*}}
\newcommand{\eeas}{\end{eqnarray*}}

\def\({\left(}
\def\){\right)}

\title{Linearization Instability in Gravity Theories}
\author{Emel Altas\footnote{altas@metu.edu.tr and emelaltas85@gmail.com}}

\numberwithin{equation}{section}

\begin{document}

%\maketitle

\begin{center}
{\LARGE \bf{\sc Linearization Instability in Gravity Theories}}\footnote{This is a Ph.D. thesis defended in METU Physics Department on the $19^{th}$ of July 2018.}
\end{center} 
 \vspace{1truecm}
\thispagestyle{empty} \centerline{
{\large \bf {\sc Emel Altas${}^{a,}$}}\footnote{E-mail address: \Comment{\href{mailto:altas@metu.edu.tr}}{\tt altas@metu.edu.tr, emelaltas85@gmail.com}}
                                                          }

\vspace{1cm}

\centerline{{\it ${}^a$ 
Department of Physics,}}
 \centerline{{\it Middle East Technical University,  06800, Ankara, Turkey.}}

\begin{abstract}
 In a nonlinear theory, such as gravity, physically relevant solutions are usually hard to find. Therefore, starting from a background exact solution with symmetries, one uses the perturbation theory, which albeit approximately, provides a lot of information regarding a physical solution. But even this approximate information comes with a price: the basic premise of a perturbative solution is that it should be improvable. Namely, by going to higher order perturbation theory, one should be able to improve and better approximate the physical problem or the solution. While this is often the case in many theories and many background solutions, there are important cases where the linear perturbation theory simply fails for various reasons. This issue is well known in the context of general relativity through the works that started in the early 1970s, but it has only been recently studied in modified gravity theories. This thesis is devoted to the study of linearization instability in generic gravity theories where there are spurious solutions to the linearized equations which do not come from the linearization of possible exact solutions. For this purpose we discuss the Taub charges, the ADT charges and the quadratic constraints on the linearized solutions. We give the three dimensional chiral gravity and the $D$ dimensional critical gravity as explicit examples and give a detailed ADM analysis of the topologically massive gravity with a cosmological constant.
\end{abstract}

\newpage

\tableofcontents
\newpage

\section{Introduction}
\parskip=5pt
\normalsize

In nonlinear theories such as Einstein's general relativity or its modifications, extensions with higher powers of curvature, physically interesting solutions with few or no symmetries are usually analytically unavailable. Therefore one relies on the perturbation theory of some sort which often involves an exact background solution (with symmetries) and perturbations or deviations from this background solution. Generally, perturbation theory works fine in the sense that it can be improved order by order in some small parameter. But there are some important cases where perturbation theory fails as a method. This thesis is devoted to a detailed study of a phenomenon called {\it``linearization instability''} which refers to the failure of the first order perturbation theory in the following sense: not every solution of the linearized equations  can be improved to get exact solutions, even in principle. In the following section we give a brief review of linearization instability. Throughout the thesis, we work both with the index free and local coordinate forms of the relevant tensors. 

\subsection{Linearization instability in brief }

A nonlinear equation $F(x)=0$  is said to be linearization stable at a solution $x_0$ if every solution $\delta x$ to the linearized equation $F^\prime(x_{0})\cdot\delta x=0 $ is tangent to a curve of solutions to the original nonlinear equation. In some nonlinear theories, not all solutions to the linearized field equations represent linearized versions of exact (nonlinear) solutions. As a common algebraic example, let us consider the following:

$Example:$ Let us consider a function $F$ from ${\rm I\!R} \times {\rm I\!R}\rightarrow {\rm I\!R}$ such that  $F(x,y)=x(x^{2}+y^{2})$. In the domain of definition, the solution set to $F(x,y)=0$ is given as $(0,y)$ which is the $y$ axis.

Linearization of this exact solution set is then tangent to the $y$ axis and can be shown as $(0,\frac{\partial}{\partial y})$ or span$\{\frac{\partial}{\partial y}\}$, which is one dimensional.

Now let us consider the particular solution $(0,0)$ and linearize the equation around it. The linearized equation is simply  $\left(3x^{2}+y^{2}\right)\delta{x}+2xy\delta{y}=0$. For $(0,0)$ there is no constraint on the linearized solutions, then the solution set is
($\delta{x}$, $\delta{y}$) where  $\delta{x}$ and $\delta{y}$ are arbitrary.
Although $\delta{x}$ can be arbitrary with this approach,
we know that it cannot be from the linearization of the exact solution. Only $(0,\frac{\partial}{\partial y})$
is allowed  to be integrable to an exact solution. Consequently, the exact solution $(0,0)$ is linearization unstable and the perturbation theory fails about it.
$F(x,y)= x( x^2 + y^2)=0$, where $x,y$ are real, exact solution space is one dimensional given as $(0, y)$, and the linearized solution space is also one dimensional $(0,  \delta y ) $ as long as $y\ne0$. But at exactly the solution $(0, 0)$, the linearized solution space is two dimensional $(\delta x,  \delta y )$ and  so there are clearly linearized solutions with $\delta x \ne 0$, which do not come from the linearization of any exact solution. The existence of such spurious solutions depends on the particular theory at hand and the background solution (with its symmetries and topology) about which linearization is carried out.  If such so called "non-integrable" solutions exist,  perturbation theory in some directions of solution space fails and we say that the theory is not linearization stable at a nonlinear exact solution. See Figure 1.1 for a depiction of the function and the solution set.

\begin{figure} 
	\begin{centering}
		\includegraphics[scale=0.6]{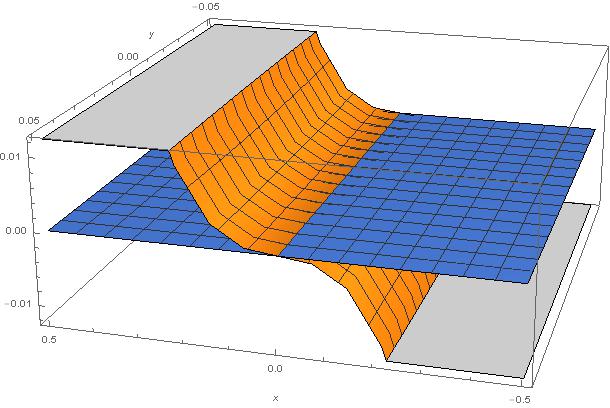} 
		\par\end{centering}
	\caption {The blue surface represents the $(x,y)$ plane and the orange surface shows the $F$ function, also the intersection of the surfaces is the $y$ axis which is the exact solution. As we mentioned above, linearization of the exact solution is  tangent to the $y$ axis. Conversely, solution of the linearized equation around the particular solution $(0,0)$ is ($\delta{x}$, $\delta{y}$). It is obvious that this solution is out of the linearized solution space. }
\end{figure}

What we have just described is not an exotic phenomenon: a {\it priori} no nonlinear theory is immune to linearization instability: one must study the problem case by case. For example, pure general relativity is linearization stable in Minkowski spacetime (with a non-compact Cauchy surface) \cite{Choquet-Bruhat}, hence perturbation theory makes sense, but it is not linearization stable on a background with compact Cauchy surfaces that possesses at least one Killing symmetry \cite{Moncrief} which is the case when the Cauchy surface is a flat 3-torus \cite{Deser_Brill}: on $T^3\times {\rm I\!R}$, at second order of the perturbation theory, one must go back and readjust the first order perturbative solution.

As gravity is our main interest here, let us consider some nonlinear gravity field equations  in a coordinate chart as $\mathscr{E}_{\mu\nu}=0$, which admits $\bar{g}_{\mu\nu}$ as an exact solution, if {\it every} solution ${h}_{\mu\nu}$ of the linearized field equations $\mathscr{E}^{(1)}(\bar{g})\cdot h=0$ is tangent to an exact solution ${g}_{\mu\nu}(\lambda)$ such that ${g}_{\mu\nu}(0)=:\bar{g}_{\mu\nu}$ and $\frac{dg_{\mu\nu}}{d\lambda}|_{\lambda=0}=:h_{\mu\nu}$ then, according to our definition above, the theory is linearization stable. Otherwise it is linearization unstable. In general, we do not have a theorem stating the {\it necessary and sufficient} conditions  for the linearization stability of a generic gravity theory about a given exact solution. We shall give a detailed discussion on generic gravity models in the next section based on our recent work \cite{emel}. For a brief note, let us consider the following: defining the second order perturbation as $\frac{d^2 g_{\mu\nu}}{d\lambda^2}|_{\lambda=0}=:k_{\mu\nu}$, if the following second order equation
\begin{equation} 
	(\mathscr{E})^{(2)}(\bar{g})\cdot [h,h]+(\mathscr{E})^{(1)}(\bar{g})\cdot k=0,
	\label{lin1}
\end{equation}
has a solution for $k_{\mu \nu}$ without a constraint on the linear solution $h_{\mu\nu}$, then the theory 
is linearization stable. Of course, at this stage it is not clear that there will arise no further constraints on the linear theory beyond the second order perturbation theory. In fact, besides Einstein's theory, this problem has not been worked out, to the best of our knowledge. But in Einstein's gravity, as the constraint equations are related to the zeros of the moment map, one knows that there will be no further constraint for the linear theory coming from higher order perturbation theory beyond the second order  \cite{Marsden_lectures}. In Einstein's gravity for compact Cauchy surfaces without a boundary, the necessary and sufficient conditions are known for linearization stability \cite{Moncrief,Marsden_Fischer,Marsden,Marsden_Arms}.  

In practice, it is very hard to show that (\ref{lin1}) is satisfied for {\it all } linearized solutions, therefore,  one resorts to a weaker condition by contracting that equation with a Killing vector field and integrates over a hypersurface to obtain  $ Q_{Taub}\left[\bar{\text{\ensuremath{\xi}}}\right] +Q_{ADT}\left[\bar{\text{\ensuremath{\xi}}}\right] =0$
where the Taub charge \cite{Taub} is defined as  
\begin{equation}
	Q_{Taub}\left[\bar{\text{\ensuremath{\xi}}}\right]:=\int_{\Sigma} d^{3}\Sigma\thinspace\sqrt{\gamma}\thinspace\hat{n}^{\nu}\thinspace\bar{\text{\ensuremath{\xi}}}^{\mu}\thinspace(\text{\ensuremath{\mathscr{E}}}_{\mu\nu})^{(2)}\cdot [h,h],
	\label{ttt}
\end{equation}
and the Abbott-Deser-Tekin (ADT) charge \cite{Abbott_Deser,Deser_Tekin} is defined as
\begin{equation}
	Q_{ADT}\left[\bar{\text{\ensuremath{\xi}}}\right] :=\int_{\Sigma} d^{3}\Sigma\thinspace\sqrt{\gamma}\thinspace\hat{n}^{\nu}\thinspace\bar{\text{\ensuremath{\xi}}}^{\mu}\left(\text{\ensuremath{\mathscr{E}}}_{\mu\nu}\right)^{(1)}\cdot k.
	\label{ADT}
\end{equation}
As it appears in the second order perturbation theory, the Taub charge is not a widely known quantity in physics, therefore a more detailed account of it will be given in the next chapter following the relevant discussion of \cite{emel}. 
The ADT charge can be expressed as a boundary integral. For the case of compact Cauchy surfaces without a boundary, $Q_{ADT} =0$, and hence one must have $Q_{Taub}=0$ which leads to the aforementioned quadratic integral constraint on the linearized perturbation $h_{\mu\nu}$ as the integral in (\ref{ttt}) should be zero. This is the case for Einstein's gravity, for example, on a flat 3-torus: $Q_{Taub}$ does not vanish automatically and so the first order perturbative result $h$ is constrained. On the other hand, for extended gravity theories (such as the ones we are interested in this thesis), $Q_{ADT}$ vanishes for a different reason, even for non-compact surfaces, as in the case of AdS. The reason is that for some tuned values of the parameters in the theory, the contribution to the conserved charges from various tensors cancel each other exact, yielding non-vacuum solutions that carry the (vanishing) charges of the vacuum. This is the source of the linearization instability.

Chapter II of this thesis is devoted to a detailed study of linearization instability in generic gravity theories. We give $D$-dimensional critical and three dimensional chiral gravity theories (which both received interest in the recent literature) as two interesting examples of theories that exhibit the kind of linearization instability we mentioned above. 

In Chapter III, we give a discussion of the initial value formulation and the ADM decomposition of topologically massive gravity, study its constraints on a spacelike surface and give a second proof of linearization instability at the chiral limit of the theory. This second proof is not based on the charge construction but is based directly on the field equations (especially the constraints of the theory). 

We relegate some of the computations to the appendices: In Appendix A, we give details of second order perturbation theory in the context of Riemannian geometry and compute the relevant expanded tensors up to second order and discuss the gauge invariance issues of the second order Einstein's tensor. Note that, in contrast to the first order Einstein's tensor, the second order one is not gauge invariant, therefore one must be very careful about any result (such as the Taub charge construction) based on the second order perturbation theory. So we discuss the gauge transformation properties of the relevant tensors under small gauge  transformations (infinitesimal diffeomorphisms). 

In Appendix B, we compute explicitly the form of the $K_{\mu\nu}$ tensor for the case of Einstein's theory in AdS and Minkowski spaces. This is relevant for the proof of the linearization stability of the Minkowski space. 

Appendix C is devoted to a detailed construction of the ADM formulation of the topologically massive gravity directly from the field equations and the action. In most of that appendix we work with nonzero lapse and shift functions, but at the end we restrict to the Gaussian normal coordinates for the particular goal of studying the linearization instability issue on the spacelike initial value surface for AdS. 

\newpage

\section[{Linearization instability for generic gravity in A\lowercase{D}S}] {Linearization instability for generic gravity in A\lowercase{D}S\footnote {This chapter was published as 
	Phys. Rev. D 97, 024028 on 24 January 2018.} }

There is an interesting conundrum in nonlinear theories, such
as Einstein's gravity or its modifications with higher curvature terms:
exact solutions without symmetries (which are physically interesting) are hard to find, hence one resorts
to symmetric "background" solutions and develops a perturbative expansion about them. But it turns out
that exactly at the symmetric solutions, namely about solutions having Killing
vector fields, naive first order perturbation theory fails under certain conditions. The set of solutions to Einstein's equations forms a smooth manifold except at the solutions with infinitesimal symmetries and spacetimes with compact Cauchy surfaces where there arise conical singularities in the solution space.  Namely, perturbation theory in non-linear theories can yield results which are simply wrong in the sense that {\it some} perturbative solutions cannot be obtained from the linearization of exact solutions. Roughly speaking, the process of first linearizing the field equations and then finding the solutions to those linearized equations; and the process of linearization of exact solutions to the non-linear equations can yield different results if certain necessary criteria are not met with regard to the background solution about which perturbation theory is carried out. Actually, the situation is more serious: linearized field equations can have spurious solutions which do not come from exact solutions. This could happen for various reasons and
the failure of the first order perturbation theory can be precisely
defined, as we shall do below.  Figure 2.1 summarizes the results. 

\begin{figure}
	\begin{centering}
		\includegraphics[scale=0.4]{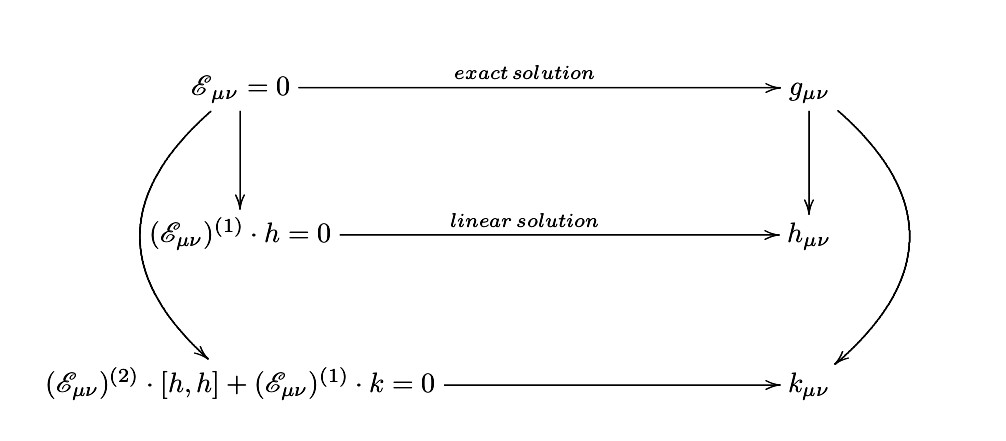} 
		\par\end{centering}
	\caption {The vertical straight arrows show first order linearization while the curved ones show second order linearization. For a linearization stable theory, the diagram makes sense and the solution to the linearized equation $h$ is not further restricted at the second order which means that there is a symmetric tensor $k$ that satisfies the second order equation in the bottom left. The details of the symbols are explained in the next section.}
\end{figure}

Let us give a couple of early observations
in this issue in the context of general relativity (GR) before we start the discussion in generic gravity.
One clear way to see the failure
of the perturbation theory is through the {\it initial value } formulation
of the theory for globally hyperbolic, oriented, time-orientable
spacetimes with  the topology ${\mathcal{M}} \approx \Sigma \times\mathbb{R}$, where $\Sigma$ is a spacelike Cauchy surface on which the induced Riemannian metric
$\gamma$ and the extrinsic curvature {\it K} (as well as matter content of the theory) are defined. [It is also common to formulate the constraint equations in terms of $\gamma$ and a tensor density of weight 1 defined as  $\pi := \sqrt{\text{det} \gamma} ( K - \gamma\text{tr}_\gamma K )$ which is the conjugate momentum of the induced metric $\gamma$.] 
We shall consider the matter-free case through-out this chapter. Since GR is nonlinear,
the initial data cannot be arbitrarily prescribed: they must satisfy
the so called Hamiltonian and momentum constraints $\Phi_{i}(\gamma,K)=0$ with $i \in \{1,2,3,4 \}$ in four dimensions. If a given initial data $(\bar{\gamma}, \bar{K})$  solving the constraints is not isolated, meaning the linearized constraint equations  $\delta \Phi_{i}(\bar \gamma,\bar K)\cdot [\delta \gamma, \delta K]=0$
allow {\it viable} linearized solutions $(\delta \gamma, \delta K)$, then the theory is said to be {\it linearization
	stable} about the initial Cauchy data. Deser and
Brill \cite{Deser_Brill} showed that in GR with a compact Cauchy surface having the topology
of a 3-torus, there are strong constraints on the perturbations of the initial data. Any such perturbation leads to contradictions in the sense that bulk integrals of conserved mass and angular momenta do not vanish, while since there is no boundary, they must vanish in this compact space: hence the background is an isolated solution. Put in another way, the linearized field equations about the background have solutions which
do not come from the linearization of exact solutions. This happens because, as
we shall see below, the linearized equations of the theory are not  {\it sufficient }
to constrain the linearized solutions: quadratic constraints on the linearized solutions, in the form of an
integral (so called {\it Taub conserved quantity} first introduced in \cite{Taub} for each Killing vector field), arise.

Most of the work regarding the
linearization stability or instability in gravity has been in the
context of GR with or without matter and with compact or with non-compact
Cauchy surfaces. A nice detailed account of all these in the context of GR is given in the book \cite{book}. See also \cite{ch1} where a chapter is devoted to this issue and the Taub conserved quantity construction which is not widely known in the physics community. Our goal here is to extend the discussion to generic gravity
theories: we show that if the field equations of the theory are defined by the Einstein tensor
plus a covariantly conserved two tensor, then a new source of linearization
instability that does not exist in GR arises, especially in de Sitter
or Anti-de Sitter backgrounds, with non-compact Cauchy surfaces. This happens because in these backgrounds
there are special {\it critical} points in the space
of parameters of the theory which conspire to cancel the conserved
charge (mass, angular momentum, {\it etc.}) of non-perturbative objects (black
holes) or the energies of the perturbative excitations. One needs to understand the origin
of this rather interesting phenomenon that non-vacuum objects have
the same charges as the vacuum. To give an example of this phenomenon let
us note that this is exactly what happens in chiral gravity  \cite{Strom1,Strom2,Strom3,Carlip}
in 2+1 dimensions where the Einstein tensor is augmented with the Cotton
tensor and the cosmological constant times the metric (namely a special limit of the cosmological topologically massive gravity \cite{djt}). In AdS,
at the chiral point, the contribution of the Cotton tensor and the
Einstein tensor in AdS cancel each other at the level of the conserved
charges. Exactly at that point, new ghost-like solutions, the so called
log modes arise \cite{Grumiller} and if the boundary conditions are not those of Brown-Henneaux type {\cite{BH}, then these modes are present in the theory with negative energies. This would mean that the theory has no vacuum. But it was argued
	in \cite{Strom2,Carlip} that chiral gravity in AdS has a linearization
	instability which would remedy this problem. A similar phenomenon occurs in critical gravity in all
	dimensions \cite{Pope, Tahsin}. Here we give a systematic discussion
	of the linearization stability and instability in generic gravity
	theories and study these two theories as examples. We will not  follow the route of defining the theory in the 3+1
	setting and considering the instability problem on the Cauchy data. The reason for this is the following: in GR for asymptotically flat spacetimes, splitting the problem into the constraints on the Cauchy data and the evolution of the 3-metric and the extrinsic curvature turns the stability problem to a problem in elliptic  operator theory which is well-developed and sufficient to rigorously prove the desired results.  In the initial value formulation setting, the problem becomes a problem of determining the surjectivity of a linear operator, namely the linearized constraint operator. But this method is not convenient for our purposes  since the source of the linearization instability in the extended gravity models that we shall discuss is quite different and so the full spacetime formulation is  much better-suited for our problem. In GR as noted in the abstract, what saves the Minkowski space from the linearization instability is its non-compact Cauchy surfaces  as was shown by Choquet-Bruhat and Deser \cite{Choquet-Bruhat}. This result is certainly consistent with the non-zero conserved charges (ADM mass or angular momentum) that can be assigned to an asymptotically-flat 3 dimensional Cauchy surface.
	
	The layout of the chapter is as follows: In section II, we discuss the linearization stability in generic gravity theory and derive the  second order constraints on the solutions of the linearized field equations. Of  course these constraints are all related to the diffeomorphism invariance and the Bianchi identities of the theory. Hence we give a careful discussion of the linearized forms of the field equations and their gauge invariance properties. As the second order perturbation theory about a generic background is quite cumbersome in the local coordinates, we carry out the index-free computations in the bulk of the chapter and relegate some parts of the component-wise computations to the appendices. In section II, we establish the relation between the Taub conserved quantities coming from the second order  perturbation theory and the Abbott-Deser-Tekin (ADT) charges coming from the first order perturbation theory.  We study the linearization stability and instability of the Minkowski space, chiral gravity and critical gravity as examples. In the forth-coming chapter, we shall give a more detailed analysis of the chiral gravity discussion in the initial value formulation context.

	\subsection{Linearization Stability in Generic Gravity}
	
	Let us consider the matter-free equation of a generic gravity theory in a $D$-dimensional spacetime, whose dynamical field is the metric tensor $g$ only. In the index-free notation the covariant two-tensor equation reads
	\begin{equation}
		\text{\ensuremath{\mathscr{E}}}(g)=0,
		\label{gen_theory}
	\end{equation}
	together with the  covariant divergence condition which comes from the diffeomorphism invariance of the theory
	\begin{equation}
		\delta_{g}\text{\ensuremath{\mathscr{E}}}(g)=0,
	\end{equation}
	where $\delta_{g}$ denotes the divergence operator with respect to the metric $g$. (As usual, one uses the musical isomorphism to extend the divergence from the contravariant tensors to the covariant ones.) Here we generalize the discussion in \cite{Marsden_Fischer,Marsden} given for Einstein's theory to generic gravity. Let us assume that there
	is a one-parameter family of solutions to (\ref{gen_theory}) denoted as $g(\lambda)$
	which is at least twice differentiable with respect to $\lambda$ parameterizing the solution set. Then we can explore the consequences
	of this assumption with the help of the following identifications : 
	\begin{equation}
		\bar{g}:=g(\lambda)\bigg\rvert_{\lambda=0}, \hskip 0.7 cm h:=\frac{d}{d\lambda}g(\lambda)\bigg\rvert_{\lambda=0},\hskip 0.7 cm k:=\frac{d^{2}}{d\lambda^{2}}g(\lambda)\bigg\rvert_{\lambda=0}.
	\end{equation}
	At this stage there is of course no immediate relation between the
	two covariant tensor fields $h$ (the first derivative of the metric)  and $k$ (the second derivative of the metric) but, as we shall see later, consistency of the theory, {\it i.e.} the first order linearized and second order linearized forms of the field equations will relate them. We would first like to find that relation. 
	
	We assume that $\bar{g}$
	exactly solves the vacuum equations $\text{\ensuremath{\mathscr{E}}}(\bar{g})=0$ and we compute the
	first derivative of the field equations with respect to $\lambda$ and evaluate it at $\lambda =0$
	as
	\begin{equation}
		\frac{d}{d\lambda}\text{\ensuremath{\mathscr{E}}}(g(\lambda))\bigg\rvert_{\lambda=0}=D\text{\ensuremath{\mathscr{E}}}(g(\lambda))\cdot\frac{dg(\lambda)}{d\lambda}\bigg\rvert_{\lambda=0}=0,
		{\label{first_lin}}
	\end{equation}
	where $D$ denotes the Fr\'echet derivative and the center-dot denotes "along the direction of the tensor that comes next" and we have used the chain rule. In local coordinates, this equation is just the first order "linearization"
	of the field equations (\ref{gen_theory}) which we shall denote as $\left(\text{\ensuremath{\mathscr{E}}}_{\mu\nu}\right)^{(1)}\cdot h=0$.
	It is important to understand that solutions of (\ref{first_lin}) yield all possible $h$ tensors (up to diffeomorphisms),
	which are tangent to the exact solution $g(\lambda)$ at $\lambda=0$ in the space of solutions. To understand if there are any {\it  further } constraints on the linearized solutions $h$, let us consider
	the second derivative of the field equation with respect to $\lambda$
	and evaluate it at $\lambda=0$ to arrive at 
	\begin{eqnarray}
		&&	\frac{d^{2}}{d\lambda^{2}}\text{\ensuremath{\mathscr{E}}}\bigg(g(\lambda)\bigg)\bigg\rvert_{\lambda=0}\nonumber\\&&=\Bigg (D^{2}\text{\ensuremath{\mathscr{E}}}(g(\lambda))\cdot\left[\frac{dg(\lambda)}{d\lambda},\frac{dg(\lambda)}{d\lambda}\right]+D\text{\ensuremath{\mathscr{E}}}(g(\lambda))\cdot\frac{d^{2}g(\lambda)}{d\lambda^{2}}\Bigg)\bigg\rvert_{\lambda=0}=0,
		\label{main1}
	\end{eqnarray}
	where we have used the common notation for the second Fr\'echet derivative in the first term and employed the chain rule when needed.  We can write (\ref{main1}) in local coordinates as
	\begin{equation}
		(\text{\ensuremath{\mathscr{E}}}_{\mu\nu})^{(2)}\cdot [h,h]+\left(\text{\ensuremath{\mathscr{E}}}_{\mu\nu}\right)^{(1)}\cdot k=0,
		{\label{main2}}
	\end{equation}
	where  again $(\text{\ensuremath{\mathscr{E}}}_{\mu\nu})^{(2)}\cdot [h,h]$
	denotes the second order linearization of the field equation about the background $\bar{g}$. Even though
	this equation is rather simple, it is important to understand its
	meaning to appreciate the rest of the discussion. This is the equation given in the bottom-left corner of Figure 2.1. Given a solution $h$ of   $\left(\text{\ensuremath{\mathscr{E}}}_{\mu\nu}\right)^{(1)}\cdot h=0,$ equation (\ref{main2})   {\it determines } 
	the tensor field $k$, which is the second order derivative of the
	metric $g(\lambda)$ at $\lambda=0.$ If such a $k$ can be found then
	there is no further constraint on the linearized solution $h$. In that case, the field equations are said to be linearization stable at the exact solution $\bar{g}$. This says that the infinitesimal deformation $h$ is tangent to a full (exact) solution and hence it is integrable to a full solution. Of course, what is tacitly assumed here is that in solving for $k$ in (\ref{main2}), one cannot change the first order solution $h$, it must be kept intact for the perturbation theory to make any sense.  
	
	We can understand these results form a more geometric vantage point as follows. For the spacetime manifold ${\cal{M}}$, let ${\mathcal{S}}$ denote the set of solutions of the field equations $\text{\ensuremath{\mathscr{E}}}(g)=0$. The obvious question is (in a suitable Sobolev topology), when does this set of solutions form a smooth manifold whose tangent space at some "point" $\bar{g}$ is the space of solutions ($h$) to the linearized equations?  The folklore in the physics literature is not to worry about this question and just assume that the perturbation theory makes sense and the linearized solution can be improved to get better solutions, or the linearized solution is assumed to be integrable to a full solution. But as we have given examples above, there are cases when the perturbation theory fails and the set ${\mathcal{S}}$ has a conical singularity instead of being a smooth manifold. 
	One should not confuse this situation with the case of dynamical instability as the latter really allows a "motion" or perturbation about a given solution. Here linearization instability refers to a literal break-down of the first order perturbation theory.   
	It is somewhat a non trivial matter to show that there are no further constraints  beyond the second order perturbation theory: In Einstein's gravity, this is related to the fact that constraint equations are related to zeros of the moment maps \cite{Marsden_lectures}. For generic gravity, this issue deserves to be further studied. 
	
	\subsection{Taub conserved quantities and ADT charges}
	
	So far, in our discussion we have not assumed anything about whether the spacetime has
	a compact Cauchy surface or not. First, let us now assume that the spacetime
	has a compact spacelike Cauchy surface and has at least one Killing vector
	field. Then we can get an \emph{integral constraint} on $h$, without
	referring to the $k$ tensor as follows. Let $\bar{\text{\ensuremath{\xi}}}$
	be a Killing vector field of the metric $\bar{g}$, then the following vector field \footnote{ For the lack of a better notation, note that $\bar{\text{\ensuremath{\xi}}}$ is contracted with the covariant background tensor with a center dot which we shall employ in what follows and it should not be confused with the center dot in the Fr\'echet derivative.}
	\begin{equation}
		T:=\bar{\text{\ensuremath{\xi}}}\cdot D^{2}\text{\ensuremath{\mathscr{E}}}(\bar{g})\cdot\left[h,h\right],
	\end{equation}
	is divergence free, since $\delta_{\bar{g}}D^{2}\text{\ensuremath{\mathscr{E}}}(\bar{g}).\left[h,h\right]=0$
	due to the linearized Bianchi identity . Then we can integrate $T$
	over a compact hypersurface $\Sigma$ and observe that the integral (for the sake of definiteness, here we consider the 3+1 dimensional case) 
	\begin{equation}
		\int_{\Sigma}d^{3}\Sigma\thinspace\sqrt{\gamma}\thinspace{T}\cdot\hat{n}_{\Sigma}
	\end{equation}
	is independent of hypersurface $\Sigma$ where $\gamma$
	is the pull-back metric on the hypersurface and $\hat{n}_{\Sigma}$ is the
	unit future pointing normal vector. Let us restate the result in a
	form that we shall use below: given {\it two} compact  disjoint hypersurfaces 
	$\Sigma_{1}$
	and $\Sigma_{2}$ (as shown in Figure 2.2) in the spacetime ${\mathcal{M}}$, we have the statement of the  "charge conservation"  as the equality of the integration over the two hypersurfaces 
	\begin{equation}
		\int_{\Sigma_{1}}d^{3}\thinspace\Sigma_{1}\thinspace\sqrt{\gamma_{\Sigma_{1}}}\thinspace{T}\cdot\hat{n}_{\Sigma_{1}}=\int_{\Sigma_{2}}d^{3}\thinspace\Sigma_{2}\thinspace\sqrt{\gamma_{\Sigma_{2}}}\thinspace{T}\cdot\hat{n}_{\Sigma_{2}}.
	\end{equation}
	We can now go to (\ref{main2}) and after contracting it with the Killing
	tensor $\bar{\text{\ensuremath{\xi}}}$, and integrating over  $\Sigma$, we obtain the identity 
	\begin{equation}
		\int_{\Sigma} d^{3}\Sigma\thinspace\sqrt{\gamma}\thinspace\bar{\text{\ensuremath{\xi}}}^{\mu}\hat{n}^{\nu}(\text{\ensuremath{\mathscr{E}}}_{\mu\nu})^{(2)}\cdot [h,h]\thinspace=-\int_{\Sigma} d^{3}\Sigma\thinspace\sqrt{\gamma}\thinspace\bar{\text{\ensuremath{\xi}}}^{\mu}\hat{n}^{\nu}\left(\text{\ensuremath{\mathscr{E}}}_{\mu\nu}\right)^{(1)}\cdot k\thinspace.
		\label{denklik}
	\end{equation}
	
	Let us study the right-hand side more carefully. In a generic theory,
	this conserved Killing charge is called the Abbott-Deser-Tekin (ADT)
	charge  ( for further details on the ADT charges, please see the recent review \cite{Adami} and the relevant references therein )when the symmetric two-tensor $k$ is the just the linearized
	two tensor $h$ \cite {Abbott_Deser, Deser_Tekin}. Once the field equations of the theory are given, it
	is possible, albeit after some lengthy computation, to show that one
	can write the integral on the right-hand side as a total derivative.
	\begin{equation}
		\bar{\text{\ensuremath{\xi}}}^{\mu}\left(\text{\ensuremath{\mathscr{E}}}_{\mu\nu}\right)^{(1)}\cdot h=\bar{\nabla}_{\alpha}\left(\text{\ensuremath{\mathscr{F}}}^{\alpha}\,_{\nu\mu}\bar{\text{\ensuremath{\xi}}}^{\mu}\right),
		\label{div_denk}
	\end{equation}
	with an anti-symmetric tensor $\mathscr{F}$ in $\alpha$ and $\nu$.  
	Hence if the Cauchy surface is compact without a boundary, the ADT charge
	vanishes identically, namely
	\begin{equation}
		Q_{ADT}\left[\bar{\text{\ensuremath{\xi}}}\right] :=\int_{\Sigma} d^{3}\Sigma\thinspace\sqrt{\gamma}\thinspace\hat{n}^{\nu}\thinspace\bar{\text{\ensuremath{\xi}}}^{\mu}\left(\text{\ensuremath{\mathscr{E}}}_{\mu\nu}\right)^{(1)}\cdot h=0,
		\label{ADT}
	\end{equation}
	which via (\ref{denklik}) says that one has the vanishing of the integral on
	the left hand-side which is called the Taub conserved quantity:
	\begin{equation}
		Q_{Taub}\left[\bar{\text{\ensuremath{\xi}}}\right]:=\int_{\Sigma} d^{3}\Sigma\thinspace\sqrt{\gamma}\thinspace\hat{n}^{\nu}\thinspace\bar{\text{\ensuremath{\xi}}}^{\mu}\thinspace(\text{\ensuremath{\mathscr{E}}}_{\mu\nu})^{(2)}\cdot [h,h]=0,
		\label{ttt}
	\end{equation}
	which must be {\it automatically} satisfied for the case when $h$ is an
	integrable deformation. Otherwise this equation is a second order constraint on
	the linearized solutions. Even though the ADT potential $\mathscr{F}$
	was explicitly found for a large family of gravity theories, such as Einstein's
	gravity \cite{Abbott_Deser}, quadratic gravity \cite{Deser_Tekin}, $f(Riem)$ theories \cite{SST}, and some examples will be given below, we can still refine
	the above argument of the vanishing of both the ADT and Taub conserved quantities
	without referring to the ADT potential (or more explicitly without
	referring to (\ref{div_denk})). The following argument was given for Einstein's
	gravity in \cite{Marsden} which immediately generalizes to the most general
	gravity as follows: consider the ADT charge (\ref{ADT}) and assume that in the spacetime one
	has two disjoint compact hypersurfaces $\Sigma_{1}$ and $\Sigma_{2}$ as above.
	Then the statement of conservation of the charge is simply
	\begin{equation}
		Q_{ADT}\left(\bar{\text{\ensuremath{\xi}}},\Sigma_{1}\right)=Q_{ADT}\left(\bar{\text{\ensuremath{\xi}}},\Sigma_{2}\right).
	\end{equation}
	\begin{figure}
		\begin{centering}
			\includegraphics[scale=0.4]{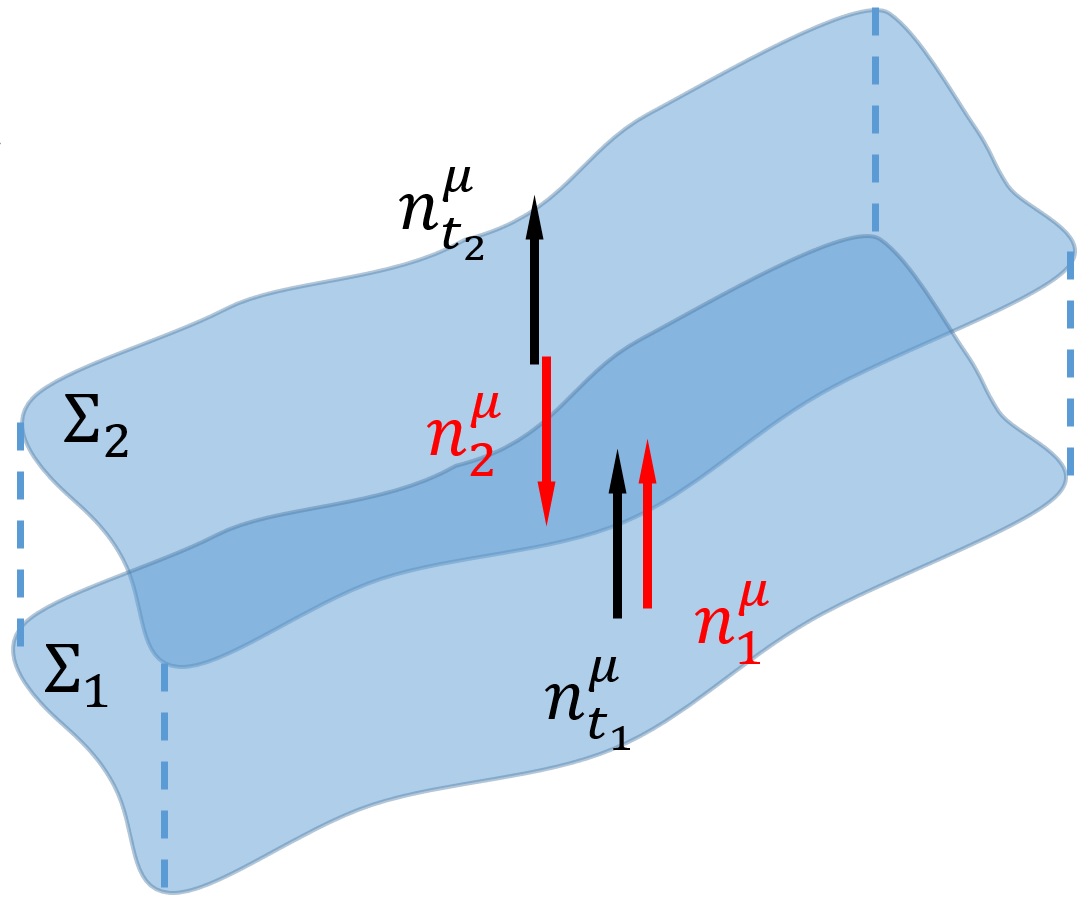} 
			\par\end{centering}
		\caption {Disjoint hypersurfaces $\Sigma_{1}$ and $\Sigma_{2}$ are shown along with their timelike unit normals. The figure is taken from \cite{Adami}.}
	\end{figure}
	Now let $k$ be a two tensor which is $k_{1}$ and non-zero on $\Sigma_{1}$
	and $k_{2}$ and zero near $\Sigma_{2}$, then $Q_{ADT}\left(\bar{\text{\ensuremath{\xi}}},\Sigma_{2}\right)=0$
	so $Q_{ADT}\left(\bar{\text{\ensuremath{\xi}}},\Sigma_{1}\right)=0$
	which in turn yields the vanishing of the Taub conserved quantities via (\ref{denklik}).
	
	To summarize the results obtained so far, let us note that assuming
	an integrable infinitesimal deformation $h$, which is by definition a solution
	to the linearized field equations about a background $\bar{g}$ solution,
	we arrived at  (\ref{main2}). And the discussion after that equation
	showed that Taub conserved quantities constructed with a Killing vector field, from the second order linearization,
	$(\text{\ensuremath{\mathscr{E}}}_{\mu\nu})^{(2)}\cdot [h,h]$,
	and the ADT charges constructed from the first order linearization,
	$(\text{\ensuremath{\mathscr{E}}}_{\mu\nu})^{(1)}\cdot h$,
	vanish identically for the case of compact Cauchy hypersurfaces without
	a boundary. If these integrals do not vanish, then there is a contradiction  and the the linearized solution 
	$h$  is further constraint. Hence it is not an integrable deformation, namely, $h$ is not in the tangent space about the point $\bar{g}$ in the space of solutions.  For Einstein's theory with compact Cauchy surfaces, it was shown that the necessary condition for linearization  stability is the absence of Killing vector fields \cite{Moncrief,Marsden_Arms}. As noted above, the interesting issue is that further study reveals that besides the quadratic constraint, there are no other constraints on the solutions to the linearized equations \cite{Marsden_lectures}.
	
	\subsection{Gauge invariance of the charges}
	
	Of course there is one major issue that we still must
	address that is the gauge-invariance (or coordinate independence)
	of the above construction which we show now. Following \cite{Marsden}, first let us consider a (not necessarily small) diffeomorphism
	$\varphi$ of the spacetime as $\varphi:{\mathcal{M}}\rightarrow {\mathcal{M}}$. Then we
	demand that having obtained our rank two tensor $\text{\ensuremath{\mathscr{E}}}(g)$
	from a diffeomorphism invariant action (or from a diffeomorphism invariant
	action up to a boundary term as in the case of topologically massive gravity) we have a global statement of diffeomorphism invariance as 
	\begin{equation}
		\text{\ensuremath{\mathscr{E}}}\left(\varphi^{*}g\right)=\varphi^{*}\text{\ensuremath{\mathscr{E}}}\left(g\right),
		\label{global}
	\end{equation}
	which states that $\text{\ensuremath{\mathscr{E}}}$ evaluated
	for the pull-back metric is equivalent to the pull-back of $\text{\ensuremath{\mathscr{E}}}$
	evaluated for $g.$ Let us now consider a one-parameter family of
	diffeomorphisms as $\varphi_\lambda$, generated by a vector field
	${X}$ well-defined on some region of spacetime. Let
	$\varphi_{0}$ be the identity diffeomorphism denoted as $\varphi_{0}=I_{\mathcal{M}}.$
	Then we can differentiate (\ref{global}) with respect to $\lambda$ once to get
	\begin{equation}
		\frac{d}{d\lambda}\text{\ensuremath{\mathscr{E}}}\left(\varphi_{\lambda}^{*}{\text{}}g\right)=\frac{d}{d\lambda}\varphi_{\lambda}^{*}\text{\ensuremath{\mathscr{E}}}\left(g\right),
	\end{equation}
	which, after making use of the chain rule, yields
	\begin{equation}
		D\text{\ensuremath{\mathscr{E}}}\left(\varphi_{\lambda}^{*}{\text{}}g\right)\cdot\frac{d}{d\lambda}\varphi_{\lambda}^{*}{\text{}}g=\varphi_{\lambda}^{*}{\text{}}\bigg (\text{\ensuremath{\mathscr{L}}}_{X}\text{\ensuremath{\mathscr{E}}}\left(g\right)\bigg),
		\label{onyedi}
	\end{equation}
	where $\text{\ensuremath{\mathscr{L}}}_{X}$ is the Lie derivative
	with respect to the vector field $X$. Taking the derivative of the last equation with respect to $g$ yields
	\begin{equation}
		D^{2}\text{\ensuremath{\mathscr{E}}}(g)\cdot\bigg(h,\text{\ensuremath{\mathscr{L}}}_{X}g\bigg)+D\text{\ensuremath{\mathscr{E}}}(g)\cdot\text{\ensuremath{\mathscr{L}}}_{X}h=\text{\ensuremath{\mathscr{L}}}_{X}\bigg(D\text{\ensuremath{\mathscr{E}}}\left(g\right)\cdot h\bigg).
		\label{onsekiz}
	\end{equation}
	In components, and after setting $\lambda=0$, equation (\ref{onyedi}) reads, respectively
	\begin{equation}
		\delta_{X}\left(\text{\ensuremath{\mathscr{E}}}_{\mu\nu}\right)^{(1)}\cdot h=\text{\ensuremath{\mathscr{L}}}_{X}{\text{\ensuremath{\mathscr{E}}}}_{\mu\nu}(\bar{g}),
		\label{birinci}
	\end{equation}
	and equation (\ref{onsekiz}) reads
	\begin{equation}
		\delta_{X}\left(\text{\ensuremath{\mathscr{E}}}_{\mu\nu}\right)^{(2)}\cdot \left[h,h\right]+\left(\text{\ensuremath{\mathscr{E}}}_{\mu\nu}\right)^{(1)}\cdot \text{\ensuremath{\mathscr{L}}}_{X}h=\text{\ensuremath{\mathscr{L}}}_{X}\left(\text{\ensuremath{\mathscr{E}}}_{\mu\nu}\right)^{(1)}\cdot h,
		\label{ikinci}
	\end{equation}
	where $\delta_{X}\left(\text{\ensuremath{\mathscr{E}}}_{\mu\nu}\right)^{(1)}\cdot h$
	denotes the variation of the background tensor $\left(\text{\ensuremath{\mathscr{E}}}_{\mu\nu}\right)^{(1)}\cdot h$
	under the flow of $X$ or under the infinitesimal diffeomorphisms. Since
	${\text{\ensuremath{\mathscr{E}}}}_{\mu\nu}(\bar{g})=0$, (\ref{birinci}) says that
	$\left(\text{\ensuremath{\mathscr{E}}}_{\mu\nu}\right)^{(1)}\cdot h$
	is gauge invariant: $\delta_{X}\left(\text{\ensuremath{\mathscr{E}}}_{\mu\nu}\right)^{(1)}\cdot h=0$.
	Similarly (\ref{ikinci}) yields 
	\begin{equation}
		\delta_{X}\left(\text{\ensuremath{\mathscr{E}}}_{\mu\nu}\right)^{(2)}\cdot \left[h,h\right]+\left(\text{\ensuremath{\mathscr{E}}}_{\mu\nu}\right)^{(1)}\cdot \text{\ensuremath{\mathscr{L}}}_{X}h=0,
		\label{ucuncu}
	\end{equation}
	since $\left(\text{\ensuremath{\mathscr{E}}}_{\mu\nu}\right)^{(1)}\cdot h=0$
	by assumption, the right hand side of (\ref{ikinci}) vanishes. It is worth stressing that since generically  $\left(\text{\ensuremath{\mathscr{E}}}_{\mu\nu}\right)^{(1)}\cdot \text{\ensuremath{\mathscr{L}}}_{X}h$ is not zero,  the second order expansion $\left(\text{\ensuremath{\mathscr{E}}}_{\mu\nu}\right)^{(2)}\cdot \left[h,h\right]$ is not gauge invariant but transforms according to (\ref{ucuncu}).
	Gauge invariance of the Taub conserved quantity and the ADT charge
	follows immediately from (\ref{ucuncu}). Contracting that equation with the Killing vector
	field $\bar{\text{\ensuremath{\xi}}}$ and integrating over the Cauchy
	surface, one finds 
	\begin{equation}
		\int_{\Sigma} d^{3}\Sigma\thinspace\sqrt{\gamma}\thinspace{n^{\nu}}\left[\bar{\text{\ensuremath{\xi}}}^{\mu}\delta_{X}\left(\text{\ensuremath{\mathscr{E}}}_{\mu\nu}\right)^{(2)}\cdot \left[h,h\right]+\bar{\text{\ensuremath{\xi}}}^{\mu}\left(\text{\ensuremath{\mathscr{E}}}_{\mu\nu}\right)^{(1)}\cdot \text{\ensuremath{\mathscr{L}}}_{X}h\right]=0.
	\end{equation}
	Since we have already shown that the second term can be written as a
	divergence we can drop it out, the remaining part is the Taub conserved quantity which is shown to be is gauge invariant, by this construction. The
	above discussion has been for a generic gravity theory based on the metric tensor as the only dynamical field, let us consider Einstein's gravity as an explicit example.
	
	\subsection{Linearization stability in Einstein's gravity}
	
	Let {\it Ein} denote the $(0,2)$ Einstein tensor, and $h$ denote a symmetric
	two tensor field as described above and $X$ be a vector field, then the effect of infinitesimal one-parameter diffeomorphisms generated by $X$ follows as
	\begin{equation}
		DEin(g)\cdot\text{\ensuremath{\mathscr{L}}}_{X}g=\text{\ensuremath{\mathscr{L}}}_{X}Ein(g),
	\end{equation}
	which in local coordinates reads
	\begin{equation}
		\delta_{X}\left(G_{\mu\nu}\right)^{(1)}\cdot h=\text{\ensuremath{\mathscr{L}}}_{X}\bar{G}_{\mu\nu},
	\end{equation}
	where $G_{\mu\nu}:=Ein(e_{\mu},e_{\nu})$ and $Ein:=Ric-\frac{1}{2}Rg.$  We have already given the proof of the above equation for a generic theory in the previous part, but it pays to do it more explicitly in Einstein's theory: so it follows as
	\begin{equation}
		\delta_{X}(G_{\mu\nu})^{(1)}\cdot h=\delta_{X}(R_{\mu\nu})^{(1)}\cdot h-\frac{1}{2}\bar{g}_{\mu\nu}\delta_{X}(R)^{(1)}\cdot h-\frac{1}{2}\bar{R}\delta_{X}h_{\mu\nu},
	\end{equation}
	which just comes from the definition of the linearized Einstein tensor.  Then one can rewrite the above expression as desired: 
	\begin{equation}
		\delta_{X}(G_{\mu\nu})^{(1)}\cdot h=\text{\ensuremath{\mathscr{L}}}_{X}\left(\bar{R}_{\mu\nu}-\frac{1}{2}\bar{g}_{\mu\nu}\bar{R}\right)=\text{\ensuremath{\mathscr{L}}}_{X}\bar{G}_{\mu\nu}.
	\end{equation}
	At the second order of linearization, one has 
	\begin{equation}
		D^{2}Ein(g)\cdot\left(h,\text{\ensuremath{\mathscr{L}}}_{X}g\right)+DEin(g)\cdot\text{\ensuremath{\mathscr{L}}}_{X}h=\text{\ensuremath{\mathscr{L}}}_{X}\bigg(DEin(g)\cdot h\bigg),
	\end{equation}
	whose local version reads 
	\begin{equation}
		\delta_{X}(G_{\mu\nu})^{(2)}\cdot [h,h]+(G_{\mu\nu})^{(1)}\cdot\text{\ensuremath{\mathscr{L}}}_{X}h=\text{\ensuremath{\mathscr{L}}}_{X}(G_{\mu\nu})^{(1)}\cdot h.
	\end{equation}
	The explicit proof of this expression is rather long, hence we relegate it to Appendix A.
	
	Now let us study the linearization stability of a particular solution to Einstein's gravity with a cosmological constant. Let $\bar{g}$ solve
	the cosmological Einstein's field equations then the equation relevant to the  study of linearization stability of this solution is (\ref{main2}) which now reads
	\begin{equation}
		(\text{\ensuremath{\mathcal{G}}}_{\mu\nu})^{(2)}\cdot [h,h]+\left(\text{\ensuremath{\mathcal{G}}}_{\mu\nu}\right)^{(1)}\cdot k=0,
		\label{eins}
	\end{equation}
	where $\left(\text{\ensuremath{\mathcal{G}}}_{\mu\nu}\right)^{(1)}\cdot k $ is a simple object but the  the second order object $(\text{\ensuremath{\mathcal{G}}}_{\mu\nu})^{(2)}\cdot [h,h]$ is quite cumbersome. It is very hard to use this equation to
	show that for a generic background $\bar{g}_{\mu\nu}$, a $k_{\mu \nu}$ can
	be found or cannot be found that satisfy (\ref{eins}). Therefore one actually
	resorts to a weaker (sufficiency) condition that the Taub charges vanish which, as
	we have seen, results from integrating this equation after contracting with
	a Killing vector field $\bar{\text{\ensuremath{\xi}}}^{\mu}.$ To set the
	stage for generic gravity theories about their AdS backgrounds, let
	us study (\ref{eins}) in AdS and flat spaces. In that case one can
	plug an explicit ansatz as follows: assume that such a $k$ exists in the form
	\begin{equation}
		k_{\mu\nu}=a\, h_{\mu\beta}h_{\nu}^{\beta}+b\, hh_{\mu\nu}+\bar{g}_{\mu\nu}(c\, h_{\alpha\beta}^{2}+d\, h^{2}),
		\label{acik}
	\end{equation}
	where $k:=k_{\mu\nu}\bar{g}^{\mu\nu}$ and $a,b,c, d$ are
	constants to  be determined and all the raising and lowering
	is done with the background AdS metric $\bar{g}.$  Here we shall work in $D$ spacetime
	dimensions. Inserting $k_{\mu\nu}$ as given in (\ref{acik}) in $\left(\text{\ensuremath{\mathcal{G}}}_{\mu\nu}\right)^{(1)}\cdot k$,
	and choosing $a=1$ and $b=-\frac{1}{2}$, one arrives at 
	\begin{equation}
		\left(\text{\ensuremath{\mathcal{G}}}_{\mu\nu}\right)^{(2)}\cdot  [h,h] + \left(\text{\ensuremath{\mathcal{G}}}_{\mu\nu}\right)^{(1)}\cdot k  =: K_{\mu\nu},
	\end{equation}
	where $K_{\mu\nu}$ is a tensor which must vanish if the background
	is linearization stable. Its explicit form is worked out in Appendix B. Let us consider the transverse traceless gauge, and make use of the field equations and the linearized field equations: Namely let us
	use $\bar{R}_{\mu\nu}=\frac{2\varLambda}{D-2}\bar{g}_{\mu\nu}$
	and $\left(\text{\ensuremath{\mathcal{G}}}_{\mu\nu}\right)^{(1)}\cdot h=0,$ which in this gauge reads 
	$\bar\square h_{\mu\nu}=\frac{4\Lambda}{(D-1)(D-2)}h_{\mu\nu}$ to arrive at 
	\begin{eqnarray}
		K_{\mu\nu}=\bar{\nabla}_{\alpha}H^{\alpha}\thinspace_{\mu\nu}+\frac{\Lambda}{D-2}\left(c(D-2)+\frac{1}{2}\right)\bar{g}_{\mu\nu}h_{\alpha\beta}^{2}-\frac{1}{4}\bar{\nabla}_{\nu}h^{\alpha\beta}\bar{\nabla}_{\mu}h_{\alpha\beta}\nonumber&  &\\-\frac{\Lambda D}{(D-1)(D-2)}h_{\mu\beta}h_{\nu}^{\beta},
		\label{div1}
	\end{eqnarray}
	where the divergence piece is given as 
	\begin{eqnarray}
		&&H^{\alpha}\thinspace_{\mu\nu}:=-\frac{1}{4}\bar{g}_{\mu\nu}h_{\sigma\beta}\bar{\nabla}^{\beta}h^{\sigma\alpha}+\left(c(2-D)-\frac{1}{2}\right)\delta_{\nu}^{\alpha}h^{\sigma\beta}\bar{\nabla}_{\mu}h_{\sigma\beta}\\&&+\left(c(D-2)+\frac{5}{8}\right)\bar{g}_{\mu\nu}h_{\sigma\beta}\bar{\nabla}^{\alpha}h^{\sigma\beta}\nonumber\\
		&&+\frac{1}{2}\left(h^{\alpha\beta}\bar{\nabla}_{\beta}h_{\nu\mu}+h_{\beta\nu}\bar{\nabla}_{\mu}h^{\alpha\beta}+h_{\beta\mu}\bar{\nabla}_{\nu}h^{\alpha\beta}-h_{\mu\beta}\bar{\nabla}^{\alpha}h_{\nu}^{\beta}-h_{\mu\beta}\bar{\nabla}^{\beta}h_{\nu}^{\alpha}\right).\nonumber
	\end{eqnarray}
	
	In the transverse-traceless gauge, the coefficient $d$ is not fixed
	and can be set to zero. $K_{\mu\nu}$ has a single parameter $c$, that
	one can choose to fix the stability of the flat spacetime (which was
	proven by \cite{Choquet-Bruhat} using the linearization of the
	constraints on a non-compact Cauchy surface in Minkowski space). Before
	looking at the flat space case, let us note that one has $\bar{\nabla}_{\mu}K^{\mu\nu}=0$
	as expected. Let us consider the flat space with $\Lambda=0$ and
	use the Cartesian coordinates so that $\bar{\nabla}_{\alpha}\rightarrow\partial_{\alpha}.$
	The corresponding linearized field equations become
	\begin{equation}
		\partial^{2}h_{\mu\nu}=0,
		\label{basit}
	\end{equation}
	together with the gauge choices $\partial_{\mu}h^{\mu\nu}=0=h.$ The
	general solution of (\ref{basit}) can be exactly constructed as a superposition
	of plane-wave solutions, hence it suffices to study the linearized
	stability of flat space against the plane-wave modes which we take
	to be the real part of 
	\begin{equation}
		h_{\mu\nu}=\text{\ensuremath{\varepsilon}}_{\mu\nu}e^{ik\cdot x},
		\label{plane}
	\end{equation}
	together with $k^{\mu}\text{\ensuremath{\varepsilon}}_{\mu\nu}=0,$
	$\text{\ensuremath{\varepsilon}}_{\mu}^{\mu}=0$ and $k^{2}=0$, which follow from the gauge condition and (\ref{basit}). In a compact space without a boundary, $k=0$ mode should also be considered, in that case one has the solution $h_{\mu \nu} =\varepsilon_{\mu \nu} (c_1 t +c_2)$  which gives rise to linearization instability \cite{Higuchi} for the case of the torus. Evaluating $K_{\mu\nu}$ for
	the solution \ref{plane}, one arrives at 
	\begin{equation}
		K_{\mu\nu}=k_{\nu}k_{\mu}\text{\ensuremath{\varepsilon}}_{\alpha\beta}\text{\ensuremath{\varepsilon}}^{\alpha\beta}e^{ik\cdot x}\left(2c(D-2)+\frac{5}{4}\right),
	\end{equation}
	which vanishes for the choice 
	\begin{equation}
		c=-\frac{5}{8(D-2)}\,.
	\end{equation}
	So  (\ref{eins}) is satisfied for 
	\begin{equation}
		k_{\mu\nu}=h_{\mu\beta}h_{\nu}^{\beta}-\frac{5}{8(D-2)}\bar{g}_{\mu\nu}h_{\alpha\beta}^{2}
	\end{equation}
	and therefore there is no further constraint on the linearized solutions (\ref{plane})
	and the Minkowski space is linearization stable. Next we move on to quadratic gravity theory.

	\subsection{Linearization instability beyond Einstein's theory}
	
	One of the reasons that lead us to study the linearization instability
	in generic gravity theories is an observation made in \cite{Deser_Tekin} where conserved charges of generic gravity theories for asymptotically AdS backgrounds were constructed.\footnote{For an earlier zero energy result in the context of asymptotically flat backgrounds for purely quadratic gravity in four dimensions, see \cite{zero_en}.}The observation was  that in AdS backgrounds, the conserved energy and angular momenta vanish in generic 
	gravity theories for all asymptotically AdS solutions at some particular
	values of the parameters defining the theory (in fact a whole section in that chapter was devoted for the zero energy issue). 
	This apparent infinite degeneracy of the vacuum for AdS spaces, is in sharp contrast to the flat space case where the unique zero energy is attained only by the Minkowski space, namely the classical ground state. Let us expound upon this a little more: for ${\it all}$ purely metric based
	theories, the energy (mass) of the space-time that asymptotically
	approaches the flat space at spatial infinity is given by the ADM
	formula \cite{adm}
	\begin{equation}
		M_{ADM}=\frac{1}{\kappa}\,\oint_{\partial \Sigma}\,dS_{i}\,(\partial_{j}h^{ij}-\partial^{i}h^{j}\,_{j})\,.\label{admass}
	\end{equation}
	It is well-known that $M_{ADM}\ge0$,
	which is known as the positive energy theorem \cite{Schoen,Witten}.
	An important part of this theorem is that the vacuum, namely the
	flat space-time with $M_{ADM}=0$, is unique (up to diffeomorphisms
	of course) \cite{Brill_Deser,Brill_Deser_Fad}.
	It should be also noted that, the ADM mass is defined in flat Cartesian
	coordinates but it was shown to be coordinate invariant. Here
	one must be very careful, if proper decaying conditions are not realized
	for $h_{ij}$, ${\it any}$ (positive, negative, finite or divergent)
	value of mass can be assigned to the flat space. It is exactly these
	properties of the ADM formula that made it a useful tool in geometry:
	without even referring to Einstein's equations, one can take (\ref{admass})
	to be a geometric invariant of an asymptotically flat manifold, modulo some decaying conditions on the first and the second fundamental forms of the spacelike surface. 
	
	Once one deviates from  asymptotic flatness, then as we have noted, for higher derivative theories 
	there are critical points which seem to make the vacuum infinitely degenerate, namely, the corresponding mass formula assigns any solution of the theory  the same zero charge.  Naively, one can try to understand the meaning of vanishing charges
	for non-vacuum solutions (namely, non-maximally symmetric solutions)
	as follows: 
	\begin{itemize}
		
		\begin{item}
			There is a confinement of the relevant perturbations (in the weak coupling), just-like
			in QCD in the strong coupling of color charge; and so a non-vacuum solution such as the proton has zero total color charge, same as the vacuum.  In the case of QCD, perturbation theory might yield spurious states
			that cannot freely exist, such as quarks, as also noted in \cite{Strom2}. In gravity confinement would mean, confinement of mass-energy or some other properties under consideration such as chirality. But this would be highly unphysical because if there are no other conserved charges to suppress the creation of confined mass, then the vacuum state of gravity would be infinitely degenerate and creating confined mass would cost nothing.  
		\end{item}
		
		\begin{item}
			
			The second possibility is that perturbation theory about a given
			background solution, be it the maximally symmetric vacuum or not, may simply fail to exist
			just because the background solution is an isolated solution in the
			solution space. Namely, the solution space may fail to be a smooth
			manifold. 
		\end{item}
		
	\end{itemize}
	
	In fact, as discussed above, linearization
	of non-linear equations such as Einstein's gravity and Yang-Mill's
	theory showed that naive first order perturbation theory fails generically
	when the background has  a Killing symmetry. To be more specific we 
	consider two recent examples: the chiral gravity in 2+1 dimensions
	which is a special case of topologically massive gravity with a cosmological
	constant and the critical gravity which is a specific example of quadratic
	gravity in AdS. These examples can be easily extended, as the phenomenon
	we discuss is quite generic and take place whenever Einstein's theory
	with a cosmological constant is modified with some curvature terms.
	
	To see how perturbation theory can fail let us go back to the necessary condition (\ref{main2}) and contract it with the Killing vector $\bar{\text{\ensuremath{\xi}}}^{\mu}$ to obtain
	\begin{equation}
		\bar{\text{\ensuremath{\xi}}}^{\mu}\left(\text{\ensuremath{\mathscr{E}}}_{\mu\nu}\right)^{(2)}\cdot \left[h,h\right]+\bar{\text{\ensuremath{\xi}}}^{\mu}\left(\text{\ensuremath{\mathscr{E}}}_{\mu\nu}\right)^{(1)}\cdot k=0.
	\end{equation}
	In some modified gravity theories one finds that the second term can be written as 
	\begin{equation}
		\bar{\text{\ensuremath{\xi}}}^{\mu}\left(\text{\ensuremath{\mathscr{E}}}_{\mu\nu}\right)^{(1)}\cdot k=c(\alpha_i, \bar{R}) \bar{\nabla}_{\alpha}{\mathcal{F}}_{1}^{\alpha}\thinspace_{\nu}+\bar{\nabla}_{\alpha}{\mathcal{F}}_{2}^{\alpha}\thinspace_{\nu},
		\label{bound}
	\end{equation}
	where $c(\alpha_i, \bar{R}) $ is a constant  determined by the parameters $\alpha_i$ of the theory as well as the curvature invariants (symbolically denoted above as $\bar{R}$) of the background metric.  ${\mathcal{F}}_{i}^{\alpha\nu}$ are antisymmetric background tensors. It turns out that for asymptotically AdS spacetimes ${\mathcal{F}}_{2}^{\alpha\nu}$ vanishes identically at the boundary as it involves higher derivative terms of the perturbation, while ${\mathcal {F}}_{1}^{\alpha\nu}$ need not if there are not so fast decaying fields such as for example the Kerr-AdS black holes. On the other hand for the particular choice of the parameters $c(\alpha_i, \bar{R}) =0$, one arrives at the constraint that again the Taub charges must  vanish identically
	\begin{equation}
		Q_{Taub}[\bar{\text{\ensuremath{\xi}}}]=\oint_{\Sigma} d^{D-1}\thinspace\Sigma\thinspace\sqrt{\gamma}\thinspace\bar{\text{\ensuremath{\xi}}}^{\mu}(\text{\ensuremath{\mathscr{E}}}_{\mu\nu})^{(2)}\cdot [h,h]=0.
		\label{nine1}
	\end{equation}
	But this time we have the additional non-trivial equation
	\begin{equation}
		\oint_{\Sigma} d^{D-1}\thinspace\Sigma\thinspace\sqrt{\gamma}\thinspace\bar{\text{\ensuremath{\xi}}}^{\mu}\left(\text{\ensuremath{\mathscr{E}}}_{\mu\nu}\right)^{(1)}\cdot h\neq0.
		\label{nine2}
	\end{equation}
	In general it is very hard to satisfy these two conditions simultaneously for all solutions. Therefore some solutions to the linearized equations $h$ turn out to be not integrable to a full solution, hence the linearization instability of the AdS background in these critical theories. Let us stress that we have not assumed that the Cauchy surfaces are compact: this type of linearization instability arises even in the non-compact  case.

	\subsection{Linearization instability in quadratic gravity}
	
	The message we would like to convey is a rather universal one in all
	generic higher derivative gravity theories, but for the sake of being
	concrete and yet sufficiently general, we shall consider the quadratic
	gravity theory with the action (in $D$ dimensions) 
	\begin{eqnarray}
		I=\int d^{D}\, x\sqrt{-g}\Big (\frac{1}{\kappa}(R -2 \Lambda_0)+\alpha R^{2}+\beta R_{\mu\nu}^{2}+\gamma(R_{\mu\nu\rho\sigma}^{2}-4R_{\mu\nu}^{2}+R^{2})\Big ),\thinspace\thinspace\thinspace\thinspace\label{action}
	\end{eqnarray}
	where the last term is organized into the Gauss-Bonnet form, which
	vanishes identically for $D=3$ and becomes a surface term for $D=4$.
	But for $D\ge5$, it contributes to the field equations with at most
	second order derivatives in the metric, just like the Einstein-Hilbert
	part. Conserved gravitational charges of this theory  in its asymptotically AdS backgrounds were constructed
	in \cite{Deser_Tekin} following the background space-time
	techniques developed in \cite{Abbott_Deser} which is an extension of the ADM approach \cite{adm}. For any theory
	with a Lagrangian density ${\cal {L}}=\frac{1}{\kappa}(R-2\Lambda_{0})+f(R_{\sigma\rho}^{\mu\nu})$,
	for a generic differentiable function $f$ of the Riemann tensor and its contractions, the conserved charges follow
	from those of (\ref{action}), as shown in \cite{SST} since any such theory can be written as a quadratic theory with effective coupling constants as far as its energy properties and particle content are concerned \cite{Tekin_particle}. In what follows, we quote some of the computations done in \cite{Deser_Tekin} here to make the ensuing discussion complete. The field equations  that follow from (\ref{action}) are 
	\begin{eqnarray}
		{\mathcal{E}}_{\mu\nu}[g]=\frac{1}{\kappa}(R_{\mu\nu}-\frac{1}{2}g_{\mu\nu}R)+2\alpha R\,(R_{\mu\nu}-\frac{1}{4}g_{\mu\nu}\,R)\nonumber\\+(2\alpha+\beta)(g_{\mu\nu}\Box-\nabla_{\mu}\nabla_{\nu})R+2\gamma\Big\{ RR_{\mu\nu}-2R_{\mu\sigma\nu\rho}R^{\sigma\rho}\nonumber\\+R_{\mu\sigma\rho\tau}R_{\nu}^{\sigma\rho\tau}-2R_{\mu\sigma}R_{\nu}^{\sigma}-\frac{1}{4}g_{\mu\nu}(R_{\tau\lambda\rho\sigma}^{2}-4R_{\sigma\rho}^{2}+R^{2})\Big\}\nonumber\\+\beta\Box(R_{\mu\nu}-\frac{1}{2}g_{\mu\nu}R)+2\beta(R_{\mu\sigma\nu\rho}-\frac{1}{4}g_{\mu\nu}R_{\sigma\rho})R^{\sigma\rho}=0.\label{eom}
	\end{eqnarray}
	As we shall study the stability/instability of the non-flat maximally symmetric solution (or solutions), let 
	$\bar{g}$ represent such a solution with the curvature  tensors normalized as
	\begin{equation}
		\bar{R}_{\mu \rho \nu \sigma}  = \frac{ 2 \Lambda}{ (D-1)(D-2)} \big ( \bar{g}_{\mu \nu}  \bar{g}_{\rho \sigma} -   \bar{g}_{\mu \sigma}  \bar{g}_{\rho \nu} \big ), 
	\end{equation}
	\begin{equation}
		\bar{R}_{\mu \nu} = \frac{2 \Lambda}{D-2} \bar{g}_{\mu \nu},
	\end{equation}
	\begin{equation}
		\bar{R} = \frac{ 2 D \Lambda}{ D-2}.
	\end{equation}
	
	The field equations reduce to a single quadratic equation :
	\begin{equation}
		\frac{\Lambda-\Lambda_{0}}{2\kappa}+k \Lambda^{2}=0,\qquad k \equiv\left(D\alpha+\beta\right)\frac{\left(D-4\right)}{\left(D-2\right)^{2}}+\gamma\frac{\left(D-3\right)\left(D-4\right)}{\left(D-1\right)\left(D-2\right)}.\label{quadratic}
	\end{equation}
	For generic values of the parameters of the theory, of course, there may not be real solution and so the theory may not posses a maximally symmetric vacuum, but here we assume that there is a real solution to this algebraic equation (so $ 8 \Lambda_0 k \kappa +1 \ge 0$)  and study the linearization stability of this solution, which we call the (classical) vacuum or the background.  One can then linearize the field equations (\ref{eom}) about the vacuum and get at the linear order
	\begin{eqnarray}
		c_1\,({\cal {G}}_{\mu\nu})^{(1)}+\left(2\alpha+\beta\right)\left(\bar{g}_{\mu\nu}\bar{\square}-\bar{\nabla}_{\mu}\bar{\nabla}_{\nu}+\frac{2\Lambda}{D-2}\bar{g}_{\mu\nu}\right)(R)^{(1)}\nonumber\\+\beta\left(\bar{\square}(\mathcal{G}_{\mu\nu})^{(1)}-\frac{2\Lambda}{D-1}\bar{g}_{\mu\nu}(R)^{(1)}\right)=0,\label{Linearized_eom}
	\end{eqnarray}
	where the constant in front of the first term is 
	\begin{equation}
		c_1\equiv\frac{1}{\kappa}+\frac{4\Lambda D}{D-2}\alpha+\frac{4\Lambda}{D-1}\beta+\frac{4\Lambda\left(D-3\right)\left(D-4\right)}{\left(D-1\right)\left(D-2\right)}\gamma,\label{eqc}\end{equation}
	and the linearized (background) tensors read
	\begin{equation}
		({\cal {G}}_{\mu\nu})^{(1)}=(R_{\mu\nu})^{(1)}-\frac{1}{2}\bar{g}_{\mu\nu}(R)^{(1)}-\frac{2\Lambda}{D-2}h_{\mu\nu},
	\end{equation}
	which is just the linearized cosmological Einstein's tensor given in terms of the linearized Ricci tensor and the linearized scalar curvature :
	\begin{eqnarray}
		(R_{\mu\nu})^{(1)}=\frac{1}{2}\Big (\bar{\nabla}^{\sigma}\bar{\nabla}_{\mu}h_{\nu\sigma}+\bar{\nabla}^{\sigma}\bar{\nabla}_{\nu}h_{\mu\sigma}-\bar{\square}h_{\mu\nu}-\bar{\nabla}_{\mu}\bar{\nabla}_{\nu}h\Big),\nonumber\\\qquad (R)^{(1)}=-\bar{\square}h+\bar{\nabla}^{\sigma}\bar{\nabla}^{\mu}h_{\sigma\mu}-\frac{2\Lambda}{D-2}h.
	\end{eqnarray}
	
	Given a background Killing vector $\bar{\xi}$, (there are $D(D+1)/2$ number of Killing vectors for this space and the arguments work for any one of these) if we had not truncated the expansion of the field equations at ${\mathcal{O}(h)}$ but collected all the non-linear terms on the right-hand side, we would have gotten 
	\begin{equation}
		\bar{\text{\ensuremath{\xi}}}^{\mu}\left(\text{\ensuremath{\mathscr{E}}}_{\mu\nu}\right)^{(1)}\cdot h := \bar{\xi}^\mu T_{\mu \nu}[h^2, h^3, ... h^n...].
		\label{lin}
	\end{equation}
	where  $T_{\mu \nu}[h^2, h^3, ...h^n...] $ represents all the higher order terms (and if there is a matter source with compact support of energy-momentum tensor, it also includes that).  The next step is the crucial step: as was shown in \cite{Deser_Tekin}, one can write (\ref{lin}) as a divergence of two pieces as described by (\ref{bound})
	\begin{equation} 
		\bar{\text{\ensuremath{\xi}}}^{\mu}\left(\text{\ensuremath{\mathscr{E}}}_{\mu\nu}\right)^{(1)}\cdot h=c \, \bar{\nabla}_{\alpha}{\mathcal{F}}_{1}^{\alpha}\thinspace_{\nu}+\bar{\nabla}_{\alpha}{\mathcal{F}}_{2}^{\alpha}\thinspace_{\nu},
		\label{bound2}
	\end{equation}
	where the constant   $c_1$  given in (\ref{eqc}) is shifted due to the $\beta$ term as 
	\begin{equation}
		c \equiv  c_1 + \frac{4\Lambda}{(D-1)(D-2)}\beta.
	\end{equation}
	The explicit forms of the ${\mathcal F}_i^{\mu \rho}$ tensors are found to be 
	\begin{eqnarray}
		{\mathcal F}_1^{\mu \rho}= 2 \bar{\xi}_\nu \bar{\nabla}^{ [ \mu}h^{\rho ]\, \nu} 
		+2 \bar{\xi}^{ [ \mu } \bar \nabla^{ \rho ] } h +2  h^{\nu [ \mu}\bar{\nabla}^{\rho ]} \bar{\xi}_\nu +2 \bar{\xi}^{[ \rho} \bar{\nabla}_{\nu}h^{\mu  ] \, \nu} +
		h\bar{\nabla}^\mu \bar{\xi}^\rho,
		\label{f1}
	\end{eqnarray}
	and
	\begin{eqnarray}
		{\mathcal{F}}_{2}^{\mu \rho} =(2\alpha+\beta)\Bigg(2 \bar{\xi}^{[\mu}\bar{\nabla}^{\rho]}(R)^{(1)}+(R)^{(1)}\bar{\nabla}^{\mu}\,\bar{\xi}^{\rho}\Bigg)\nonumber\\
		+2\beta\Bigg(\bar{\xi}^{\sigma}\bar{\nabla}^{[\rho}({\cal {G}}^{\mu]}\,_{\sigma})^{(1)}+({\cal {G}}^{[\rho\sigma})^{(1)}\bar{\nabla}^{\mu ]}\bar{\xi}_{\sigma}\Bigg).
	\end{eqnarray}
	For asymptotically AdS spacetimes, ${\mathcal{F}}_{2}^{\mu \rho} $ vanishes at spatial infinity  due to the vanishing of both of $ (R)^{(1)}$ and $({\cal {G}}_{\mu\sigma})^{(1)}$. As discussed in the previous section, vanishing of the constant $c$ leads to two strong constraints (\ref{nine1}) and (\ref{nine2}) on the linearized solution $h$ which is a statement of the instability of the background solution.  
	Note that, for this higher order theory, we have not assumed that the spatial hypersurface is compact. (In fact, to be more accurate, AdS is not globally hyperbolic and does not  have a Cauchy surface but one can work in the double cover which does). 
	
	The point at which $c=0$ is the point when the mass of the spin-2 massive mode also vanishes and further, assuming $4\alpha(D-1) + D\beta=0$, one can also decouple the massive spin-0 mode in this theory and arrive at the so called {\it critical gravity } defined in $D=4$  \cite{Pope} for generic $D$ in \cite{Tahsin}.
	All these conditions are compatible with the existence of a maximally symmetric vacuum.  
	For critical gravity, the apparent  mass and angular momenta of all black holes and perturbative excitations with asymptotically AdS conditions vanish.\footnote{The energy of the perturbative bulk excitations can be constructed using the Ostrogradsky Hamiltonian \cite{Tahsin}.} But as we have seen here, perturbation theory used for both the excitations and construction of conserved quantities does not work exactly at the critical point: namely, the theory for the AdS background is not linearization stable. At the chiral point, there arise exact log-modes in chiral gravity \cite{Alisah, GGST} which are of the wave type but they do not correspond to the linearized log-modes of \cite{Grumiller}.
	
	Just for the sake of completeness, let us note that if $c \ne 0$, then  the perturbation theory makes sense and the conserved charges of the theory for any asymptotically AdS solutions (such as the Kerr-AdS black holes) are simply given in terms of the conserved charges of the same solution in Einstein's gravity as 
	\begin{align}
		\frac{Q_{\text{quad}}(\bar{\xi})}{Q_{\text{Einstein}}(\bar{\xi})} = -\beta m_g^2 ,\label{Quad_charge}
	\end{align}
	where $m_g$ is the mass of the spin-2 graviton given as 
	\begin{equation}
		-\beta m_g^2=\frac{1}{\kappa}+\frac{4\Lambda(D\alpha+\beta)}{D-2}+\frac{4\Lambda\left(D-3\right)\left(D-4\right)}{\left(D-1\right)\left(D-2\right)}\gamma.
	\end{equation}
	In (\ref{Quad_charge}), $Q_{\text{Einstein}}(\bar{\xi})$ refers to (with $\kappa_{\text{Newton}}=1$)
	the conserved charge (mass, angular momenta) in the cosmological
	Einstein's theory.
	
	\subsection{Linearization instability in chiral gravity}
	
	A model of quantum gravity even in the simpler $2+1$ dimensional setting has been rather elusive. One of the 
	latest promising proposals was the so called chiral gravity \cite{Strom1} which is a specific limit of topologically massive gravity (TMG) \cite{djt} with the asymptotically AdS boundary conditions. TMG, as opposed to Einstein's gravity has non-trivial local dynamics hence in this respect, it might be more relevant  to the four dimensional gravity both at the classical and quantum level. The crux of the arguments of the quantum version chiral gravity is that the bulk theory is dual to a unitary and chiral conformal field theory (CFT) on the two dimensional boundary, whose symmetry is known to be one of the two copies of the Virasoro algebra {\cite{BH}.  Finding the correct conformal field theory would amount to defining the quantum gravity via the AdS/CFT duality \cite{Maldacena}. But immediately after the proposal of chiral gravity, it was realized that the theory has arbitrarily negative energy log modes that appear exactly at the chiral point and not only the dual CFT is not unitary (but a logarithmic one), but apparently  chiral gravity does not have even a classical vacuum \cite{Grumiller}.  If true, this of course would be disastrous  for chiral gravity. But later it was argued in \cite{Strom2,Carlip} that chiral gravity has  linearization instability against these log modes in AdS: namely, these perturbative negative energy solutions do not actually come from the linearization of any exact solution. If that is the case, then linearization instability saves chiral gravity certainly at the classical level and perhaps at the quantum level. Here we give further arguments of the existence of linearization  
		instability in chiral gravity. 
		
		The field equations of topologically massive gravity \cite{djt} with a negative cosmological constant ($\Lambda := -\frac{1}{\ell^2}$) is 
		\begin{equation}
			R_{\mu\nu}-\frac{1}{2}g_{\mu\nu}R -\frac{1}{\ell^2} g_{\mu\nu}
			+\frac{1}{\mu}C_{\mu\nu}=0,
			\label{fieldeqns1}
		\end{equation}
		where  the Cotton tensor in terms of the anti-symmetric tensor and the covariant derivative of the Schouten tensor reads
		\begin{equation}
			C_{\mu\nu}=\eta_{\mu}\,^{\alpha\beta}\nabla_{\alpha}S_{\beta\nu}, \hskip 0.5 cm S_{\mu\nu}=R_{\mu\nu}-\frac{1}{4}g_{\mu\nu}R.
		\end{equation}
		The boundary theory has two copies of the Virasoro algebra \cite{BH} for asymptotically AdS  boundary conditions given as 
		\begin{equation}
			c_{R/L}=\frac{3\text{\ensuremath{l}}}{2G_{3}}\left(1 \pm\frac{1}{\mu\text{\ensuremath{l}}}\right), 
			\label{central}
		\end{equation}
		and the bulk theory has a single helicity 2 mode with a mass-square 
		\begin{equation}
			m_{g}^{2}=\mu^{2}-\frac{1}{l^{2}}.
		\end{equation}
		It was shown in \cite{adt} that the contraction of the Killing vector $(\bar{\xi})$ with the linearized equations coming from (\ref{fieldeqns1}) yields 
		\begin{equation}
			\bar{\xi}^\mu\Bigg (({\mathcal{G}}_{\mu\nu})^{(1)}+\frac{1}{\mu}(C_{\mu\nu})^{(1)} \Bigg) 
			= \, \bar{\nabla}_{\alpha}{\mathcal{F}}_{1}^{\alpha}\thinspace_{\nu} [\bar{\Xi}]+\bar{\nabla}_{\alpha}{\mathcal{F}}_{3}^{\alpha}\thinspace_{\nu}[\bar \xi],
			\label{lineareq}
		\end{equation}
		where ${\mathcal{F}}^{\mu \rho}_1$ was given in (\ref{f1}) whereas one finds ${\mathcal{F}}^{\mu \rho}_3$ to be
		\begin{eqnarray}
			{\mathcal{F}}^{\mu \rho}_3 [\bar{\xi}] = 
			\eta^{\mu  \rho \beta} \, ({\mathcal G}_{\nu\beta})^{(1)} \, \bar{\xi}^{\nu}
			+ \eta^{\nu \rho \beta} \, ({\mathcal G}^{\mu}\,_{\beta})^{(1)} \, \bar{\xi}_{\nu}
			+ \eta^{\mu\nu\beta} \, ({\mathcal G}^{\rho}\,_{\beta})^{(1)} \, \bar{\xi}_{\nu},
			\label{f3}
		\end{eqnarray}
		where  a new (twisted) Killing vector ($\bar{\Xi}$) appears:
		\begin{equation}
			\bar{\Xi}^{\alpha} := \bar \xi^\alpha+ \frac{1}{2 \mu}\eta^{ \alpha \beta\nu} \, 
			\bar{\nabla}_{\beta} \, \bar{\xi}_{\nu}.
		\end{equation}
		The conserved charges of TMG for asymptotically AdS backgrounds read as an integral over the circle at infinity as
		\begin{equation}
			Q [\bar{\xi}] = \frac{1}{8 \pi G_3} \, \oint_{\partial {\cal M}} \,
			dS_{i} \, \left( {\mathcal{F}}^{0 i}_1 [\bar{\Xi}] + \frac{1}{2 \mu} \,{\mathcal{F}}^{0 i}_3 [\bar{\xi}]
			\right) .
		\end{equation}
		
		Once again for the asymptotically  AdS cases ${\mathcal{F}}_{3}^{\alpha}\thinspace_{\nu}[\bar \xi]$ vanishes identically on the boundary as it involves the linearized Einstein tensor at infinity. For generic values of $\mu$ and $\ell$, the first term, that is $ {\mathcal{F}}_{1}^{\alpha}\thinspace_{\nu} [\bar{\Xi}]$ gives the conserved charges for the corresponding Killing vector. But, for $\mu^2 \ell^2 =1$, as was shown in \cite{Serkay} the angular momentum and the energy of the rotating black hole solutions with the rotation parameter ($j$) and the mass ($m$) related as ($j = m \ell$) (the  extremal BTZ black hole) vanishes identically. This particular point was further studied in \cite{Strom1} where it was argued and conjectured that the theory, so called {\it chiral} gravity, as one of the central charges noted above (\ref{central}) becomes zero, makes sense both classically and quantum mechanically. 
		
		Classically the theory should have a stable vacuum and  quantum mechanically, it should have a  dual healthy boundary conformal field theory. In \cite{Strom1} it was shown that all the bulk excitations have vanishing energy exactly at the chiral point. Later new log modes that were not accounted for were found in \cite{Grumiller} which violated the existence of a ground state (namely, these modes have arbitrarily large negative energy compared to the zero energy of the vacuum).  For further work on chiral gravity, see \cite{Wise,Porrati}.
		In \cite{Strom2} and \cite{Carlip} it was argued that the AdS has linearization instability in chiral gravity against these log modes. Here, our construction lends support to these arguments.
		
		For the sake of concreteness, let us consider the background metric as
		\begin{equation}
			\bar{g} = - \bigg(1+\frac{r^2}{\ell^2}\bigg) dt^2 + \frac{ dr^2}{1+\frac{r^2}{\ell^2}}+ r^2 d\phi^2,
		\end{equation} 
		then for $\bar{\xi} = ( -1, 0, 0)$ referring to the time-like energy Killing vector, one finds the twisted Killing vector to be
		\begin{equation}
			\Xi =  ( -1, 0,  - \frac{1}{\ell^2 \mu}).
		\end{equation} 
		For this $\Xi$ to be a time-like Killing vector for {\it all} $r$ including the boundary at $r \rightarrow \infty$, one can see that (excluding the trivial $\mu \rightarrow \infty$ case) one must set $\mu^2 \ell^2 =1$, which is the chiral gravity limit. To further see this chiral gravity limit, let us recast  ${\mathcal{F}}^{\mu \rho}_1 [\Xi]$ using the  the superpotential ${\mathcal{K}}^{\mu\alpha\nu\beta}$ is defined by \cite{Abbott_Deser}
		\begin{eqnarray}
			{\mathcal{K}}^{\mu\nu\alpha\beta} :=\frac{1}{2}\big(\bar{g}^{\mu\beta}\tilde h^{\nu\alpha}+\bar{g}^{\nu\alpha}\tilde h^{\mu\beta}-\bar{g}^{\mu\nu}\tilde h^{\alpha\beta}-\bar{g}^{\alpha\beta}\tilde h^{\mu\nu}\big),\hskip0.3cm\tilde h^{\mu\nu} :=h^{\mu\nu}-\frac{1}{2}\bar{g}^{\mu\nu}h,\thinspace\thinspace\thinspace\thinspace
		\end{eqnarray}
		which yields 
		\begin{equation}
			{\mathcal{F}}^{\mu \rho}_1 [\Xi] =  \Xi_\nu \bar{\nabla}_{\beta} {\mathcal{K}}^{\mu \rho \nu \beta}-
			{\mathcal{K}}^{ \mu \sigma \nu \rho }\bar{\nabla}_{\sigma}\bar{\Xi}_\nu.
		\end{equation}
		For all asymptotically AdS solutions with the Brown-Henneaux boundary conditions, one can show that  
		\begin{equation}
			{\mathcal{F}}^{\mu \rho}_1 [\Xi] = \bigg(1- \frac{1}{\ell^2 \mu^2}\bigg){\mathcal{F}}^{\mu \rho}_1 [\bar{\xi}],
		\end{equation}
		which vanishes at the chiral point. So exactly at this point, there exists second order integral constraints on the linearized solutions as discussed in the previous section. The log-modes of \cite{Grumiller} do not satisfy these integral constraints and so fail to be integrable to full solutions.\footnote{See \cite{log}  a nice compilation of possible applications  of logarithmic field theories in the context of holography and gravity. }   
		
		Let us compute the value of the Taub conserved quantity for the log solution which was given in the background with the global coordinates for which the  metric reads
		\begin{equation}
			ds^2 = \ell^2\big(-\cosh^2{\rho}\, d\tau^2 +\sinh^2{\rho}\,d\phi^2+d\rho^2\big).
		\end{equation}
		For the coordinates $u=\tau+\phi$, $v=\tau-\phi$, at exactly in the chiral point, one has the following additional solution 
		\begin{eqnarray}
			h_{ \mu\nu} &=&	-\tanh^2\!\!{\rho}\,\big(\sin{(2u)}\,\tau+\cos{(2u)}\,\ln{\cosh{\rho}}\big)\left( \begin{array}{ccl}
				1 & 1 & \qquad \qquad 0 \\
				1 & 1 & \qquad \qquad 0 \\
				0 & 0 & -4 \sinh^{-2}{2 \rho}
			\end{array} \right)_{\mu\nu}  \nonumber \\ &&+\frac{\sinh{\rho}}{\cosh^3{\rho}}\,\big(\cos{(2u)}\,\tau-\sin{(2u)}\,\ln{\cosh{\rho}}\big)\left( \begin{array}{ccc}
				0 & 0 & 1 \\
				0 & 0 & 1 \\
				1 & 1 & 0
			\end{array} \right)_{\mu\nu}
			\label{grumillermetric}
		\end{eqnarray}
		
		Considering the Killing vector $\bar{\xi} = (-1,0,0)$ one finds the result of the integral in (\ref{ttt}) to be  non-vanishing 
		\begin{equation}
			Q_{Taub}\left[\bar{\text{\ensuremath{\xi}}}\right] = \frac{\pi}{2 \ell} \bigg( 3 \tau^2 - \frac{ 161}{72} \bigg),
		\end{equation}
		which shows that this log mode is not in the tangent space of the solution space of chiral gravity around the $AdS_3$ metric.

		In this chapter, we have shown that at certain {\it critical parameter } values of extended gravity theories in constant curvature backgrounds, perturbation theory fails. Our arguments provide  support to the discussion given by \cite{Strom2, Carlip} regarding the linearization instability in three dimensional chiral gravity and extend the discussion to generic gravity theories in a somewhat former form. The crucial point is that even in spacetimes with non-compact Cauchy surfaces, linearization instability can exist for background metrics with at least one Killing vector field.  Our computation also sheds light on the earlier observations \cite{Deser_Tekin} that at certain critical values of the parameters defining the theory, conserved charges of all solutions, such as black holes, excitations  vanish identically.\footnote{For a recent review of conserved charges in generic gravity theories see the book \cite{kitap}.} For example,  Kerr-AdS black hole metrics  have the same mass and angular momentum as the AdS background.  This leads to a rather non-physical infinite degeneracy of the vacuum: for example, creating back holes costs nothing which is unacceptable. With our discussion above, it is now clear that,  perturbation theory which is used to define boundary integrals of the conserved Killing charges does not make sense exactly at the critical values of the parameters. Therefore  one really needs a new method to find/define conserved charges in these  theories at their critical points. One such method was proposed in  for quadratic theories \cite{Deser_Tekin_2007} and in \cite{sezginefendi} for TMG. 
		
		We must note that, for asymptotically flat spacetimes, the ADM mass is the correct definition of mass-energy for any metric-based theory of gravity. Therefore, the stability of the Minkowski space as was shown for Einstein's theory by Choquet-Bruhat and Deser \cite{Choquet-Bruhat}  is valid for all higher derivative models as long as one considers the non-compact  Cauchy surfaces and asymptotically flat boundary conditions.  But once a cosmological  constant is introduced, the problem changes dramatically as we have shown: the ADM mass-energy (or angular momentum) expressions are modified and conserved charges get contributions from each covariant  tensors added to the field equations. Once such a construction is understood, it is clear that 
		some theories will have identically vanishing charges for all solutions with some fixed boundary conditions, which is a signal of linearization instability.  
		
		It is also important to realize that, linearization instability of certain background solutions in some theories is not bad as it sounds: for example chiral gravity is a candidate   both as a non-trivial classical and quantum gravity theory in $AdS_3$ with a two dimensional chiral conformal field theory induced on the boundary. But it has log-mode solutions which appear as ghosts in the classical theory and negative norm states in the quantum theory.  It just turns out that chiral gravity in $AdS_3$ has linearization instability along these log-modes: namely, they do not have vanishing Taub conserved quantities which is a constraint for all integrable solutions. 
		Therefore, they cannot come from linearization of exact solutions.  A similar phenomenon takes place for the minimal massive gravity \cite{Bergshoeff} which was proposed as a possible solution to the bulk-boundary unitarity clash in three dimensional gravity theories and as a viable model that has a healthy dual conformal field theory on the boundary of $AdS_3$. It was shown recently in \cite{Ercan} that this theory  only makes sense at the chiral point \cite{Tekin, mom} and hence linearization instability arises at that point which can save the theory from its log-modes. Let us note that we have also computed the second order constraint in the minimal massive gravity, namely the Taub conserved quantity and found that it is non-vanishing. In the discussion of linearization stability and instability of a given exact solution in the context of general relativity, we noted that to make use of the powerful techniques of elliptic operator theory, on rewrites the 
		four dimensional Einstein's theory as a dynamical system with constraints on a spacelike Cauchy surface and  
		the evolution equations.  As the constraints are intact, initial Cauchy data uniquely defines a spacetime 
		(modulo some technical assumptions). Therefore, to study the linearization stability one can simply study the linearization stability of the constraints on the surface where the metric  tensor field is positive-definite. 
		All these arguments boil down to showing that the initial background metric is not a singular point and that the space of solutions around the initial metric is an open subset (in fact a submanifold) of all solutions. This can be shown by proving the surjectivity of the operators that appear in the linearized constraints. A similar construction, dynamical formulation of the  higher derivative models studied here in AdS and the surjectivity of the relevant linear maps would be highly valuable.  
		For the case of the cosmological Einstein's theory, such a construction was carried out in \cite{2009_hyperbolic} where it was observed that certain strong decays lead to linearization instability even for non-compact Cauchy surfaces with hyperbolic asymptotics.

\newpage		

\section[Linearization instability in chiral gravity in detail ]{Linearization instability in chiral gravity in detail \footnote{This chapter appeared as Phys. Rev. D 97, 124068 on 27 June 2018.} }

 Quantum gravity 
is elusive not mainly because we lack computational tools, but because we do not know {\it what} to compute and so how to define the theory for a generic spacetime. One possible exception and a promising path is the case of asymptotically anti-de Sitter (AdS) spacetimes for which a dual  quantum conformal field theory that lives on the boundary of a bulk spacetime with gravity would amount to a definition of quantum gravity. But, even for this setting, we do not have a realistic four dimensional example. In three dimensions, the situation is slightly better: the cosmological Einstein's theory (with $\Lambda <0$) has a black hole solution \cite{BTZ} and possesses the right boundary symmetries (a double copy of the centrally extended Virasoro algebra \cite{BH}) for a unitary two dimensional conformal field theory. But as the theory has no local dynamics (namely gravitons), it is not clear exactly how much one can learn from this model as far as quantum gravity is concerned. Having said that, even for this ostensibly simple model, we still do not yet have a quantum gravity theory. Recasting Einstein's gravity in terms of a solvable Chern-Simons gauge theory is a possible avenue \cite{Witten88}, but this only works for non-invertible dreibein  which cannot be coupled to generic matter. 

A more realistic gravity in three dimensions is the topologically massive gravity (TMG) \cite{djt} which has black hole solutions as well as a dynamical massive graviton. But the apparent problem with TMG is that the bulk graviton and the black hole cannot be made to have positive energy generally.  This obstruction to a viable classical and perhaps quantum theory was observed to disappear in an important work \cite{Strom1}, where it was realized that at a "chiral point"  defined by a tuned topological mass in terms of the AdS radius, one of the Virasoro algebras has a vanishing central charge (and so admits a trivial unitary representation) and the other has a positive nonzero central charge with unitary nontrivial representations,  the theory has a positive energy black hole and zero energy bulk gravitons. This tuned version of TMG, called "chiral gravity", seems to be a viable candidate for a well-behaved classical and quantum gravity.

One of the main objections raised against the chiral gravity is that it possesses a negative energy perturbative log-mode about the AdS vacuum which ruins the unitarity of the putative boundary CFT \cite{Grumiller}. Of course if this is the case, chiral gravity is not even viable at the classical level, since it does not have a vacuum.  It was argued in \cite{Strom2,Carlip} that chiral gravity could survive if the theory is linearization unstable about its AdS solution. This means that there would be perturbative modes which cannot be obtained from any exact solution of the theory.  In fact, these arguments were supported with the computations given in \cite{emel} where it was shown that the Taub charges which are functionals quadratic in the perturbative modes that must vanish identically due to background diffeomorphism invariance, do not vanish for the log-mode that ruins the chiral gravity.  This means that the log-mode found from the linearized field equations is an artifact of the linearized equations and does not satisfy the global constraints coming from the Bianchi identities.

In this part, we give a direct proof of the linearization instability of chiral gravity in AdS using the constraint analysis of the full TMG equations defined on a spacelike hypersurface. The crux of the argument that we shall lay out below is the following: the linearized constraint equations of TMG show that there are inconsistencies exactly at the chiral point. Namely perturbed matter fields do not determine the perturbations of the metric components on the  spacelike hypersurface and there are unphysical constraints on matter perturbations besides the usual covariant conservation. 

To support our local analysis on the hypersurface, we compute  the symplectic structure (that carries all the information about the phase space of the theory) for all perturbative solutions of the linearized field equations and find that the symplectic 2-form is degenerate and so non-invertible hence these modes do not approximate ({\it i.e.}\thinspace they are not tangent to) actual nonlinear solutions.
The symplectic 2-form evaluated for the log-mode is time-dependent (hence not coordinate-invariant) and vanishes at the initial value surface and grows unbounded in the future. 

To carry out the constraint analysis and their linearizations (which will yield possible nearby solutions to exact solution), we shall use the field equations instead of the TMG action as  the latter is not diffeomorphism invariant which complicates the discussion via the introduction of tensor densities (momenta) instead of tensors. We shall also work in the metric formulation instead of the first order one as there can be significant differences between the two formulations. Before we indulge into the analysis, let us note that the linearization instability that arises in the perturbative treatment of nonlinear theories and can be confused with dynamical or structural instability, as both are determined with the same linearization techniques.The difference is important: the latter refers to a real instability of a system such as the instability of the vacuum in a theory with ghosts such as the $R+\beta R_{\mu\nu}^2$ theory with $\beta\neq0$, this is simply not physically acceptable. On the other hand linearization instability refers to the failure of perturbation theory for a given background solution and one should resort to another method to proceed. From the point of view of the full solution space of the theory, this means that this (possibly infinite dimensional) space is not a smooth manifold but it has conical singularities around certain solutions. Let us expound on this a little bit.

\subsection{ADM decomposition of TMG}

Before restricting to the chiral gravity limit, we first study the full TMG field equations coupled with matter fields as an initial value problem, hence we take 
\begin{equation}
	\mathscr{E}_{\mu\nu}=G_{\mu\nu}+\Lambda g_{\mu\nu}+\frac{1}{\mu}C_{\mu\nu}=\kappa\tau_{\mu\nu}.
\end{equation}
The ADM \cite{adm} decomposition of the metric reads
\begin{equation}
	ds^{2}=-(n^{2}-n_{i}n^{i})dt^{2}+2n_{i}dtdx^{i}+\gamma_{ij}dx^{i}dx^{j},
\end{equation}
where ($n$, $n_{i}$) are lapse and shift  functions and $\gamma_{ij}$ is the $2D$
spatial metric. From now on, the Greek indices will run over the full spacetime, while the Latin indices will run over the  hypersurface  $\varSigma$, as $i,j...=1,2$. The spatial indices will be raised and lowered by the $2D$ metric. The extrinsic curvature ($k_{i j}$) of the surface is given as 
\begin{equation}
	2 n k_{ij}=\dot{\gamma}_{ij}-2 D_{(i}n_{j)},
\end{equation}
where $D$ is the covariant derivative compatible with $\gamma_{ij}$ and $\dot{\gamma}_{ij} :=\partial_{0}\gamma_{ij}$ and the round brackets denote symmetrization with a factor of 1/2. With the convention 
$R_{\rho\sigma}=\partial_{\mu}\Gamma_{\rho\sigma}^{\mu}-\partial_{\rho}\Gamma_{\mu\sigma}^{\mu}+\Gamma_{\mu\nu}^{\mu}\Gamma_{\rho\sigma}^{\nu}-\Gamma_{\sigma\nu}^{\mu}\Gamma_{\mu\rho}^{\nu}$, one finds  the hypersurface components of the three dimensional Ricci tensor as 
\begin{eqnarray}
	&&R_{ij}={^{(2)}R}_{ij}+k k_{ij}-2k_{ik}k_{j}^{k}  \\ 
	&&+\frac{1}{n}(\dot{k}_{ij}
	-n^{k}D_{k}k_{ij}  -D_{i}\partial_{j}n-2 k_{k(i}D_{j)}n^{k}), \nonumber
\end{eqnarray}
where ${^{(2)}R}_{ij}$ is the Ricci tensor of the hypersurface and $k\equiv \gamma^{i j} k_{ij}$. Similarly one find the twice projection to the normal of the surface as 
\begin{align}
	R_{00}=&\frac{n^{i}n^{j}}{n}(\dot{k}_{ij}-n^{k}D_{k}k_{ij}-D_{i}\partial_{j}n-2k_{kj}D_{i}n^{k}) \nonumber\\&-n^{2}k_{ij}^{2}
	+n^{i}n^{j}(^{(2)}R_{ij}+k k_{ij}-2k_{ik}k_{j}^{k})\\&+n(D_{k}\partial^{k}n-\dot{k}-n^{k}D_{k}k+2n^{k}D_{m}k_{k}^{m}). \nonumber
\end{align}
On the other hand, projecting once to the surface and once normal to the surface yields 
\begin{align}
	R_{0i}&=\frac{n^{j}}{n}(\dot{k}_{ij}-n^{k}D_{k}k_{ij}-D_{i}\partial_{j}n-2k_{k(i}D_{j)}n^{k})\\&+n^{j} ({^{(2)}R}_{ij}+kk_{ij}-2k_{ik}k_{j}^{k})+n(D_{i}k+D_{m}k_{i}^{m}). \nonumber
\end{align}
We also need the 3D scalar curvature in terms of the hypersurface quantities which can be found as 
\begin{equation}
	R={}^{(2)}R+k^{2}-k_{ij}^{2}+\frac{2}{n}(\dot{k}+n k_{ij}^{2}-D_{i}D^{i}n-n^{i}D_{i}k).
	{\label{3DR}}
\end{equation}
Given the Schouten tensor $S_{\mu\nu}:=R_{\mu\nu}-\frac{1}{4}Rg_{\mu\nu}$, the Cotton tensor is defined as
\begin{equation}
	C_{\mu\nu} :=\frac{1}{2}\epsilon{}^{\rho\alpha\beta}(g_{\mu\rho}\nabla_{\alpha}S_{\beta\nu}+g_{\nu\rho}\nabla_{\alpha}S_{\beta\mu}),
\end{equation}
where  $\epsilon{}^{\rho\alpha\beta}$ is the totally antisymmetric tensor which splits as $\epsilon{}^{0mn}=\frac{1}{n}\epsilon^{mn}=\frac{1}{n}\gamma^{-\frac{1}{2}}\varepsilon^{mn}$ where $\varepsilon^{mn}$ is the antisymmetric symbol. Just as we have done the ADM decomposition of the Ricci tensor, a rather lengthy computation yields the following expressions, for the projections of the Cotton tensor
\begin{align}
	2 n C_{ij}=&\epsilon{}^{mn}n_{i}( D_{m}S_{nj}-k_{mj}(D_{r}k_{n}^{r}-\partial_{n}k)) \nonumber\\
	&+\epsilon{}^m\,_i\bigg \{\dot{S}_{mj}-n k_{j}^{k}S_{mk}-S_{mk}D_{j}n^{k} \nonumber \\
	&-(\partial_{j}n+n^{r}k_{rj})(D_{s}k_{m}^{s}-\partial_{m}k) \nonumber \\
	&-D_{m}(n^{r}S_{rj}+n(D_{r}k_{j}^{r}-D_{j}k) ) \nonumber \\
	&+k_{mj}(D_{k}\partial^{k}n-\dot{k}+n^{k}D_{s}k_{k}^{s}+n(\frac{R}{4}-k_{rs}^{2})) \bigg \} \nonumber \\
	&+i\leftrightarrow j,
\end{align}
and 
\begin{equation}
	C_{i0}=n^{j}C_{ij}-\frac{\epsilon{}^{mn}}{2}(n A_{mni}-n_{i}B_{mn}-\gamma_{in}( C_{m}+n E_{m}))
\end{equation}
and
\begin{align}
	&C_{00}=n^{i}n^{j}C_{ij} \\
	&-\epsilon{}^{mn}(nn^{i}A_{mni}-(n_{i}n^{i}-n^{2})B_{mn}-n_{n}(C_{m}+n E_{m})),
	\nonumber
\end{align}
where we have defined the following tensors
\begin{equation}
	A_{mni}\equiv D_{m}S_{ni}-k_{mi}\left(D_{r}k_{n}^{r}-\partial_{n}k\right),
\end{equation}
\begin{equation}
	B_{mn}\equiv D_{m}D_{r}k_{n}^{r}-k_{m}^{k}S_{kn}, 
\end{equation}
\begin{equation}
	E_{m}\equiv 2k_{rs}D_{m}k^{rs}-\frac{1}{4}\partial_{m}R+k_{m}^{k}\left(D_{r}k_{k}^{r}-\partial_{k}k\right),
\end{equation}
\begin{eqnarray}
	&&C_{m}\equiv \partial_{0}D_{r}k_{m}^{r}-S_{m}^{k}\left(\partial_{k}n+n^{r}k_{rk}\right)-D_{m}D_{k}\partial^{k}n\nonumber\\
	&&\,\,\,\,\,\,\,-D_{m}\left(n^{k}D_{s}k_{k}^{s}\right)+k_{m}^{k}S_{kr}n^{r}+\partial_{m}n(k_{rs}^{2}-\frac{R}{4}).
\end{eqnarray}

Using  the above decomposition, we can recast the ADM form of the full TMG equations as
\begin{equation}
	\mathscr{E}_{ij}=\kappa\tau_{ij}=S_{ij}-\frac{1}{4}\gamma_{ij}R+\Lambda\gamma_{ij}+\frac{1}{\mu}C_{ij}
\end{equation}
and
\begin{align}
	\mathscr{E}_{0i}=&\kappa\tau_{0i}=n^{j}\mathscr{E}_{ij}+n(D_{r}k_{i}^{r}-\partial_{i}k)\\-&\frac{1}{2\mu}\epsilon{}^{mn}(n A_{mni}-n_{i}B_{mn}-\gamma_{in}(C_{m}+n E_{m})) \nonumber
\end{align}
and
\begin{align}
	\mathscr{E}_{00}&=\kappa\tau_{00}=2n^{i}\mathscr{E}_{0i}-n^{i}n^{j}\mathscr{E}_{ij}-\Lambda n^{2}-\frac{1}{\mu}\epsilon{}^{mn}n^{2}B_{mn}\nonumber \\&+n(D_{k}\partial^{k}n-\dot{k}+n^{k}D_{k}k+n(\frac{R}{2}-k_{rs}^{2})).
\end{align}
From $\mathscr{E}_{0i}$, we get the momentum constraint as
\begin{align}
	\Phi_{i}&=\kappa(\tau_{0i}-n^{j}\tau_{ij})=n(D_{r}k_{i}^{r}-\partial_{i}k)\\+&\frac{1}{2\mu}\epsilon{}^{mn}(n_{i}B_{mn}-nA_{mni}+\gamma_{in}C_{m}+n\gamma_{in}E_{m}) \nonumber
\end{align}
and from $\mathscr{E}_{00}$ we get the Hamiltonian constraint as 
\begin{align}
	\Phi=&\frac{\kappa}{n^{2}}(\tau_{00}-2n^{i}\tau_{0i}+n^{i}n^{j}\tau_{ij}) \nonumber \\
	+&\frac{1}{2}(^{(2)}R+k^{2}-k_{ij}^{2}-2\Lambda) \nonumber \\
	-&\frac{1}{\mu}\epsilon{}^{mn}\left(D_{m}D_{r}k_{n}^{r}-k_{m}^{k}S_{kn}\right),
\end{align}
where in the last equation we made use of the explicit form of $R$ given in (\ref{3DR}) which for TMG is  $R=6\Lambda-2 \kappa\tau $. From now on, for our purposes, it will suffice to work in the Gaussian normal coordinates with $n=1$ and $n_{i}=0$ for which $k_{ij}=\frac{1}{2}\dot{\gamma}_{ij}$ and the constraints reduce to
\begin{align}
	&\frac{\epsilon^{mn}}{4\mu}(\dot{\gamma}_{i m}\gamma^{i k}(^{(2)}R_{kn}-\dot{\gamma}_{kp}\dot{\gamma}_{sn}\gamma^{ps}-\ddot{\gamma}_{kn})-2 D_{m}D^{k}\dot{\gamma}_{kn}) \nonumber \\
	&-\frac{1}{8}\dot{\gamma}_{ij}\left(\dot{\gamma}_{ab}\gamma^{ab}\gamma^{ij}+\dot{\gamma}^{ij}\right)=\kappa\tau_{00}+\Lambda-\frac{^{(2)}R}{2}
\end{align}
and
\begin{align}
	&\frac{\epsilon^{m}\thinspace_{i}}{8\mu}\left( \dot{\gamma}^{kp}(2 D_{k}\dot{\gamma}_{pm}-D_{m}\dot{\gamma}_{kp})+2D^{k}\ddot{\gamma}_{km}-\dot{\gamma}_{mk}\gamma^{kl}D^{p}\dot{\gamma}_{pl}\right) \nonumber\\
	&-\frac{\epsilon^{mn}}{8\mu}\bigg(\dot{\gamma}_{ab}\gamma^{ab}D_{m}\dot{\gamma}_{in}-2\gamma^{ks}D_{m}(\dot{\gamma}_{kn}\dot{\gamma}_{si}) \nonumber  \\
	&+2D_{m}\ddot{\gamma}_{in}-\dot{\gamma}_{mi}D^{k}\dot{\gamma}_{kn}\bigg) \\
	&+\frac{1}{2}\left(D^{k}\dot{\gamma}_{ki}-\gamma^{ab}D_{i}\dot{\gamma}_{ab}\right)=\kappa\tau_{0i}+\frac{1}{2\mu}\epsilon^{mn}D_{m}{}^{(2)}R_{ni}. \nonumber
\end{align}
Furthermore, taking a conformally flat $2D$ metric on $\Sigma$, we have $\gamma_{ij}=e^{\varphi}\delta_{ij}$, where $\varphi=\varphi(t,x_i)$, $k_{ij}=\frac{1}{2}\dot{\varphi}\gamma_{ij}$ and the $2D$ Ricci tensor becomes
\begin{equation}
	^{(2)}R_{ij}=-\frac{1}{4}\gamma_{ij}e^{-\varphi}\left(2D_{k}\partial_{k}\varphi+\partial_{k}\varphi\partial_{k}\varphi\right),
\end{equation}
whereas the $3D$ Ricci tensor reads
\begin{equation}
	R_{ij}=\frac{1}{2}\gamma_{ij}(-D^{k}\partial_{k}\varphi+\dot{\varphi}^{2}+\ddot{\varphi}-\frac{1}{2}\partial^{k}\varphi\partial_{k}\varphi)
\end{equation}
and the $3D$ scalar curvature is 
\begin{equation}
	R=-D^{k}\partial_{k}\varphi+\frac{3}{2}\dot{\varphi}^{2}+2\ddot{\varphi}-\frac{1}{2}\partial^{k}\varphi\partial_{k}\varphi.
\end{equation}
With all these results in hand, one can obtain from the constraint equations the following relation
\begin{equation}
	\partial_{i}\dot{\varphi}=-J_{i}+\frac{1}{2\mu}\epsilon^{m}\thinspace_{i}\dot{\varphi}\partial_{m}\dot{\varphi},
	\label{kolay}
\end{equation}
where we have introduced the "source current"  which, on the hypersurface, reads
\begin{equation}
	J_{i} := 2\kappa\tau_{0i}+\frac{\kappa}{\mu}\epsilon^{m}\thinspace_{i}\partial_m\tau_{00}.
\end{equation}
Contracting (\ref{kolay}) with the epsilon-tensor, one arrives at
\begin{equation}
	\frac{2\mu}{\dot{\varphi}}\epsilon^{mi}\partial_{m}\dot{\varphi}\left(1+\frac{\dot{\varphi}^{2}}{4\mu^{2}}\right)=-\frac{2\mu}{\dot{\varphi}}\epsilon^{mi}J_{m}+J^{i}.
	{\label{main_equation}}
\end{equation}
In the case of vacuum, $\tau_{\mu\nu}=0$, and so $J_{i}=0$,  the unique solution to (\ref{main_equation}) is of the form $\varphi_{0}= c t$, where $c $ is a constant which can be found from the trace equation that reads  $R=6\Lambda$. So $\ensuremath{c}=2\sqrt{\Lambda} \equiv \frac{2}{\ell}$, which is the de Sitter (dS) solution and $\ell >0$ is its radius. Turning on a compactly supported matter perturbation with $\delta\tau_{\mu\nu}\ne0$, one has $\delta J_i\ne0$ and perturbing the constraint equations about $\varphi_{0}$ as $\varphi=\varphi_{0}+\delta\varphi$, we find a linearized constraint equation 
\begin{align}\label{pert}
	&\mu(1+\frac{1}{\mu^{2} \ell^2})\epsilon^{m}\thinspace_{i}\partial_{m}\delta\dot{\varphi}  \\=
	&(\partial_{i}+\frac{1}{\mu \ell}\epsilon^{m}\thinspace_{i}\partial_{m})\kappa\delta\tau_{00}+2\mu(\epsilon_{i}\thinspace^{m}+\frac{1}{\mu \ell}\delta^{m}\thinspace_{i})\kappa\delta\tau_{0m}\nonumber,
\end{align}
from which, for the dS case, one can solve the perturbation ($\delta \varphi$) and hence the perturbed metric  by integration
in terms of the perturbed matter fields  on the  hypersurface. Hence dS is linearization stable in TMG for any finite value of $\mu \ell$. The other linearized constraints are compatible with this solution. Our computation has been analytic in $\ell$, hence, we can do the following "Wick" rotation to study the  AdS case: $x^i\rightarrow ix^i$, $t\rightarrow it$, $ \ell \rightarrow i \ell$ yielding  $\Lambda=-\frac{1}{\ell^{2}}$ with the Gaussian normal form of the (signature changed) metric 
$ds^{2}=dt^{2}-e^{-{2 t/ \ell}}\left(dx^{2}+dx^{2}\right).$ Then for AdS,  (\ref{pert}) becomes 
\begin{align}\label{pert2}
	&\mu (1-\frac{1}{\mu^{2} \ell^2})\epsilon^{m}\thinspace_{i}\partial_{m}\delta\dot{\varphi}  \\=
	&-(\partial_{i}-\frac{1}{\mu \ell}\epsilon^{m}\thinspace_{i}\partial_{m})\kappa\delta\tau_{00}-2\mu (\epsilon_{i}\thinspace^{m}+\frac{1}{\mu \ell}\delta^{m}\thinspace_{i})\kappa\delta\tau_{0m}\nonumber
\end{align}
and once again the perturbation theory is valid for {\it generic } values of $\mu \ell$ in AdS as in the case of dS. But at the chiral point, $\mu \ell =1$,  the left-hand side vanishes identically and there is an unphysical constraint on the matter perturbations $\delta\tau_{0m}$ and $\delta\tau_{00}$ in addition to their background covariant conservation. Moreover, the metric perturbation is not determined by the matter perturbation. What this says is that in the chiral gravity limit of TMG, for AdS, the exact AdS solution is linearization unstable. The above computation has been a local one, and does not depend on the fact that AdS does not have a Cauchy surface on which one can define the initial value problem. AdS requires initial and boundary values together, but what we have computed is a necessary condition for such a formulation (not a sufficient one) and AdS in chiral gravity does not satisfy the necessary conditions for the initial-boundary value problem.

\subsection{Symplectic structure of TMG}
Let us give another argument for the linearization instability of AdS making use of the symplectic structure of TMG which was found in \cite{caner} following \cite{w} as 
$\omega := \int_\Sigma d \Sigma_\alpha  \sqrt{|g|} {\cal{J}}^\alpha$,
where $\Sigma$ is the hypersurface. $\omega$ is a closed ($\delta w=0$)  non-degenerate (except for gauge directions) 2-form for full TMG including chiral gravity.  Here the on-shell covariantly conserved symplectic current reads
\begin{align}
	{\cal{J}}^\alpha&=\delta  \Gamma^\alpha_{\ \mu\nu} \wedge  ( \delta  g^{\mu \nu} + \frac{1}{2}  g^{\mu \nu} \delta \ln g ) \nonumber \\
	&-  \delta  \Gamma^\nu_{\ \mu\nu} \wedge  ( \delta  g^{\alpha \mu} + \frac{1}{2}  g^{\alpha \mu} \delta \ln g ) \nonumber \\
	&+   \frac{1}{\mu}\epsilon^{\alpha \nu \sigma} ( \delta  {S}^\rho_{\ \sigma} \wedge \delta g_{\nu \rho}
	+ \frac{1}{2} \delta \Gamma^\rho_{\ \nu \beta} \wedge  \delta \Gamma^\beta_{\ \sigma \rho}  ).
	\label{symplectic_two}
\end{align}
What is important to understand is that  $\omega$ is a gauge invariant object on the solution space, say ${\mathcal {Z}}$, and also on the (more relevant) quotient ${\mathcal {Z}}/Diff$ which is the phase space and $Diff$ is the group of diffeomorphisms. Therefore, even without knowing the full space of solutions, by studying the symplectic structure, one gains a lot of information for both classical and quantum versions of the theory. Perturbative solutions live in the tangent space of the phase space and hence they are crucial in the discussion. We refer the reader to \cite{caner} for a full discussion of this.

Let us show that for the linearized solutions of chiral gravity given in \cite{Strom1}   the symplectic 2-form is degenerate and hence not invertible. In the global coordinates, the background metric reads
\begin{equation}
	ds^2 = \ell^2\big(-\cosh^2{\rho}\, d\tau^2 +\sinh^2{\rho}\,d\phi^2+d\rho^2\big),
\end{equation}
defining  $u=\tau+\phi$, $v=\tau-\phi$, making use of the $SL(2, R)\times SL(2, R)$, \cite{Strom1} found all  the primary states (but one) and their descendants. The primary solutions are
\begin{equation}
	h_{\mu \nu} = \Re \left\{e^{ -i  \Delta \tau -i S \phi } F_{\mu \nu}(\rho)\right\},
\end{equation}
where the real part is taken and the background tensor reads
\begin{eqnarray}
	F_{\mu\nu}(\rho)=f(\rho)\left(\begin{array}{ccc}
		1 & {S\over2}& {2i\over\sinh 2\rho} \\
		{S\over2} & 1 & {i S\over\sinh 2\rho} \\
		{2i\over\sinh 2 \rho} &{i S\over \sinh 2 \rho}   & - {4\over\sinh^2 2\rho} \\
	\end{array}\right)
\end{eqnarray}
and $f(\rho)=(\cosh{\rho})^{-\Delta }\sinh^2{\rho}$,
where $\Delta \equiv  h+\bar{h}$ and $S \equiv  h-\bar{h}$.
Components of the symplectic current for these modes (for generic $\mu \ell$) can be found as
\begin{align}
	&{\cal{J}}^\tau =\frac{( 4-S^2)(S + 2 \mu \ell)\Delta}{8 \mu \ell^7 (\cosh\rho)^{2(1+\Delta)}}\sin \left( 2 \Delta \tau +2 S \phi \right), \nonumber \\
	& {\cal{J}}^\phi= -\frac{2 \coth^2 \rho}{S+ 2 \mu \ell} {\cal{J}}^\tau, \\
	&{\cal{J}}^\rho =-\frac{(S \Delta +4\mu \ell)\coth \rho +(\Delta -2)\mu \ell \sinh 2 \rho}
	{\Delta ( S+ 2 \mu \ell)}{\cal{J}}^\tau, \nonumber
\end{align}
which yield a vanishing  $\omega$ at the chiral limit since for left, right and massive modes we have $S^2 = 4$ and the relevant symplectic current ${\cal{J}}^\tau$ vanishes identically, hence the solution is not viable. Moreover, one can show that its Taub charge diverges, while its ADT charge is for the background Killing vector $(-1,0,0)$ is
\begin{equation}
	Q_{ADT}=-\lim_{r \rightarrow \infty}\frac{ \sin (\pi S) \cos ( 2  \pi S + \Delta t)}{ 4 \pi S 2^{2-\Delta}\ell}\Delta ( 2 \Delta + S-2)
	e^{ r(2-\Delta)},  
\end{equation}
which vanishes for the massive mode $\Delta = S=2$. 
In addition to the above solutions, there is an additional  the log-mode given in \cite{Grumiller} which reads
\begin{align}
	&h_{ \mu\nu} = f_1(\tau,\rho)\left( \begin{array}{ccc}
		0 & 0 & 1 \\
		0 & 0 & 1 \\ 
		1 & 1 & 0
	\end{array} \right)_{\mu\nu} 
	&+f_2(\tau, \rho)\left( \begin{array}{ccl}
		1 & 1 & \qquad \qquad 0 \\
		1 & 1 & \qquad \qquad 0 \\
		0 & 0 & - {4\over\sinh^2 2\rho}
	\end{array} \right)_{\mu\nu},
	\label{grumillermetric}
\end{align}where the two functions are given as 
\begin{eqnarray}
	&&f_1(\tau, \rho) =  \frac{\sinh{\rho}}{\cosh^3{\rho}}\,(\tau \cos{2u}-\sin{2u}\,\ln{\cosh{\rho}}) , \\
	&&f_2(\tau, \rho) =-\tanh^2\!\!{\rho}\,(\tau \sin{2u}+\cos{2u}\,\ln{\cosh{\rho}}). 
\end{eqnarray}
The components of the symplectic current for this mode  read
\begin{align}
	&{\cal{J}}^\tau = \frac{1}{\mu \ell^7}\tau ((1-\mu \ell ) \cosh 2 \rho+1)\text{sech}^{10}\rho , \nonumber \\
	& {\cal{J}}^\phi= - \frac{2}{\mu \ell^7}\tau (1- \mu \ell) \text{sech}^8 \rho, \\
	&{\cal{J}}^\rho=\frac{1}{ \ell^6}\tanh \rho\, \text{sech}^8\rho(4
	(\log ^2\cosh \rho+\tau ^2)+\log \text{sech}\,\rho) \nonumber ,
\end{align}
which yield a linearly growing $\omega$ in $\tau$ and vanishes on the initial value surface.
What all these say is that first order perturbation theory simply fails in chiral gravity limit of TMG. If the theory makes any sense at the classical and/or quantum level one must resort to a new method to carry out computations.  This significantly affects its interpretation in the context of AdS/CFT as the perturbed metric couples to the energy-momentum tensor of the  boundary CFT. This of course does not say anything about the solutions of the theory which are not globally AdS and one might simply have to define the theory in a different background.

In this chapter we have studied a frequently recurring problem \cite{kastor}, for example it also appears in critical gravity \cite{Pope, Tahsin}. Linearized solutions by definition satisfy the linearized equations but this is not sufficient; they should also satisfy a quadratic constraint to actually represent linearized versions of exact solutions. This deep result comes from the Bianchi identities and their linearizations and it is connected to the conserved quantities. With the observation of gravity waves, research in general relativity and its modifications, extensions has entered an exciting era in which many theories might be possibly tested. One major tool of computation in nonlinear theories, such as gravity, is perturbation theory from which one obtains a lot of information and the gravitational wave physics is no exception as one uses the tools of perturbation theory to obtain the wave profile far away from the sources. Therefore, the issue of linearization instability arises in any use of perturbation theory as the examples provided here and before \cite{emel} show even for the ostensibly safe case of spacetimes with noncompact Cauchy surfaces. 

\vspace{0.4cm}

\section{Conclusions}

In Einstein's general relativity, perturbation theory about a background  exact solution
fails if the background spacetime has a Killing symmetry and a
compact (without a boundary) spacelike Cauchy surface. This failure, dubbed as {\it linearization instability}, shows itself as a nonintegrability of the perturbative infinitesimal deformation to a finite deformation
of the background. Namely, the linearized field equations have spurious solutions which cannot be obtained from the linearization of some exact solutions. In absence of the knowledge of exact solutions, in practice, one can show the failure of the linear perturbation theory  by showing that a certain quadratic (integral) constraint, that is the vanishing of the so-called Taub charge, on the linearized solutions is not satisfied. This is the case for compact Cauchy surfaces ( or in the absence of Cauchy surfaces as in the case of AdS, for spacelike surfaces which constitute part of the initial-boundary value problem). 

For noncompact Cauchy surfaces, the situation is different and for example, Minkowski space, having a noncompact Cauchy surface, is linearization stable. If this were not the case, one could not trust perturbation theory in a Minkowski background, including all the computations related to the gravitational waves.  In this thesis, we have studied the linearization instability in generic metric theories 
of gravity where Einstein's theory is modified with additional curvature terms. Of course the problem of validity of perturbation theory becomes much more complicated as one usually lacks the tools of the elliptic operator theory on a spacelike hypersurface. Our main finding is that, unlike the case of general relativity, for modified theories even in the noncompact Cauchy surface cases (or spacelike surfaces), there are some theories which show linearization instability about their anti-de Sitter backgrounds. Recent $D$ dimensional critical and three dimensional chiral gravity theories are two such examples. We have discussed them, especially the chiral gravity case, in great detail. This observation sheds light on the paradoxical behavior of vanishing conserved charges (mass, angular momenta) for nonvacuum solutions, such as black holes, in these extended theories. This vanishing of conserved charges for nonvacuum solutions for certain theories in asymptotically AdS spacetimes was discussed at length in \cite{Deser_Tekin}. At the time this zero-energy problem was not properly understood. Let us explain why in some extended theories the conserved charges vanish identically. This vanishing happens as the charge is a global quantity constructed with the help of the Stokes' theorem and asymptotic Killing symmetries where the perturbation theory is sufficient. In perturbation theory, about a constant curvature background (such as the AdS spacetime) all higher order curvature terms in the field equations contribute to the conserved charges in an additive manner which always vanishes for some particular combination of parameters of the theory.  Here we have shown that, exactly at that point, the perturbation theory fails and therefore one needs other methods to define conserved quantities.

As a second, perhaps more direct proof of the linearization instability in chiral gravity, we have carried out a detailed analysis of the constraints and their linearizations on a spacelike hypersurface, we have shown that the topologically massive gravity (which is a dynamical theory of gravity in three dimensions) has a linearization instability at the chiral gravity limit about its $AdS_3$ vacuum \cite{emel2}. We have also calculated the symplectic structure (that is built from the perturbative tensors and that carries all the information about the classical theory as well as its linearizations) for all the known perturbative modes, including the log-mode, for the linearized field equations and find it to be degenerate (non-invertible). Hence these modes do not approximate any possible exact solutions and so do not belong  to the linearized phase space of the theory. Naive perturbation theory fails: the linearized field equations are necessary but not sufficient in finding viable linearized solutions. This proof supports the construction given in \cite{Strom2} where it was shown that the linearization of all exact solutions of chiral gravity around $AdS_3$ has positive energy. This has important consequences for both classical and possible quantum versions of the theory which need further scrutiny. 

In the linearization stability problem of general relativity, one can show that the first order perturbation theory can receive constraints at most from the second order perturbation theory. Namely, there will be no more constraints from the higher order perturbation theory. This issue is an outstanding problem in generic modified gravity theories which needs to be studied further. For example, we do not know what are the necessary and sufficient conditions for linearization stability of a third order theory such as topologically massive gravity.

\newpage

\appendix

\section{Appendix: Second order perturbation theory and gauge invariance issues}
\label{chp:appendixa}
\subsection{Second order perturbation theory}
Let us summarize some results about the second order perturbation
theory (see also \cite{Tahsin_born}). By definition one has 
\begin{equation}
	g_{\mu\nu}:=\bar{g}_{\mu\nu}+\tau h_{\mu\nu},
\end{equation}
where $\tau$ is introduced to keep the order of the expansions. The inverse of the metric tensor up to and including the second order is 
\begin{equation}
	g^{\mu\nu}=\bar{g}^{\mu\nu}+\tau h^{\mu\nu}+\tau^{2}h_{\alpha}^{\mu}h^{\alpha\nu}+O(\tau^{3}).
\end{equation}
Let $T$ be a generic tensor or a geometrical object, then it can be expanded as 
\begin{equation}
	T=\bar{T}+\tau{T}^{(1)}+\tau^{2}{T}^{(2)}+O(\tau^{3}).
\end{equation}
For the Christoffel connection we have 
\begin{equation}
	\varGamma_{\mu\nu}\thinspace^{\gamma}=\bar{\varGamma}_{\mu\nu}\thinspace^{\gamma}+\tau(\varGamma_{\mu\nu}\thinspace^{\gamma})^{(1)}+\tau^{2}(\varGamma_{\mu\nu}\thinspace^{\gamma})^{(2)},
\end{equation}
where the first order term is 
\begin{equation}
	(\Gamma_{\mu\nu}\thinspace^{\gamma})^{(1)}=\frac{1}{2}\big(\bar{\nabla}_{\mu}h_{\nu}^{\gamma}+\bar{\nabla}_{\nu}h_{\mu}^{\gamma}-\bar{\nabla}^{\gamma}h_{\mu\nu}),\label{a5}
\end{equation}
and the second order one is 
\begin{equation}
	(\Gamma_{\mu\nu}\thinspace^{\gamma})^{(2)}=-h^{\gamma\delta}(\Gamma_{\mu\nu\delta})^{(1)}.
\end{equation}
As is clear from (\ref{a5}) the linearized Christoffel connection is a background tensor (so is the second order one), we can raise and lower the indices
with the background metric and its inverse 
\begin{equation}
	(\Gamma_{\mu\nu\delta})^{(1)}=\bar{g}_{\gamma\delta}(\Gamma_{\mu\nu}\thinspace^{\gamma})^{(1)}.
\end{equation}
Please note that the up index is lowered to the third position, there
is a symmetry only in the first two indices. The first order linearized
Riemann tensor is 
\begin{equation}
	(R^{\rho}\thinspace_{\mu\sigma\nu})^{(1)}=\bar{\nabla}_{\sigma}(\Gamma_{\nu\mu}\thinspace^{\rho})^{(1)}-\bar{\nabla}_{\nu}(\Gamma_{\sigma\mu}\thinspace^{\rho})^{(1)},\label{eq:firstorderriemann}
\end{equation}
and the second order linearized Riemann tensor is 
\begin{eqnarray}
	(R^{\rho}\thinspace_{\mu\sigma\nu})^{(2)}=\bar{\nabla}_{\sigma}(\Gamma_{\nu\mu}\thinspace^{\rho})^{(2)}-\bar{\nabla}_{\nu}(\Gamma_{\sigma\mu}\thinspace^{\rho})^{(2)}+(\Gamma_{\mu\nu}\thinspace^{\alpha})^{(1)}(\Gamma_{\sigma\alpha}\thinspace^{\rho})^{(1)}\nonumber \\
	-(\Gamma_{\mu\sigma}\thinspace^{\alpha})^{(1)}(\Gamma_{\nu\alpha}\thinspace^{\rho})^{(1)}.
\end{eqnarray}
The first order linearized Ricci tensor is 
\begin{equation}
	(R_{\mu\nu})^{(1)}=\bar{\nabla}_{\sigma}(\Gamma_{\mu\nu}\thinspace^{\sigma})^{(1)}-\bar{\nabla}_{\nu}(\Gamma_{\sigma\mu}\thinspace^{\sigma})^{(1)},
\end{equation}
and the second order linearized Ricci tensor is 
\begin{eqnarray}
	(R_{\mu\nu})^{(2)}=\bar{\nabla}_{\sigma}(\Gamma_{\nu\mu}\thinspace^{\sigma})^{(2)}-\bar{\nabla}_{\nu}(\Gamma_{\sigma\mu}\thinspace^{\sigma})^{(2)}+(\Gamma_{\mu\nu}\thinspace^{\alpha})^{(1)}(\Gamma_{\sigma\alpha}\thinspace^{\sigma})^{(1)}\nonumber \\
	-(\Gamma_{\mu\sigma}\thinspace^{\alpha})^{(1)}(\Gamma_{\nu\alpha}\thinspace^{\sigma})^{(1)}.
\end{eqnarray}
We shall need the explicit form of it in terms of the $h_{\mu\nu}$
field, which reads 
\begin{eqnarray}
	&&(R_{\mu\nu})^{(2)}=-\frac{1}{2}\bar{\nabla}_{\rho}\left(h^{\rho\beta}(\bar{\nabla}_{\mu}h_{\nu\beta}+\bar{\nabla}_{\nu}h_{\mu\beta}-\bar{\nabla}_{\beta}h_{\nu\mu})\right)+\frac{1}{2}\bar{\nabla}_{\nu}\left(h^{\rho\beta}\bar{\nabla}_{\mu}h_{\rho\beta}\right)\nonumber \\
	&&-\frac{1}{4}\left(\bar{\nabla}_{\mu}h_{\rho\beta}\right)\bar{\nabla}_{\nu}h^{\rho\beta}+\frac{1}{4}\left(\bar{\nabla}^{\beta}h\right)(\bar{\nabla}_{\mu}h_{\nu\beta}+\bar{\nabla}_{\nu}h_{\mu\beta}-\bar{\nabla}_{\beta}h_{\nu\mu})\nonumber \\
	&&+\frac{1}{2}(\bar{\nabla}_{\beta}h_{\nu\alpha})\bar{\nabla}^{\beta}h_{\mu}^{\alpha}-\frac{1}{2}(\bar{\nabla}_{\beta}h_{\nu\alpha})\bar{\nabla}^{\alpha}h_{\mu}^{\beta}.\thinspace\thinspace\thinspace\thinspace\label{a12}
\end{eqnarray}
The first order linearized scalar curvature is 
\begin{equation}
	(R)^{(1)}=\bar{\nabla}_{\alpha}\bar{\nabla}_{\beta}h^{\alpha\beta}-\bar{\square}h-\bar{R}_{\mu\nu}h^{\mu\nu},
\end{equation}
and the second order linearized scalar curvature is 
\begin{equation}
	(R)^{(2)}=\bar{R}_{\mu\nu}h_{\alpha}^{\mu}h^{\alpha\nu}-(R_{\mu\nu})^{(1)}h^{\mu\nu}+\bar{g}^{\mu\nu}(R_{\mu\nu})^{(2)}.
\end{equation}
Explicitly we have a rather cumbersome result for the last expression which reads 
\begin{eqnarray}
	&&(R)^{(2)}=\frac{1}{4}\left(\bar{\nabla}_{\sigma}h_{\rho\beta}\right)\bar{\nabla}^{\sigma}h^{\rho\beta}-\frac{1}{2}\left(\bar{\nabla}_{\sigma}h_{\rho\beta}\right)\bar{\nabla}^{\rho}h^{\sigma\beta}\nonumber \\
	&&+\frac{1}{2}\bar{\nabla}^{\sigma}\left(h^{\rho\beta}\bar{\nabla}_{\sigma}h_{\rho\beta}\right)+\frac{1}{4}\left(\bar{\nabla}^{\beta}h\right)(2\bar{\nabla}_{\sigma}h_{\beta}^{\sigma}-\bar{\nabla}_{\beta}h)-\frac{1}{2}\bar{\nabla}_{\rho}\left(h^{\rho\beta}(2\bar{\nabla}_{\sigma}h_{\beta}^{\sigma}-\bar{\nabla}_{\beta}h)\right)\nonumber \\
	&&-\frac{1}{2}h^{\rho\beta}\left(2\bar{\nabla}_{\sigma}\bar{\nabla}_{\rho}h_{\beta}^{\sigma}-\bar{\square}h_{\rho\beta}-\bar{\nabla}_{\rho}\bar{\nabla}_{\beta}h\right)+\bar{R}_{\rho\beta}h^{\rho\alpha}h_{\alpha}^{\beta}.\thinspace\thinspace\thinspace\thinspace\thinspace\thinspace\thinspace\thinspace
\end{eqnarray}

Here without going into too much detail, let us summarize some of the
relevant formulas that we use in the bulk of the second chapter to
show various expressions, such as the gauge transformation of the
linearized tensors, second order forms of the tensors \textit{etc.}

\subsection{Some useful identities that involve Lie and covariant derivatives }
Lie and covariant derivatives do not commute so we shall need the
following expressions. Let $X$ be a vector field on our spacetime manifold
with a metric $\bar{g}$ and let $T$ be a $(0,2)$ background tensor
field. Then in components one has the Lie derivative of $T$ with
respect to the vector field $X$ as 
\begin{equation}
	\text{\ensuremath{\mathscr{L}}}_{X}T_{\rho\sigma}=X^{f}\bar{\nabla}_{f}T_{\rho\sigma}+\left(\bar{\nabla}_{\rho}X^{f}\right)T_{f\sigma}+\left(\bar{\nabla}_{\sigma}X^{f}\right)T_{\rho f}.
\end{equation}
Taking the covariant derivative of the last expression gives 
\begin{eqnarray}
	&  & \bar{\nabla}_{\mu}\text{\ensuremath{\mathscr{L}}}_{X}T_{\rho\sigma}=\left(\bar{\nabla}_{\mu}X^{f}\right)\bar{\nabla}_{f}T_{\rho\sigma}+X^{f}\bar{\nabla}_{\mu}\bar{\nabla}_{f}T_{\rho\sigma}+\left(\bar{\nabla}_{\mu}\bar{\nabla}_{\rho}X^{f}\right)T_{f\sigma}\nonumber \\
	&  & +\left(\bar{\nabla}_{\rho}X^{f}\right)\bar{\nabla}_{\mu}T_{f\sigma}+\left(\bar{\nabla}_{\mu}\bar{\nabla}_{\sigma}X^{f}\right)T_{\rho f}+\left(\bar{\nabla}_{\sigma}X^{f}\right)\bar{\nabla}_{\mu}T_{\rho f}.
\end{eqnarray}
Now let us change the order of the differentiations, we have 
\begin{eqnarray}
	&&\text{\ensuremath{\mathscr{L}}}_{X}\bar{\nabla}_{\mu}T_{\rho\sigma}=X^{f}\bar{\nabla}_{f}\bar{\nabla}_{\mu}T_{\rho\sigma}+\left(\bar{\nabla}_{\mu}X^{f}\right)\bar{\nabla}_{f}T_{\rho\sigma}+\left(\bar{\nabla}_{\rho}X^{f}\right)\bar{\nabla}_{\mu}T_{f\sigma}\nonumber\\&&+\left(\bar{\nabla}_{\sigma}X^{f}\right)\bar{\nabla}_{\mu}T_{\rho f}.
\end{eqnarray}
Subtracting the last two expressions gives 
\begin{eqnarray}
	&  & \bar{\nabla}_{\mu}\text{\ensuremath{\mathscr{L}}}_{X}T_{\rho\sigma}-\text{\ensuremath{\mathscr{L}}}_{X}\bar{\nabla}_{\mu}T_{\rho\sigma}=X^{f}\bar{\nabla}_{\mu}\bar{\nabla}_{f}T_{\rho\sigma}-X^{f}\bar{\nabla}_{f}\bar{\nabla}_{\mu}T_{\rho\sigma}+\left(\bar{\nabla}_{\mu}\bar{\nabla}_{\rho}X^{f}\right)T_{f\sigma}\nonumber \\
	&  & +\left(\bar{\nabla}_{\mu}\bar{\nabla}_{\sigma}X^{f}\right)T_{\rho f},
\end{eqnarray}
where 
\begin{equation}
	\left[\bar{\nabla}_{\mu},\bar{\nabla}_{f}\right]T_{\rho\sigma}=\bar{R}_{\mu f\rho}\thinspace^{\lambda}T_{\lambda\sigma}+\bar{R}_{\mu f\sigma}\thinspace^{\lambda}T_{\lambda\rho}.
\end{equation}
Then we obtain 
\begin{eqnarray}
	\bar{\nabla}_{\mu}\text{\ensuremath{\mathscr{L}}}_{X}T_{\rho\sigma}-\text{\ensuremath{\mathscr{L}}}_{X}\bar{\nabla}_{\mu}T_{\rho\sigma}=\left(\bar{\nabla}_{\mu}\bar{\nabla}_{\rho}X^{f}+X^{\lambda}\bar{R}_{\mu\lambda\rho}\thinspace^{f}\right)T_{f\sigma}\nonumber \\
	+\left(\bar{\nabla}_{\mu}\bar{\nabla}_{\sigma}X^{f}+X^{\lambda}\bar{R}_{\mu\lambda\sigma}\thinspace^{f}\right)T_{f\rho}.\label{a21}
\end{eqnarray}
Let $\delta_{X}$ denote the gauge transformation generated by the vector field $X$,
such that 
\begin{equation}
	\delta_{X}h_{\mu\nu}=\bar{\nabla}_{\mu}X_{\nu}+\bar{\nabla}_{\nu}X_{\mu}.
\end{equation}
From the definition of the first order linearized Christoffel connection,
we have
\begin{eqnarray}
	\delta_{X}(\Gamma_{\mu\nu}\thinspace^{\gamma})^{\left(1\right)}=\frac{1}{2}\left(\bar{\nabla}_{\mu}\bar{\nabla}_{\nu}X^{\gamma}+\bar{\nabla}_{\mu}\bar{\nabla}^{\gamma}X_{\nu}+\bar{\nabla}_{\nu}\bar{\nabla}_{\mu}X^{\gamma}+\bar{\nabla}_{\nu}\bar{\nabla}^{\gamma}X_{\mu}\right.\nonumber\\
	\left.-\bar{\nabla}^{\gamma}\bar{\nabla}_{\mu}X_{\nu}-\bar{\nabla}^{\gamma}\bar{\nabla}_{\nu}X_{\mu}\right),
\end{eqnarray}
which can be expressed as
\begin{eqnarray}
	&&\delta_{X}(\Gamma_{\mu\nu}\thinspace^{\gamma})^{\left(1\right)}=\bar{\nabla}_{\mu}\bar{\nabla}_{\nu}X^{\gamma}\nonumber\\&&+\frac{1}{2}\left(\left[\bar{\nabla}_{\mu},\bar{\nabla}^{\gamma}\right]X_{\nu}+\left[\bar{\nabla}_{\nu},\bar{\nabla}_{\mu}\right]X^{\gamma}+\left[\bar{\nabla}_{\nu},\bar{\nabla}^{\gamma}\right]X_{\mu}\right).
\end{eqnarray}
By using the following identity
\begin{equation}
	\left[\bar{\nabla}_{\nu},\bar{\nabla}_{\mu}\right]X^{\gamma}=\bar{R}_{\nu\mu}\thinspace^{\gamma\sigma}X_{\sigma},
\end{equation}
and the first Bianchi identity
\begin{equation}
	\bar R_{\alpha\beta\gamma\delta}+\bar R_{\beta\gamma\alpha\delta}+\bar R_{\gamma\alpha\beta\delta}=0,
\end{equation}
the gauge transformation of the first order linearized Christoffel connection reads,
\begin{equation}
	\delta_{X}(\Gamma_{\mu\nu}\thinspace^{\gamma})^{(1)}=\bar{\nabla}_{\mu}\bar{\nabla}_{\nu}X^{\gamma}+\bar{R}^{\gamma}\thinspace_{\nu\sigma\mu}X^{\sigma}.\label{eq:gaugetranschristoffel}
\end{equation}
The symmetry in $\mu$ and $\nu$ is not explicit on the right hand side, but it is in fact symmetric. Making use of this in (\ref{a21}) one finds 
\begin{equation}
	\bar{\nabla}_{\mu}\text{\ensuremath{\mathscr{L}}}_{X}T_{\rho\sigma}=\text{\ensuremath{\mathscr{L}}}_{X}\bar{\nabla}_{\mu}T_{\rho\sigma}+T_{\alpha\sigma}\delta_{X}(\Gamma_{\mu\rho}\thinspace^{\alpha})^{(1)}+T_{\rho\alpha}\delta_{X}(\Gamma_{\mu\sigma}\thinspace^{\alpha})^{(1)}.\label{id1}
\end{equation}
Similarly, the Lie derivative of a tensor with one upper and one
lower indices is 
\begin{equation}
	\text{\ensuremath{\mathscr{L}}}_{X}T_{\rho}\thinspace^{\sigma}=X^{f}\bar{\nabla}_{f}T_{\rho}\thinspace^{\sigma}+\left(\bar{\nabla}_{\rho}X^{f}\right)T_{f}\thinspace^{\sigma}-\left(\bar{\nabla}_{f}X^{\sigma}\right)T_{\rho}\thinspace^{f},
\end{equation}
and the covariant differentiation of the result gives 
\begin{eqnarray}
	&  & \bar{\nabla}_{\mu}\text{\ensuremath{\mathscr{L}}}_{X}T_{\rho}\thinspace^{\sigma}=\left(\bar{\nabla}_{\mu}X^{f}\right)\bar{\nabla}_{f}T_{\rho}\thinspace^{\sigma}+X^{f}\bar{\nabla}_{\mu}\bar{\nabla}_{f}T_{\rho}\thinspace^{\sigma}+\left(\bar{\nabla}_{\mu}\bar{\nabla}_{\rho}X^{f}\right)T_{f}\thinspace^{\sigma}\nonumber \\
	&  & +\left(\bar{\nabla}_{\rho}X^{f}\right)\bar{\nabla}_{\mu}T_{f}\thinspace^{\sigma}-\left(\bar{\nabla}_{\mu}\bar{\nabla}_{f}X^{\sigma}\right)T_{\rho}\thinspace^{f}-\left(\bar{\nabla}_{f}X^{\sigma}\right)\bar{\nabla}_{\mu}T_{\rho}\thinspace^{f}.
\end{eqnarray}
Reversing the order of the derivatives, we have 
\begin{eqnarray}
	&&\text{\ensuremath{\mathscr{L}}}_{X}\bar{\nabla}_{\mu}T_{\rho}\thinspace^{\sigma}=X^{f}\bar{\nabla}_{f}\bar{\nabla}_{\mu}T_{\rho}\thinspace^{\sigma}+\left(\bar{\nabla}_{\mu}X^{f}\right)\bar{\nabla}_{f}T_{\rho}\thinspace^{\sigma}\nonumber\\&&+\left(\bar{\nabla}_{\rho}X^{f}\right)\bar{\nabla}_{\mu}T_{f}\thinspace^{\sigma}-\left(\bar{\nabla}_{f}X^{\sigma}\right)\bar{\nabla}_{\mu}T_{\rho}\thinspace^{f},
\end{eqnarray}
and subtracting the last two expressions we obtain
\begin{eqnarray}
	&&\bar{\nabla}_{\mu}\text{\ensuremath{\mathscr{L}}}_{X}T_{\rho}\thinspace^{\sigma}-\text{\ensuremath{\mathscr{L}}}_{X}\bar{\nabla}_{\mu}T_{\rho}\thinspace^{\sigma}=X^{f}\left[\bar{\nabla}_{\mu},\bar{\nabla}_{f}\right]T_{\rho}\thinspace^{\sigma}\nonumber\\&&+\left(\bar{\nabla}_{\mu}\bar{\nabla}_{\rho}X^{f}\right)T_{f}\thinspace^{\sigma}-\left(\bar{\nabla}_{\mu}\bar{\nabla}_{f}X^{\sigma}\right)T_{\rho}\thinspace^{f}.
\end{eqnarray}
By using 
\begin{equation}
	\left[\bar{\nabla}_{\mu},\bar{\nabla}_{f}\right]T_{\rho}\thinspace^{\sigma}=\bar{R}_{\mu f\rho}\thinspace^{\lambda}T_{\lambda}\thinspace^{\sigma}+\bar{R}_{\mu f}\thinspace^{\sigma}\thinspace_{\lambda}T_{\rho}\thinspace^{\lambda},
\end{equation}
and the expression for the gauge transformation of the first order linearized Christoffel
connection (\ref{eq:gaugetranschristoffel}), we can rewrite the result as 
\begin{equation}
	\bar{\nabla}_{\mu}\text{\ensuremath{\mathscr{L}}}_{X}T_{\rho}\thinspace^{\sigma}=\text{\ensuremath{\mathscr{L}}}_{X}\bar{\nabla}_{\mu}T_{\rho}\thinspace^{\sigma}+T_{\alpha}\thinspace^{\sigma}\delta_{X}(\Gamma_{\mu\rho}\thinspace^{\alpha})^{(1)}-T_{\rho}\thinspace^{\alpha}\delta_{X}(\Gamma_{\mu\alpha}\thinspace^{\rho})^{(1)}.
\end{equation}
Applying the same procedure for the case of any three lower
index tensor, we arrive at the relations 
\begin{eqnarray}
	\bar{\nabla}_{\mu}\text{\ensuremath{\mathscr{L}}}_{X}T_{\rho\sigma\gamma}=\text{\ensuremath{\mathscr{L}}}_{X}\bar{\nabla}_{\mu}T_{\rho\sigma\gamma}+T_{\alpha\sigma\gamma}\delta_{X}(\Gamma_{\mu\rho}\thinspace^{\alpha})^{(1)}+T_{\rho\alpha\gamma}\delta_{X}(\Gamma_{\mu\sigma}\thinspace^{\alpha})^{(1)}\nonumber \\
	+T_{\rho\sigma\alpha}\delta_{X}(\Gamma_{\mu\gamma}\thinspace^{\alpha})^{(1)},\label{id2}
\end{eqnarray}
and 
\begin{eqnarray}
	&  & \bar{\nabla}_{\mu}\text{\ensuremath{\mathscr{L}}}_{X}T_{\rho\gamma}\thinspace^{\sigma}=\text{\ensuremath{\mathscr{L}}}_{X}\bar{\nabla}_{\mu}T_{\rho\gamma}\thinspace^{\sigma}+T_{\alpha\gamma}\thinspace^{\sigma}\delta_{X}(\Gamma_{\mu\rho}\thinspace^{\alpha})^{\left(1\right)}+T_{\rho\alpha}\thinspace^{\sigma}\delta_{X}(\Gamma_{\mu\gamma}\thinspace^{\alpha})^{\left(1\right)}\nonumber \\
	&  & -T_{\rho\gamma}\thinspace^{\alpha}\delta_{X}(\Gamma_{\mu\alpha}\thinspace^{\rho})^{\left(1\right)}.
\end{eqnarray}
In fact in the most general case of a $(m,n)$ tensor $T$ we have
\begin{eqnarray}
	&&\bar{\nabla}_{\mu}\text{\ensuremath{\mathscr{L}}}_{X}T_{\rho_{1}\rho_{2}...\rho_{n}}\thinspace^{\sigma_{1}\sigma_{2}...\sigma_{m}}=\text{\ensuremath{\mathscr{L}}}_{X}\bar{\nabla}_{\mu}T_{\rho_{1}\rho_{2}...\rho_{n}}\thinspace^{\sigma_{1}\sigma_{2}...\sigma_{m}}\\
	&&+T_{\alpha\rho_{2}...\rho_{n}}\thinspace^{\sigma_{1}\sigma_{2}...\sigma_{m}}\delta_{X}(\Gamma_{\mu\rho_{1}}\thinspace^{\alpha})^{\left(1\right)}+T_{\rho_{1}\alpha...\rho_{n}}\thinspace^{\sigma_{1}\sigma_{2}...\sigma_{m}}\delta_{X}(\Gamma_{\mu\rho_{2}}\thinspace^{\alpha})^{\left(1\right)}\nonumber \\
	&&+...+T_{\rho_{1}\rho_{2}...\alpha}\thinspace^{\sigma_{1}\sigma_{2}...\sigma_{m}}\delta_{X}(\Gamma_{\mu\rho_{n}}\thinspace^{\alpha})^{\left(1\right)}-T_{\rho_{1}\rho_{2}...\rho_{n}}\thinspace^{\alpha\sigma_{2}...\sigma_{m}}\delta_{X}(\Gamma_{\mu\alpha}\thinspace^{\sigma_{1}})^{\left(1\right)}\nonumber \\
	&&-T_{\rho_{1}\rho_{2}...\rho_{n}}\thinspace^{\sigma_{1}\alpha...\sigma_{m}}\delta_{X}(\Gamma_{\mu\alpha}\thinspace^{\sigma_{2}})^{\left(1\right)}-...-T_{\rho_{1}\rho_{2}...\rho_{n}}\thinspace^{\sigma_{1}\sigma_{2}...\alpha}\delta_{X}(\Gamma_{\mu\sigma_{m}}\thinspace^{\rho})^{\left(1\right)}.\nonumber 
\end{eqnarray}

\subsection{Gauge transformation issues}
Using the formulas we discussed above, let us find how the first order linearized Einstein tensor changes
under the gauge transformations generated by the flow of the vector field $X$. In
the index free notation one has
\begin{equation}
	DEin(g)\cdot\left(\text{\ensuremath{\mathscr{L}}}_{X} g\right)=\text{\ensuremath{\mathscr{L}}}_{X}Ein(g),
\end{equation}
and in a local coordinate chart it can be expressed as
\begin{equation}
	\delta_{X}\left(G_{\mu\nu}\right)^{(1)}\cdot h=\text{\ensuremath{\mathscr{L}}}_{X}\bar{G}_{\mu\nu}.\label{eq:aimfirst}
\end{equation}
The center dot notation means that the operator on the left is evaluated at $h$. This notation seems a little bit redundant at the first order, but it becomes useful in the second order. Let us compute the left hand side of the last equation explicitly,
we can write
\begin{equation}
	\delta_{X}(G_{\mu\nu})^{(1)}\cdot h=\delta_{X}(R_{\mu\nu})^{(1)}\cdot h-\frac{1}{2}\bar{g}_{\mu\nu}\delta_{X}(R)^{(1)}\cdot h-\frac{1}{2}\bar{R}\delta_{X}h_{\mu\nu}.\label{eq:gaugefirsteinstein}
\end{equation}
In order to find the change of the first order linearized Ricci tensor
under the gauge transformation, let us compute the transformation
of the linear order perturbation of the Riemann tensor under the gauge
transformation. Using the previous equations (\ref{eq:firstorderriemann},
\ref{eq:gaugetranschristoffel}), we have
\begin{equation}
	\delta_{X}(R^{\rho}\thinspace_{\mu\sigma\nu})^{\left(1\right)}\cdot h =\bar{\nabla}_{\sigma}(\bar{\nabla}_{\nu}\bar{\nabla}_{\mu}X^{\rho}+\bar{R}^{\rho}\thinspace_{\mu\alpha\nu}X^{\alpha})-\bar{\nabla}_{\nu}(\bar{\nabla}_{\sigma}\bar{\nabla}_{\mu}X^{\rho}+\bar{R}^{\rho}\thinspace_{\mu\alpha\sigma}X^{\alpha}),
\end{equation}
which can be reexpressed as
\begin{eqnarray}
	&&\delta_{X}(R^{\rho}\thinspace_{\mu\sigma\nu})^{\left(1\right)}\cdot h=\left[\bar{\nabla}_{\sigma},\bar{\nabla}_{\nu}\right]\bar{\nabla}_{\mu}X^{\rho}+\bar{R}^{\rho}\thinspace_{\mu\alpha\nu}\bar{\nabla}_{\sigma}X^{\alpha}-\bar{R}^{\rho}\thinspace_{\mu\alpha\sigma}\bar{\nabla}_{\nu}X^{\alpha}\nonumber\\
	&&+X^{\alpha}(\bar{\nabla}_{\sigma}\bar{R}^{\rho}\thinspace_{\mu\alpha\nu}-\bar{\nabla}_{\nu}\bar{R}^{\rho}\thinspace_{\mu\alpha\sigma}).
\end{eqnarray}
By using the identity
\begin{equation}
	\left[\bar{\nabla}_{\sigma},\bar{\nabla}_{\nu}\right]\bar{\nabla}_{\mu}X^{\rho}=\bar{R}_{\sigma\nu\mu}\thinspace^{\alpha}\bar{\nabla}_{\alpha}X^{\rho}+\bar{R}_{\sigma\nu}\thinspace^{\rho}\thinspace_{\alpha}\bar{\nabla}_{\mu}X^{\alpha},
\end{equation}
and the second Bianchi identity
\begin{equation}
	\bar\nabla_{\mu}\bar R_{\nu\rho\alpha\beta}+\bar\nabla_{\alpha}\bar R_{\nu\rho\beta\mu}+\bar\nabla_{\beta}\bar R_{\nu\rho\mu\alpha}=0,
\end{equation}
we can write
\begin{eqnarray}
	&&\delta_{X}(R^{\rho}\thinspace_{\mu\sigma\nu})^{(1)}\cdot h=X^{\alpha}\bar{\nabla}_{\alpha}\bar{R}^{\rho}\thinspace_{\mu\sigma\nu}+(\bar{\nabla}_{\mu}X^{\alpha})\bar{R}^{\rho}\thinspace_{\alpha\sigma\nu}+(\bar{\nabla}_{\sigma}X^{\alpha})\bar{R}^{\rho}\thinspace_{\mu\alpha\nu}\nonumber\\
	&&+(\bar{\nabla}_{\nu}X^{\alpha})\bar{R}^{\rho}\thinspace_{\mu\sigma\alpha}-(\bar{\nabla}_{\alpha}X^{\rho})\bar{R}^{\alpha}\thinspace_{\mu\sigma\nu},
\end{eqnarray}
where the right hand side of the equation is the Lie derivative of the
Riemann tensor with respect to vector field $X$. So we can express
the gauge transformation of the first order linearized Riemann tensor
as
\begin{equation}
	\delta_{X}(R^{\rho}\thinspace_{\mu\sigma\nu})^{\left(1\right)}\cdot h=\text{\ensuremath{\mathscr{L}}}_{X}\bar{R}^{\rho}\thinspace_{\mu\sigma\nu}.
\end{equation}
Contracting and renaming the indices we obtain the change of the Ricci
tensor under the gauge transformation as
\begin{equation}
	(R_{\mu\nu})^{\left(1\right)}\cdot h=\text{\ensuremath{\mathscr{L}}}_{X}\bar{R}_{\mu\nu}.
\end{equation}
Using the previous results, for the first order linearized scalar curvature
we have
\begin{equation}
	\delta_{X}(R)^{\left(1\right)}\cdot h=\bar{g}^{\mu\nu}\text{\ensuremath{\mathscr{L}}}_{X}\bar{R}_{\mu\nu}+\bar{R}_{\mu\nu}\text{\ensuremath{\mathscr{L}}}_{X}\bar{g}^{\mu\nu},
\end{equation}
which becomes
\begin{equation}
	\delta_{X}(R)^{\left(1\right)}\cdot h=\text{\ensuremath{\mathscr{L}}}_{X}\bar{R}.
\end{equation}
Inserting these results in equation (\ref{eq:gaugefirsteinstein}), the gauge
transformation of the first order linearized Einstein tensor can be
expressed as
\begin{equation}
	\delta_{X}(G_{\mu\nu})^{\left(1\right)}\cdot h=\text{\ensuremath{\mathscr{L}}}_{X}\bar{R}_{\mu\nu}-\frac{1}{2}\bar{g}_{\mu\nu}\text{\ensuremath{\mathscr{L}}}_{X}\bar{R}-\frac{1}{2}\bar{R}\text{\ensuremath{\mathscr{L}}}_{X}\bar{g}_{\mu\nu}.
\end{equation}
By combining the Lie derivative terms, it also can be written as
\begin{equation}
	\delta_{X}(G_{\mu\nu})^{\left(1\right)}\cdot h=\text{\ensuremath{\mathscr{L}}}_{X}\left(\bar{R}_{\mu\nu}-\frac{1}{2}\bar{g}_{\mu\nu}\bar{R}\right),
\end{equation}
where the right hand side of the last expression shows the Lie derivative
of the Einstein tensor which is evaluated at the background metric. Then this construction proves equation (\ref{eq:aimfirst}). Let us note that for $\bar{G}_{\mu\nu}=0$, that is for solutions of Einstein's theory, $\delta_{X}(G_{\mu\nu})^{\left(1\right)}\cdot h$, that is the first order linearized Einstein's tensor is gauge invariant.

Using the above results, let us find how the second order linearized
form of the Einstein tensor transforms under the gauge transformations
generated by the flow of the vector field $X$. In the index-free notation one has
\begin{eqnarray}
	D^{2}Ein(g)\cdot\left(h,\text{\ensuremath{\mathscr{L}}}_{X} g\right)+DEin(g)\cdot\text{\ensuremath{\mathscr{L}}}_{X}h\nonumber \\
	=\text{\ensuremath{\mathscr{L}}}_{X}\left(DEin(g)\cdot h\right),
\end{eqnarray}
which reads in local coordinates as 
\begin{equation}
	\delta_{X}(G_{\mu\nu})^{(2)}\cdot[h,h]+(G_{\mu\nu})^{(1)}\cdot\text{\ensuremath{\mathscr{L}}}_{X}h=\text{\ensuremath{\mathscr{L}}}_{X}(G_{\mu\nu})^{(1)}\cdot h.\label{eq:denklem1}
\end{equation}
Let us prove this. By definition we have 
\begin{eqnarray}
	\delta_{X}(G_{\mu\nu})^{(2)}\cdot[h,h]=\delta_{X}(R_{\mu\nu})^{(2)}\cdot[h,h]-\frac{1}{2}\bar{g}_{\mu\nu}\delta_{X}(R)^{(2)}\cdot[h,h]\nonumber \\
	-\frac{1}{2}(R)^{(1)}\cdot h\,\delta_{X}h_{\mu\nu}-\frac{1}{2}h_{\mu\nu}\delta_{X}(R)^{(1)}\cdot h.\label{eq:denklem2}
\end{eqnarray}
Let us calculate the right hand side of the equation term by term. The first term reads
\begin{eqnarray}
	\delta_{X}(R{}_{\mu\nu})^{(2)}\cdot[h,h]=-\left(\delta_{X}h^{\rho\beta}\right)\left(\bar{\nabla}_{\rho}(\Gamma_{\nu\mu\beta})^{(1)}-\bar{\nabla}_{\nu}(\Gamma_{\rho\mu\beta})^{(1)}\right)\nonumber \\
	-h^{\rho\beta}\delta_{X}\left(\bar{\nabla}_{\rho}(\Gamma_{\nu\mu\beta})^{(1)}-\bar{\nabla}_{\nu}(\Gamma_{\rho\mu\beta})^{(1)}\right)\nonumber \\
	-\delta_{X}\left((\Gamma_{\mu\nu}\thinspace^{\alpha})^{(1)}(\Gamma_{\sigma}\thinspace^{\sigma}\thinspace_{\alpha})^{(1)}-(\Gamma_{\mu\sigma}\thinspace^{\alpha})^{(1)}(\Gamma_{\nu}\thinspace^{\sigma}\thinspace_{\alpha})^{(1)}\right).\label{a55}
\end{eqnarray}
Since one has 
\begin{equation}
	\delta_{X}h^{\rho\beta}=-\text{\ensuremath{\mathscr{L}}}_{X}\bar{g}^{\rho\beta},
\end{equation}
the first two terms on the right hand side of the (\ref{a55}) can be
written as 
\begin{eqnarray}
	&  & -\left(\delta_{X}h^{\rho\beta}\right)\left(\bar{\nabla}_{\rho}(\Gamma_{\nu\mu\beta})^{\left(1\right)}-\bar{\nabla}_{\nu}(\Gamma_{\rho\mu\beta})^{\left(1\right)}\right)\nonumber \\
	&  & =\text{\ensuremath{\mathscr{L}}}_{X}(R_{\mu\nu})^{\left(1\right)}.h-\bar{g}^{\rho\beta}\text{\ensuremath{\mathscr{L}}}_{X}\left(\bar{\nabla}_{\rho}(\Gamma_{\nu\mu\beta})^{\left(1\right)}-\bar{\nabla}_{\nu}(\Gamma_{\rho\mu\beta})^{\left(1\right)}\right).
\end{eqnarray}
By using the definition of the linearized Riemann tensor, the terms
on the second line of the (\ref{a55}) yield 
\begin{eqnarray}
	&&h^{\rho}\thinspace_{\beta}\delta_{X}\left(\bar{\nabla}_{\rho}(\Gamma_{\nu\mu}\thinspace^{\beta})^{\left(1\right)}-\bar{\nabla}_{\nu}(\Gamma_{\rho\mu}\thinspace^{\beta})^{\left(1\right)}\right)=h^{\rho}\thinspace_{\beta}\delta_{X}\left(R^{\beta}\thinspace_{\mu\rho\nu}\right)^{\left(1\right)}\cdot h\nonumber\\&&=h^{\rho}\thinspace_{\beta}\text{\ensuremath{\mathscr{L}}}_{X}\bar{R}^{\beta}\thinspace_{\mu\rho\nu}.
\end{eqnarray}
Collecting these expressions we have 
\begin{eqnarray}
	&  & \delta_{X}(R{}_{\mu\nu})^{(2)}\cdot[h,h]=\text{\ensuremath{\mathscr{L}}}_{X}(R_{\mu\nu})^{\left(1\right)}\cdot h-\bar{g}^{\rho\beta}\text{\ensuremath{\mathscr{L}}}_{X}\left(\bar{\nabla}_{\rho}(\Gamma_{\nu\mu\beta})^{\left(1\right)}-\bar{\nabla}_{\nu}(\Gamma_{\rho\mu\beta})^{\left(1\right)}\right)\nonumber \\
	&  & -\delta_{X}\left((\Gamma_{\mu\nu}\thinspace^{\alpha})^{\left(1\right)}(\Gamma_{\sigma}\thinspace^{\sigma}\thinspace_{\alpha})^{\left(1\right)}-(\Gamma_{\mu\sigma}\thinspace^{\alpha})^{\left(1\right)}(\Gamma_{\nu}\thinspace^{\sigma}\thinspace_{\alpha})^{\left(1\right)}\right)\nonumber \\
	&  & -h^{\rho}\thinspace_{\beta}\delta_{X}\left(R^{\beta}\thinspace_{\mu\rho\nu}\right)^{\left(1\right)}\cdot h.
\end{eqnarray}
Using the identity given in the equation (\ref{id2}) we can write
the following equalities 
\begin{eqnarray}
	\text{\ensuremath{\mathscr{L}}}_{X}\bar{\nabla}_{\rho}(\Gamma_{\nu\mu\beta})^{\left(1\right)}=\bar{\nabla}_{\rho}\text{\ensuremath{\mathscr{L}}}_{X}(\Gamma_{\nu\mu\beta})^{\left(1\right)}-(\Gamma_{\sigma\mu\beta})^{\left(1\right)}\delta_{X}(\Gamma_{\rho\nu}\thinspace^{\sigma})^{\left(1\right)}\nonumber \\
	-(\Gamma_{\nu\sigma\beta})^{\left(1\right)}\delta_{X}(\Gamma_{\rho\mu}\thinspace^{\sigma})^{\left(1\right)}-(\Gamma_{\nu\mu\sigma})^{\left(1\right)}\delta_{X}(\Gamma_{\rho\beta}\thinspace^{\sigma})^{\left(1\right)},
\end{eqnarray}
\begin{eqnarray}
	\text{\ensuremath{\mathscr{L}}}_{X}\bar{\nabla}_{\nu}(\Gamma_{\rho\mu\beta})^{\left(1\right)}=\bar{\nabla}_{\nu}\text{\ensuremath{\mathscr{L}}}_{X}(\Gamma_{\rho\mu\beta})^{\left(1\right)}-(\Gamma_{\sigma\mu\beta})^{\left(1\right)}\delta_{X}(\Gamma_{\nu\rho}\thinspace^{\sigma})^{\left(1\right)}\nonumber \\
	-(\Gamma_{\rho\sigma\beta})^{\left(1\right)}\delta_{X}(\Gamma_{\nu\mu}\thinspace^{\sigma})^{\left(1\right)}-(\Gamma_{\rho\mu\sigma})^{\left(1\right)}\delta_{X}(\Gamma_{\nu\beta}\thinspace^{\sigma})^{\left(1\right)}.
\end{eqnarray}
Inserting the results we have 
\begin{eqnarray}
	&  & \delta_{X}(R{}_{\mu\nu})^{(2)}\cdot[h,h]=\text{\ensuremath{\mathscr{L}}}_{X}(R_{\mu\nu})^{\left(1\right)}\cdot h\\
	&  & -\bar{g}^{\rho\beta}\left(\bar{\nabla}_{\rho}\text{\ensuremath{\mathscr{L}}}_{X}(\Gamma_{\nu\mu\beta})^{\left(1\right)}-\bar{\nabla}_{\nu}\text{\ensuremath{\mathscr{L}}}_{X}(\Gamma_{\rho\mu\beta})^{\left(1\right)}\right)+(\Gamma_{\nu\sigma}\thinspace^{\rho})^{\left(1\right)}\delta_{X}(\Gamma_{\rho\mu}\thinspace^{\sigma})^{\left(1\right)}\nonumber \\
	&  & +(\Gamma_{\nu\mu\sigma})^{\left(1\right)}\delta_{X}(\Gamma_{\rho}\thinspace^{\rho\sigma})^{\left(1\right)}-(\Gamma_{\rho\sigma}\thinspace^{\rho})^{\left(1\right)}\delta_{X}(\Gamma_{\nu\mu}\thinspace^{\sigma})^{\left(1\right)}-(\Gamma_{\rho\mu\sigma})^{\left(1\right)}\delta_{X}(\Gamma_{\nu}\thinspace^{\rho\sigma})^{\left(1\right)}\nonumber \\
	&  & -h^{\rho}\thinspace_{\beta}\delta_{X}\left(R^{\beta}\thinspace_{\mu\rho\nu}\right)^{\left(1\right)}.h-\delta_{X}\left((\Gamma_{\mu\nu}\thinspace^{\alpha})^{\left(1\right)}(\Gamma_{\sigma}\thinspace^{\sigma}\thinspace_{\alpha})^{\left(1\right)}-(\Gamma_{\mu\sigma}\thinspace^{\alpha})^{\left(1\right)}(\Gamma_{\nu}\thinspace^{\sigma}\thinspace_{\alpha})^{\left(1\right)}\right).\nonumber 
\end{eqnarray}
Decomposing the first order linearized Christoffel connection as 
\begin{equation}
	(\Gamma_{\nu\sigma}\thinspace^{\rho})^{\left(1\right)}=\bar{\nabla}_{\nu}h_{\sigma}\thinspace^{\rho}-(\Gamma_{\nu}\thinspace^{\rho}\thinspace_{\sigma})^{\left(1\right)},
\end{equation}
and 
\begin{equation}
	(\Gamma_{\rho\sigma}\thinspace^{\rho})^{\left(1\right)}=\bar{\nabla}_{\rho}h_{\sigma}\thinspace^{\rho}-(\Gamma_{\rho}\thinspace^{\rho}\thinspace_{\sigma})^{\left(1\right)},
\end{equation}
we obtain 
\begin{eqnarray}
	&  & \delta_{X}(R{}_{\mu\nu})^{(2)}\cdot[h,h]=\text{\ensuremath{\mathscr{L}}}_{X}(R_{\mu\nu})^{\left(1\right)}\cdot h\\
	&  & -\bar{g}^{\rho\beta}\left(\bar{\nabla}_{\rho}\text{\ensuremath{\mathscr{L}}}_{X}(\Gamma_{\nu\mu\beta})^{\left(1\right)}-\bar{\nabla}_{\nu}\text{\ensuremath{\mathscr{L}}}_{X}(\Gamma_{\rho\mu\beta})^{\left(1\right)}\right)\nonumber \\
	&  & -(\Gamma_{\nu}\thinspace^{\rho}\thinspace_{\sigma})^{\left(1\right)}\delta_{X}(\Gamma_{\rho\mu}\thinspace^{\sigma})^{\left(1\right)}+\left(\bar{\nabla}_{\nu}h_{\sigma}\thinspace^{\rho}\right)\delta_{X}(\Gamma_{\rho\mu}\thinspace^{\sigma})^{\left(1\right)}+(\Gamma_{\nu\mu\sigma})^{\left(1\right)}\delta_{X}(\Gamma_{\rho}\thinspace^{\rho\sigma})^{\left(1\right)}\nonumber \\
	&  & -\left(\bar{\nabla}_{\rho}h_{\sigma}\thinspace^{\rho}\right)\delta_{X}(\Gamma_{\nu\mu}\thinspace^{\sigma})^{\left(1\right)}+(\Gamma_{\rho}\thinspace^{\rho}\thinspace_{\sigma})^{\left(1\right)}\delta_{X}(\Gamma_{\nu\mu}\thinspace^{\sigma})^{\left(1\right)}-(\Gamma_{\rho\mu\sigma})^{\left(1\right)}\delta_{X}(\Gamma_{\nu}\thinspace^{\rho\sigma})^{\left(1\right)}\nonumber \\
	&  & -h^{\rho}\thinspace_{\beta}\delta_{X}\left(R^{\beta}\thinspace_{\mu\rho\nu}\right)^{\left(1\right)}\cdot h-\delta_{X}\left((\Gamma_{\mu\nu}\thinspace^{\alpha})^{\left(1\right)}(\Gamma_{\sigma}\thinspace^{\sigma}\thinspace_{\alpha})^{\left(1\right)}-(\Gamma_{\mu\sigma}\thinspace^{\alpha})^{\left(1\right)}(\Gamma_{\nu}\thinspace^{\sigma}\thinspace_{\alpha})^{\left(1\right)}\right).\nonumber 
\end{eqnarray}
Since we have 
\begin{eqnarray}
	-(\Gamma_{\nu}\thinspace^{\rho}\thinspace_{\sigma})^{\left(1\right)}\delta_{X}(\Gamma_{\rho\mu}\thinspace^{\sigma})^{\left(1\right)}+(\Gamma_{\nu\mu\sigma})^{\left(1\right)}\delta_{X}(\Gamma_{\rho}\thinspace^{\rho\sigma})^{\left(1\right)}+(\Gamma_{\rho}\thinspace^{\rho}\thinspace_{\sigma})^{\left(1\right)}\delta_{X}(\Gamma_{\nu\mu}\thinspace^{\sigma})^{\left(1\right)}\nonumber \\
	-(\Gamma_{\rho\mu\sigma})^{\left(1\right)}\delta_{X}(\Gamma_{\nu}\thinspace^{\rho\sigma})^{\left(1\right)}=\delta_{X}\left((\Gamma_{\rho}\thinspace^{\rho}\thinspace_{\sigma})^{\left(1\right)}(\Gamma_{\nu\mu}\thinspace^{\sigma})^{\left(1\right)}-(\Gamma_{\rho\mu\sigma})^{\left(1\right)}(\Gamma_{\nu}\thinspace^{\rho\sigma})^{\left(1\right)}\right),
\end{eqnarray}
we can write 
\begin{eqnarray}
	&  & \delta_{X}(R{}_{\mu\nu})^{(2)}\cdot[h,h]=\text{\ensuremath{\mathscr{L}}}_{X}(R_{\mu\nu})^{\left(1\right)}\cdot h\\
	&  & -\bar{g}^{\rho\beta}\left(\bar{\nabla}_{\rho}\text{\ensuremath{\mathscr{L}}}_{X}(\Gamma_{\nu\mu\beta})^{\left(1\right)}-\bar{\nabla}_{\nu}\text{\ensuremath{\mathscr{L}}}_{X}(\Gamma_{\rho\mu\beta})^{\left(1\right)}\right)\nonumber \\
	&  & +\left(\bar{\nabla}_{\nu}h_{\sigma}\thinspace^{\rho}\right)\delta_{X}(\Gamma_{\rho\mu}\thinspace^{\sigma})^{\left(1\right)}-\left(\bar{\nabla}_{\rho}h_{\sigma}\thinspace^{\rho}\right)\delta_{X}(\Gamma_{\nu\mu}\thinspace^{\sigma})^{\left(1\right)}-h^{\rho}\thinspace_{\beta}\delta_{X}\left(R^{\beta}\thinspace_{\mu\rho\nu}\right)^{\left(1\right)}\cdot h.\nonumber 
\end{eqnarray}
Expressing the terms which involve the Lie derivative of the Christoffel
connection in terms of the linear order perturbation of the metric
tensor, we can write 
\begin{eqnarray}
	&  & \delta_{X}(R{}_{\mu\nu})^{(2)}\cdot[h,h]=\text{\ensuremath{\mathscr{L}}}_{X}(R_{\mu\nu})^{\left(1\right)}\cdot h+\left(\bar{\nabla}_{\nu}h_{\sigma}\thinspace^{\rho}\right)\delta_{X}(\Gamma_{\rho\mu}\thinspace^{\sigma})^{\left(1\right)}\nonumber \\
	&  & -\left(\bar{\nabla}_{\rho}h_{\sigma}\thinspace^{\rho}\right)\delta_{X}(\Gamma_{\nu\mu}\thinspace^{\sigma})^{\left(1\right)}-h^{\rho}\thinspace_{\beta}\delta_{X}\left(R^{\beta}\thinspace_{\mu\rho\nu}\right)^{\left(1\right)}\cdot h\nonumber \\
	&  & -\frac{1}{2}\bar{g}^{\rho\beta}\left(\bar{\nabla}_{\rho}\text{\ensuremath{\mathscr{L}}}_{X}\left(\bar{\nabla}_{\nu}h_{\mu\beta}+\bar{\nabla}_{\mu}h_{\nu\beta}-\bar{\nabla}_{\beta}h_{\mu\nu}\right)-\bar{\nabla}_{\nu}\text{\ensuremath{\mathscr{L}}}_{X}\bar{\nabla}_{\mu}h_{\rho\beta}\right),
\end{eqnarray}
and using the identity given in the equation (\ref{id1}) we have
\begin{eqnarray}
	&  & \delta_{X}(R{}_{\mu\nu})^{(2)}\cdot[h,h]=\text{\ensuremath{\mathscr{L}}}_{X}(R_{\mu\nu})^{\left(1\right)}\cdot h+h_{\sigma}\thinspace^{\rho}\left(\bar{\nabla}_{\rho}\delta_{X}(\Gamma_{\nu\mu}\thinspace^{\sigma})^{\left(1\right)}-\bar{\nabla}_{\nu}\delta_{X}(\Gamma_{\rho\mu}\thinspace^{\sigma})^{\left(1\right)}\right)\nonumber \\
	&  & -\frac{1}{2}\bar{g}^{\rho\beta}\left(\bar{\nabla}_{\rho}\bar{\nabla}_{\nu}\text{\ensuremath{\mathscr{L}}}_{X}h_{\mu\beta}+\bar{\nabla}_{\rho}\bar{\nabla}_{\mu}\text{\ensuremath{\mathscr{L}}}_{X}h_{\nu\beta}-\bar{\nabla}_{\rho}\bar{\nabla}_{\beta}\text{\ensuremath{\mathscr{L}}}_{X}h_{\mu\nu}-\bar{\nabla}_{\nu}\bar{\nabla}_{\mu}\text{\ensuremath{\mathscr{L}}}_{X}h_{\rho\beta}\right)\nonumber \\
	&  & -h^{\rho}\thinspace_{\beta}\delta_{X}\left(R^{\beta}\thinspace_{\mu\rho\nu}\right)^{\left(1\right)}\cdot h,
\end{eqnarray}
where the second line is the first order linearization of the Ricci
tensor, with a minus sign, which is evaluated at $\text{\ensuremath{\mathscr{L}}}_{X}h$. Also,
from the definition of the first order linearized Riemann tensor we
get 
\begin{equation}
	h_{\sigma}\thinspace^{\rho}\left(\bar{\nabla}_{\rho}\delta_{X}(\Gamma_{\nu\mu}\thinspace^{\sigma})^{\left(1\right)}-\bar{\nabla}_{\nu}\delta_{X}(\Gamma_{\rho\mu}\thinspace^{\sigma})^{\left(1\right)}\right)=h_{\sigma}\thinspace^{\rho}\delta_{X}\left(R^{\sigma}\thinspace_{\mu\rho\nu}\right)^{\left(1\right)}\cdot h,
\end{equation}
and finally one can express the following equation 
\begin{equation}
	\delta_{X}(R{}_{\mu\nu})^{(2)}\cdot[h,h]=\text{\ensuremath{\mathscr{L}}}_{X}(R_{\mu\nu})^{(1)}\cdot h-\left(R_{\mu\nu}\right)^{(1)}\cdot\text{\ensuremath{\mathscr{L}}}_{X}h.
\end{equation}
By using the expression for the second order linearized scalar curvature,
we can write 
\begin{eqnarray}
	\delta_{X}\left(R\right)^{(2)}\cdot[h,h] & =\bar{R}_{\rho\sigma}h^{\sigma\lambda}\delta_{X}h_{\lambda}\thinspace^{\rho}+\bar{R}_{\rho\sigma}h_{\lambda}\thinspace^{\rho}\delta_{X}h^{\sigma\lambda}-\delta_{X}h^{\sigma\rho}(R_{\rho\sigma})^{\left(1\right)}\cdot h\nonumber \\
	& -h^{\sigma\rho}\delta_{X}(R_{\rho\sigma})^{\left(1\right)}\cdot h+\bar{g}^{\sigma\lambda}\delta_{X}(R_{\rho\sigma})^{(2)}\cdot[h,h].
\end{eqnarray}
Using the result for the gauge transformation of the second order
linearized Ricci tensor we have 
\begin{eqnarray}
	\delta_{X}\left(R\right)^{(2)}\cdot[h,h]=\bar{R}_{\rho\sigma}h^{\sigma\lambda}\delta_{X}h_{\lambda}\thinspace^{\rho}+\bar{R}_{\rho\sigma}h_{\lambda}\thinspace^{\rho}\delta_{X}h^{\sigma\lambda}+\text{\ensuremath{\mathscr{L}}}_{X}\bar{g}^{\sigma\rho}(R_{\rho\sigma})^{\left(1\right)}\cdot h\nonumber \\
	-h^{\sigma\rho}\text{\ensuremath{\mathscr{L}}}_{X}\bar{R}_{\rho\sigma}+\bar{g}^{\sigma\lambda}\left(\text{\ensuremath{\mathscr{L}}}_{X}(R_{\rho\sigma})^{\left(1\right)}-\left(R_{\rho\sigma}\right)^{\left(1\right)}\cdot\text{\ensuremath{\mathscr{L}}}_{X}h\right).
\end{eqnarray}
This expression can be recast as 
\begin{eqnarray}
	&  & \delta_{X}\left(R\right)^{(2)}\cdot[h,h]=\bar{R}_{\rho\sigma}h^{\sigma\lambda}\delta_{X}h_{\lambda}\thinspace^{\rho}+\bar{R}_{\rho\sigma}h_{\lambda}\thinspace^{\rho}\delta_{X}h^{\sigma\lambda}+\text{\ensuremath{\mathscr{L}}}_{X}\left(\bar{g}^{\sigma\rho}(R_{\rho\sigma})^{\left(1\right)}\cdot h\right)\nonumber \\
	&  & -\bar{g}^{\sigma\rho}\text{\ensuremath{\mathscr{L}}}_{X}(R_{\rho\sigma})^{\left(1\right)}\cdot h-\text{\ensuremath{\mathscr{L}}}_{X}\left(h^{\sigma\rho}\bar{R}_{\rho\sigma}\right)+\left(\text{\ensuremath{\mathscr{L}}}_{X}h^{\sigma\rho}\right)\bar{R}_{\rho\sigma}+\bar{g}^{\sigma\lambda}\text{\ensuremath{\mathscr{L}}}_{X}(R_{\rho\sigma})^{\left(1\right)}\cdot h\nonumber \\
	&  & -\bar{g}^{\sigma\rho}\left(R_{\rho\sigma}\right)^{\left(1\right)}\cdot\text{\ensuremath{\mathscr{L}}}_{X}h.
\end{eqnarray}
Combining the Lie derivative terms we obtain 
\begin{eqnarray}
	&  & \delta_{X}\left(R\right)^{(2)}\cdot[h,h]=\bar{R}_{\rho\sigma}h^{\sigma\lambda}\delta_{X}h_{\lambda}\thinspace^{\rho}+\bar{R}_{\rho\sigma}h_{\lambda}\thinspace^{\rho}\delta_{X}h^{\sigma\lambda}\\
	&  & +\text{\ensuremath{\mathscr{L}}}_{X}\left(\bar{g}^{\sigma\rho}(R_{\rho\sigma})^{\left(1\right)}\cdot h-h^{\sigma\rho}\bar{R}_{\rho\sigma}\right)+\left(\text{\ensuremath{\mathscr{L}}}_{X}h^{\sigma\rho}\right)\bar{R}_{\rho\sigma}-\bar{g}^{\sigma\rho}\left(R_{\rho\sigma}\right)^{\left(1\right)}\cdot\text{\ensuremath{\mathscr{L}}}_{X}h,\nonumber 
\end{eqnarray}
where 
\begin{equation}
	\bar{g}^{\sigma\rho}(R_{\rho\sigma})^{\left(1\right)}\cdot h-h^{\sigma\rho}\bar{R}_{\rho\sigma}=\left(R\right)^{\left(1\right)}\cdot h.
\end{equation}
After a straightforward calculation, for the remaining terms we can
write 
\begin{equation}
	\bar{R}_{\rho\sigma}h^{\sigma\lambda}\delta_{X}h_{\lambda}\thinspace^{\rho}+\bar{R}_{\rho\sigma}h_{\lambda}\thinspace^{\rho}\delta_{X}h^{\sigma\lambda}+\left(\text{\ensuremath{\mathscr{L}}}_{X}h^{\sigma\rho}\right)\bar{R}_{\rho\sigma}=\bar{R}^{\rho\sigma}\text{\ensuremath{\mathscr{L}}}_{X}h_{\sigma\rho},
\end{equation}
and then we obtain 
\begin{equation}
	\delta_{X}\left(R\right)^{(2)}\cdot[h,h]=\text{\ensuremath{\mathscr{L}}}_{X}\left(R\right)^{\left(1\right)}\cdot h-\left(\bar{g}^{\sigma\rho}\left(R_{\rho\sigma}\right)^{\left(1\right)}\cdot\text{\ensuremath{\mathscr{L}}}_{X}h-\bar{R}^{\rho\sigma}\text{\ensuremath{\mathscr{L}}}_{X}h_{\sigma\rho}\right),
\end{equation}
where the terms in the parenthesis forms the linearized scalar curvature
evaluated at $\text{\ensuremath{\mathscr{L}}}_{X}h$. As a result
the gauge transformation of the second order linearized scalar curvature
becomes 
\begin{equation}
	\delta_{X}\left(R\right)^{(2)}\cdot[h,h]=\text{\ensuremath{\mathscr{L}}}_{X}\left(R\right)^{\left(1\right)}\cdot h-\left(R\right)^{\left(1\right)}\cdot\text{\ensuremath{\mathscr{L}}}_{X}h.
\end{equation}
Using our results in the equation (\ref{eq:denklem2}) we have 
\begin{eqnarray}
	&  & \delta_{X}(G_{\mu\nu})^{(2)}\cdot[h,h]=\text{\ensuremath{\mathscr{L}}}_{X}(R_{\mu\nu})^{\left(1\right)}\cdot h-\left(R_{\mu\nu}\right)^{\left(1\right)}\cdot\text{\ensuremath{\mathscr{L}}}_{X}h-\frac{1}{2}h_{\mu\nu}\delta_{X}(R)^{\left(1\right)}\cdot h\nonumber \\
	&  & -\frac{1}{2}\bar{g}_{\mu\nu}\left(\text{\ensuremath{\mathscr{L}}}_{X}\left(R\right)^{\left(1\right)}\cdot h-\left(R\right)^{\left(1\right)}\cdot\text{\ensuremath{\mathscr{L}}}_{X}h\right)-\frac{1}{2}\delta_{X}h_{\mu\nu}(R)^{\left(1\right)}\cdot h,
\end{eqnarray}
and it can be expressed as 
\begin{eqnarray}
	&  & \delta_{X}(G_{\mu\nu})^{(2)}\cdot[h,h]=\text{\ensuremath{\mathscr{L}}}_{X}\left((R_{\mu\nu})^{\left(1\right)}\cdot h-\frac{1}{2}\bar{g}_{\mu\nu}\left(R\right)^{\left(1\right)}\cdot h-\frac{1}{2}h_{\mu\nu}\bar{R}\right)\nonumber \\
	&  & -\left(\left(R_{\mu\nu}\right)^{\left(1\right)}\cdot\text{\ensuremath{\mathscr{L}}}_{X}h-\frac{1}{2}\bar{g}_{\mu\nu}\left(R\right)^{\left(1\right)}\cdot\text{\ensuremath{\mathscr{L}}}_{X}h-\frac{1}{2}\bar{R}\text{\ensuremath{\mathscr{L}}}_{X}h_{\mu\nu}\right).
\end{eqnarray}
Here the total Lie derivative terms express the Lie derivative of
the first order linearized Einstein tensor and the second line is
the first order linearized Einstein tensor evaluated at $\text{\ensuremath{\mathscr{L}}}_{X}h$.
We can express the final result as 
\begin{equation}
	\delta_{X}(G_{\mu\nu})^{(2)}\cdot[h,h]=\text{\ensuremath{\mathscr{L}}}_{X}(G_{\mu\nu})^{\left(1\right)}\cdot h-\left(G_{\mu\nu}\right)^{\left(1\right)}\cdot\text{\ensuremath{\mathscr{L}}}_{X}h,
\end{equation}
which is the desired formula that was used in the text to study the
gauge-invariance properties of the Taub charges. Observe that if $h_{\mu\nu}$ solve the linearized equations, the first term on the right hand side vanishes, but the second one does not vanish. Therefore, unlike the first order linearized Einstein tensor, the second order linearized Einstein tensor is not gauge invariant.

\newpage

\section{Appendix: Explicit form of the K-tensor in AdS}
\label{chp:appendixb}
Here let us depict some of the intermediate steps leading to (\ref{div1}).
Assuming a general form for the second order perturbation of the metric tensor, $k_{\mu\nu}$, as 
\begin{equation}
	k_{\mu\nu}=a\,h_{\mu\beta}h_{\nu}^{\beta}+b\,hh_{\mu\nu}+\bar{g}_{\mu\nu}(c\,h_{\alpha\beta}^{2}+d\,h^{2}).
\end{equation}
The trace of the $k_{\mu\nu}$ tensor is
\begin{equation}
	k=\bar g^{\mu\nu}k_{\mu\nu}=(a+cD)h_{\alpha\beta}^{2}+(b+dD)h^{2},
\end{equation}
where the constants $a,b,c,d$ are to be determined. The first order
Ricci operator evaluated at the $k_{\mu\nu}$ tensor is 
\begin{equation}
	(R_{\mu\nu})^{(1)}\cdot k=\frac{1}{2}(\bar{\nabla}_{\alpha}\bar{\nabla}_{\mu}k_{\nu}^{\alpha}+\bar{\nabla}_{\alpha}\bar{\nabla}_{\nu}k_{\mu}^{\alpha}-\bar{\square}k_{\mu\nu}-\bar{\nabla}_{\mu}\bar{\nabla}_{\nu}k),
\end{equation}
whose explicit form follows as 
\begin{eqnarray}
	&&(R_{\mu\nu})^{(1)}\cdot k=\frac{a}{2}(\bar{\nabla}_{\alpha}\bar{\nabla}_{\mu}h^{\alpha\beta}h_{\beta\nu}+\bar{\nabla}_{\alpha}\bar{\nabla}_{\nu}h^{\alpha\beta}h_{\beta\mu}-\bar{\square}h_{\nu}^{\beta}h_{\beta\mu}-\bar{\nabla}_{\mu}\bar{\nabla}_{\nu}h_{\alpha\beta}^{2})\nonumber \\
	&&+\frac{b}{2}(\bar{\nabla}_{\alpha}\bar{\nabla}_{\mu}hh_{\nu}^{\alpha}+\bar{\nabla}_{\alpha}\bar{\nabla}_{\nu}hh_{\mu}^{\alpha}-\bar{\square}hh_{\mu\nu}-\bar{\nabla}_{\mu}\bar{\nabla}_{\nu}h^{2})\nonumber \\
	&&+\frac{c}{2}(\bar{\nabla}_{\nu}\bar{\nabla}_{\mu}h_{\alpha\beta}^{2}+\bar{\nabla}_{\mu}\bar{\nabla}_{\nu}h_{\alpha\beta}^{2}-\bar{g}_{\mu\nu}\bar{\square}h_{\alpha\beta}^{2}-D\bar{\nabla}_{\mu}\bar{\nabla}_{\nu}h_{\alpha\beta}^{2})\nonumber \\
	&&+\frac{d}{2}(\bar{\nabla}_{\nu}\bar{\nabla}_{\mu}h^{2}+\bar{\nabla}_{\mu}\bar{\nabla}_{\nu}h^{2}-\bar{g}_{\mu\nu}\bar{\square}h^{2}-D\bar{\nabla}_{\mu}\bar{\nabla}_{\nu}h^{2}).
\end{eqnarray}
We should set $a=1$ and $b=-1/2$ to get the second order linearized Ricci tensor given in (\ref{a12})
\begin{eqnarray}
	&&(R_{\mu\nu})^{(1)}\cdot k=\frac{1}{2}\bar{\nabla}_{\alpha}\left(h^{\alpha\beta}\left(\bar{\nabla}_{\mu}h_{\nu\beta}+\bar{\nabla}_{\nu}h_{\mu\beta}\right)\right)\nonumber \\
	&&+\frac{1}{2}\bar{\nabla}_{\alpha}\left(h_{\beta\nu}\bar{\nabla}_{\mu}h^{\alpha\beta}+h_{\beta\mu}\bar{\nabla}_{\nu}h^{\alpha\beta}+h^{\alpha\beta}\bar{\nabla}_{\beta}h_{\nu\mu}-\bar{\nabla}^{\alpha}\left(h_{\nu}^{\beta}h_{\beta\mu}\right)\right)\nonumber \\
	&&-\bar{\nabla}_{\nu}\left(h^{\alpha\beta}\bar{\nabla}_{\mu}h_{\alpha\beta}\right)-\frac{1}{4}h\bar{\nabla}_{\alpha}\left(\bar{\nabla}_{\mu}h_{\nu}^{\alpha}+\bar{\nabla}_{\nu}h_{\mu}^{\alpha}-\bar{\nabla}^{\alpha}h_{\mu\nu}\right)+\frac{1}{2}\bar{\nabla}_{\nu}\left(h\bar{\nabla}_{\mu}h\right)\nonumber \\
	&&-\frac{1}{4}\bar{\nabla}_{\alpha}(h_{\nu}^{\alpha}\bar{\nabla}_{\mu}h+h_{\mu}^{\alpha}\bar{\nabla}_{\nu}h-h_{\mu\nu}\bar{\nabla}^{\alpha}h)-\frac{1}{4}\bar{\nabla}^{\beta}h\left(\bar{\nabla}_{\mu}h_{\nu\beta}+\bar{\nabla}_{\nu}h_{\mu\beta}-\bar{\nabla}_{\beta}h_{\mu\nu}\right)\nonumber \\
	&&+\frac{c}{2}((2-D)\bar{\nabla}_{\nu}\bar{\nabla}_{\mu}h_{\alpha\beta}^{2}-\bar{g}_{\mu\nu}\bar{\square}h_{\alpha\beta}^{2})+\frac{d}{2}((2-D)\bar{\nabla}_{\nu}\bar{\nabla}_{\mu}h^{2}-\bar{g}_{\mu\nu}\bar{\square}h^{2}).
\end{eqnarray}
Using the expression for $(R_{\mu\nu})^{(2)}\cdot[h,h]$, finally
the Ricci tensor evaluated at the $k_{\mu\nu}$ tensor becomes
\begin{eqnarray}
	&&(R_{\mu\nu})^{(1)}\cdot k=-(R_{\mu\nu})^{(2)}\cdot[h,h]-\frac{3}{4}\bar{\nabla}_{\nu}h^{\alpha\beta}\bar{\nabla}_{\mu}h_{\alpha\beta}+\frac{1}{2}\bar{\nabla}_{\alpha}h_{\mu\beta}\bar{\nabla}^{\alpha}h_{\nu}^{\beta}\nonumber \\
	&&+\frac{1}{2}\bar{\nabla}_{\alpha}\left(h_{\beta\nu}\bar{\nabla}_{\mu}h^{\alpha\beta}+h_{\beta\mu}\bar{\nabla}_{\nu}h^{\alpha\beta}+h^{\alpha\beta}\bar{\nabla}_{\beta}h_{\nu\mu}-\bar{\nabla}^{\alpha}\left(h_{\nu}^{\beta}h_{\beta\mu}\right)\right)\nonumber \\
	&&-\frac{h}{2}(R_{\mu\nu})^{(1)}\cdot h+\frac{1}{2}\bar{\nabla}_{\nu}\left(h\bar{\nabla}_{\mu}h\right)-\frac{1}{4}\bar{\nabla}_{\alpha}(h_{\nu}^{\alpha}\bar{\nabla}_{\mu}h+h_{\mu}^{\alpha}\bar{\nabla}_{\nu}h-h_{\mu\nu}\bar{\nabla}^{\alpha}h)\nonumber \\
	&&-\frac{1}{2}h^{\alpha\beta}\bar{\nabla}_{\nu}\bar{\nabla}_{\mu}h_{\alpha\beta}+\frac{c}{2}((2-D)\bar{\nabla}_{\nu}\bar{\nabla}_{\mu}h_{\alpha\beta}^{2}-\bar{g}_{\mu\nu}\bar{\square}h_{\alpha\beta}^{2})\nonumber \\
	&&-\frac{1}{4}h\bar{\nabla}_{\nu}\bar{\nabla}_{\mu}h-\frac{1}{2}\bar{\nabla}_{\alpha}h_{\mu\beta}\bar{\nabla}^{\beta}h_{\nu}^{\alpha}+\frac{d}{2}((2-D)\bar{\nabla}_{\nu}\bar{\nabla}_{\mu}h^{2}-\bar{g}_{\mu\nu}\bar{\square}h^{2}).
\end{eqnarray}
For the first order linearized scalar curvature which is evaluated
at the $k_{\mu\nu}$ tensor we have
\begin{equation}
	(R)^{\left(1\right)}\cdot k=\bar{g}^{\mu\nu}(R_{\mu\nu})^{\left(1\right)}\cdot k-\bar{R}_{\mu\nu}k^{\mu\nu},
\end{equation}
which is explicitly
\begin{eqnarray}
	&&(R)^{\left(1\right)}\cdot k=-\bar{g}^{\mu\nu}(R_{\mu\nu})^{\left(2\right)}\cdot\left[h,h\right]-\frac{3}{4}\bar{\nabla}^{\mu}h^{\alpha\beta}\bar{\nabla}_{\mu}h_{\alpha\beta}+\frac{1}{2}\bar{\nabla}_{\alpha}h_{\mu\beta}\bar{\nabla}^{\alpha}h^{\beta\mu}\nonumber\\
	&&-\frac{1}{2}\bar{\nabla}_{\alpha}h_{\mu\beta}\bar{\nabla}^{\beta}h^{\mu\alpha}+\frac{1}{2}\bar{\nabla}_{\alpha}\left(2h_{\beta}^{\mu}\bar{\nabla}_{\mu}h^{\alpha\beta}+h^{\alpha\beta}\bar{\nabla}_{\beta}h-\bar{\nabla}^{\alpha}\left(h_{\beta\mu}^{2}\right)\right)\nonumber\\
	&&-\frac{1}{2}h^{\alpha\beta}\bar{\square}h_{\alpha\beta}-\frac{h}{2}\bar{g}^{\mu\nu}(R_{\mu\nu})^{\left(1\right)}\cdot h-\frac{1}{4}h\bar{\square}h+\frac{1}{2}\bar{\nabla}^{\mu}\left(h\bar{\nabla}_{\mu}h\right)\nonumber\\
	&&+c(1-D)\bar{\square}h_{\alpha\beta}^{2}+d(1-D)\bar{\square}h^{2}-\bar{R}_{\mu\nu}\left(h_{\beta}^{\mu}h^{\beta\nu}-\frac{1}{2}hh^{\mu\nu}+\bar{g}^{\mu\nu}\left(ch_{\alpha\beta}^{2}+dh^{2}\right)\right)\nonumber\\&&-\frac{1}{4}\bar{\nabla}_{\alpha}(2h^{\alpha\mu}\bar{\nabla}_{\mu}h-h\bar{\nabla}^{\alpha}h).
\end{eqnarray}
By using
\begin{equation}
	(R)^{(2)}\cdot[h,h]=\bar{R}_{\mu\nu}h_{\alpha}^{\mu}h^{\alpha\nu}-h^{\mu\nu}(R_{\mu\nu})^{\left(1\right)}+\bar{g}^{\mu\nu}(R_{\mu\nu})^{(2)}\cdot[h,h],
\end{equation}
we obtain
\begin{eqnarray}
	&&(R)^{\left(1\right)}\cdot k=-(R)^{(2)}\cdot[h,h]-h^{\mu\nu}(R_{\mu\nu})^{\left(1\right)}\cdot h-\frac{5}{4}\bar{\nabla}^{\mu}h^{\alpha\beta}\bar{\nabla}_{\mu}h_{\alpha\beta}\nonumber\\
	&&+\frac{1}{2}\bar{\nabla}_{\alpha}h_{\mu\beta}\bar{\nabla}^{\beta}h^{\mu\alpha}+\frac{1}{2}h^{\mu\beta}\left(2\bar{\nabla}_{\alpha}\bar{\nabla}_{\mu}h_{\beta}^{\alpha}-\bar{\nabla}_{\alpha}\bar{\nabla}^{\alpha}h_{\mu\beta}-\bar{\nabla}_{\mu}\bar{\nabla}_{\beta}h\right)\nonumber\\
	&&+\frac{1}{2}h^{\alpha\beta}\bar{\nabla}_{\alpha}\bar{\nabla}_{\beta}h-h^{\alpha\beta}\bar{\square}h_{\alpha\beta}-\frac{h}{2}(R)^{\left(1\right)}\cdot h+\frac{1}{2}h\bar{\square}h+\frac{3}{4}\bar{\nabla}^{\mu}h\bar{\nabla}_{\mu}h\nonumber\\
	&&+c(1-D)\bar{\square}h_{\alpha\beta}^{2}+d(1-D)\bar{\square}h^{2}-\bar{R}(ch_{\alpha\beta}^{2}+dh^{2}),
\end{eqnarray}
where
\begin{equation}
	\frac{1}{2}h^{\mu\beta}\left(2\bar{\nabla}_{\alpha}\bar{\nabla}_{\mu}h_{\beta}^{\alpha}-\bar{\square}h_{\mu\beta}-\bar{\nabla}_{\mu}\bar{\nabla}_{\beta}h\right)=h^{\mu\beta}(R_{\mu\beta})^{\left(1\right)}\cdot h.
\end{equation}
Finally the Ricci tensor evaluated at the $k_{\mu\nu}$ tensor becomes:
\begin{eqnarray}
	(R)^{(1)}\cdot k=-(R)^{(2)}\cdot[h,h]-\frac{5}{4}\bar{\nabla}^{\mu}h^{\alpha\beta}\bar{\nabla}_{\mu}h_{\alpha\beta}+\frac{1}{2}\bar{\nabla}_{\alpha}h_{\mu\beta}\bar{\nabla}^{\beta}h^{\mu\alpha}\nonumber \\
	+\frac{1}{2}h^{\alpha\beta}\bar{\nabla}_{\alpha}\bar{\nabla}_{\beta}h-h^{\alpha\beta}\bar{\square}h_{\alpha\beta}-\frac{h}{2}(R)^{(1)}\cdot h+\frac{1}{2}h\bar{\square}h+\frac{3}{4}\bar{\nabla}^{\mu}h\bar{\nabla}_{\mu}h\nonumber \\
	+c(1-D)\bar{\square}h_{\alpha\beta}^{2}+d(1-D)\bar{\square}h^{2}-\bar{R}(ch_{\alpha\beta}^{2}+dh^{2}).
\end{eqnarray}
We can express the first order linearized cosmological Einstein tensor
evaluated at the $k_{\mu\nu}$ tensor as
\begin{equation}
	\left(\text{\ensuremath{\mathcal{G}}}_{\mu\nu}\right)^{\left(1\right)}\cdot k=\left(R_{\mu\nu}\right)^{\left(1\right)}\cdot k-\frac{1}{2}\bar{g}_{\mu\nu}\left(R\right)^{\left(1\right)}\cdot k-\frac{2\Lambda}{(D-2)}k_{\mu\nu},
\end{equation}
where we made use of the fact that the background is an AdS spacetime
with the normalization $\bar{R}_{\mu\nu}=\frac{2\Lambda}{D-2}\bar{g}_{\mu\nu}$.
Inserting our results in the last expression we have
\begin{eqnarray}
	&&\left(\text{\ensuremath{\mathcal{G}}}_{\mu\nu}\right)^{\left(1\right)}\cdot k=-(R_{\mu\nu})^{(2)}\cdot\left[h,h\right]+\frac{1}{2}\bar{g}_{\mu\nu}(R)^{(2)}\cdot[h,h]\nonumber\\
	&&-\frac{h}{2}\left((R_{\mu\nu})^{\left(1\right)}\cdot h-\frac{1}{2}\bar{g}_{\mu\nu}(R)^{\left(1\right)}\cdot h-\frac{2\Lambda}{(D-2)}h_{\mu\nu}\right)+\frac{1}{2}\bar{\nabla}_{\alpha}h_{\mu\beta}\bar{\nabla}^{\alpha}h_{\nu}^{\beta}\nonumber\\
	&&+\frac{1}{2}\bar{\nabla}_{\alpha}\left(h_{\beta\nu}\bar{\nabla}_{\mu}h^{\alpha\beta}+h_{\beta\mu}\bar{\nabla}_{\nu}h^{\alpha\beta}+h^{\alpha\beta}\bar{\nabla}_{\beta}h_{\nu\mu}-\bar{\nabla}^{\alpha}\left(h_{\nu}^{\beta}h_{\beta\mu}\right)\right)\nonumber\\
	&&-\frac{3}{4}\bar{\nabla}_{\nu}h^{\alpha\beta}\bar{\nabla}_{\mu}h_{\alpha\beta}-\frac{1}{2}h^{\alpha\beta}\bar{\nabla}_{\nu}\bar{\nabla}_{\mu}h_{\alpha\beta}-\frac{1}{4}h\bar{\nabla}_{\nu}\bar{\nabla}_{\mu}h+\frac{1}{2}\bar{\nabla}_{\nu}\left(h\bar{\nabla}_{\mu}h\right)\nonumber\\
	&&-\frac{1}{4}\bar{\nabla}_{\alpha}(h_{\nu}^{\alpha}\bar{\nabla}_{\mu}h+h_{\mu}^{\alpha}\bar{\nabla}_{\nu}h-h_{\mu\nu}\bar{\nabla}^{\alpha}h)+\frac{c}{2}(2-D)\bar{\nabla}_{\nu}\bar{\nabla}_{\mu}h_{\alpha\beta}^{2}\nonumber\\
	&&+\frac{d}{2}(2-D)\bar{\nabla}_{\nu}\bar{\nabla}_{\mu}h^{2}-\frac{1}{2}\bar{g}_{\mu\nu}\left(-\frac{5}{4}\bar{\nabla}^{\sigma}h^{\alpha\beta}\bar{\nabla}_{\sigma}h_{\alpha\beta}+\frac{1}{2}\bar{\nabla}_{\alpha}h_{\sigma\beta}\bar{\nabla}^{\beta}h^{\sigma\alpha}\right)\nonumber\\
	&&-\frac{1}{2}\bar{g}_{\mu\nu}\left(\frac{1}{2}h^{\alpha\beta}\bar{\nabla}_{\alpha}\bar{\nabla}_{\beta}h-h^{\alpha\beta}\bar{\square}h_{\alpha\beta}+\frac{1}{2}h\bar{\square}h+\frac{3}{4}\bar{\nabla}^{\sigma}h\bar{\nabla}_{\sigma}h\right)\nonumber\\
	&&-\frac{1}{2}\bar{g}_{\mu\nu}\left(c(2-D)\bar{\square}h_{\alpha\beta}^{2}+d(2-D)\bar{\square}h^{2})-\bar{R}(ch_{\alpha\beta}^{2}+dh^{2})\right)\nonumber\\
	&&-\frac{1}{2}\bar{\nabla}_{\alpha}h_{\mu\beta}\bar{\nabla}^{\beta}h_{\nu}^{\alpha}-\frac{2\Lambda}{(D-2)}\left(h_{\mu\beta}h_{\nu}^{\beta}+\bar{g}_{\mu\nu}(ch_{\alpha\beta}^{2}+dh^{2})\right),
\end{eqnarray}
where
\begin{equation}
	-\left(R_{\mu\nu}\right)^{(2)}\cdot[h,h]+\frac{1}{2}\bar{g}_{\mu\nu}\left(R\right)^{(2)}\cdot[h,h]=-\left(\text{\ensuremath{\mathcal{G}}}_{\mu\nu}\right)^{(2)}\cdot[h,h]-\frac{1}{2}h_{\mu\nu}\left(R\right)^{\left(1\right)}\cdot h.
\end{equation}
Then the linearized Einstein tensor evaluated at the $k_{\mu\nu}$ tensor can be found
as 
\begin{eqnarray}
	&  & \left(\text{\ensuremath{\mathcal{G}}}_{\mu\nu}\right)^{(1)}\cdot k=-\left(\text{\ensuremath{\mathcal{G}}}_{\mu\nu}\right)^{(2)}\cdot[h,h]-\frac{1}{2}h_{\mu\nu}\left(R\right)^{(1)}\cdot h-\frac{h}{2}(\mathcal{G}{}_{\mu\nu})^{(1)}\cdot h\nonumber \\
	&  & -\frac{3}{4}\bar{\nabla}_{\nu}h^{\alpha\beta}\bar{\nabla}_{\mu}h_{\alpha\beta}+\frac{1}{2}\bar{\nabla}_{\alpha}h_{\mu\beta}\bar{\nabla}^{\alpha}h_{\nu}^{\beta}-\frac{1}{2}\bar{\nabla}_{\alpha}h_{\mu\beta}\bar{\nabla}^{\beta}h_{\nu}^{\alpha}\nonumber \\
	&  & +\frac{1}{2}\bar{\nabla}_{\alpha}\left(h_{\beta\nu}\bar{\nabla}_{\mu}h^{\alpha\beta}+h_{\beta\mu}\bar{\nabla}_{\nu}h^{\alpha\beta}+h^{\alpha\beta}\bar{\nabla}_{\beta}h_{\nu\mu}-\bar{\nabla}^{\alpha}\left(h_{\nu}^{\beta}h_{\beta\mu}\right)\right)\nonumber \\
	&  & -\frac{1}{2}h^{\alpha\beta}\bar{\nabla}_{\nu}\bar{\nabla}_{\mu}h_{\alpha\beta}-\frac{1}{4}h\bar{\nabla}_{\nu}\bar{\nabla}_{\mu}h+\frac{1}{2}\bar{\nabla}_{\nu}\left(h\bar{\nabla}_{\mu}h\right)+\frac{d}{2}(2-D)\bar{\nabla}_{\nu}\bar{\nabla}_{\mu}h^{2}\nonumber \\
	&  & -\frac{1}{4}\bar{\nabla}_{\alpha}(h_{\nu}^{\alpha}\bar{\nabla}_{\mu}h+h_{\mu}^{\alpha}\bar{\nabla}_{\nu}h-h_{\mu\nu}\bar{\nabla}^{\alpha}h)+\frac{c}{2}(2-D)\bar{\nabla}_{\nu}\bar{\nabla}_{\mu}h_{\alpha\beta}^{2}\nonumber \\
	&  & -\frac{1}{2}\bar{g}_{\mu\nu}\left(-\frac{5}{4}\bar{\nabla}^{\sigma}h^{\alpha\beta}\bar{\nabla}_{\sigma}h_{\alpha\beta}+\frac{1}{2}\bar{\nabla}_{\alpha}h_{\sigma\beta}\bar{\nabla}^{\beta}h^{\sigma\alpha}\right)\nonumber \\
	&  & -\frac{1}{2}\bar{g}_{\mu\nu}\left(\frac{1}{2}h^{\alpha\beta}\bar{\nabla}_{\alpha}\bar{\nabla}_{\beta}h-h^{\alpha\beta}\bar{\square}h_{\alpha\beta}+\frac{1}{2}h\bar{\square}h+\frac{3}{4}\bar{\nabla}^{\sigma}h\bar{\nabla}_{\sigma}h\right)\nonumber \\
	&  & -\frac{1}{2}\bar{g}_{\mu\nu}\left(c(2-D)\bar{\square}h_{\alpha\beta}^{2}+d(2-D)\bar{\square}h^{2})-\bar{R}(ch_{\alpha\beta}^{2}+dh^{2})\right)\nonumber \\
	&  & -\frac{2\Lambda}{D-2}\left(h_{\mu\beta}h_{\nu}^{\beta}+\bar{g}_{\mu\nu}(ch_{\alpha\beta}^{2}+dh^{2})\right).
\end{eqnarray}
We can express the last equation in a more compact form as
\begin{equation}
	\left(\text{\ensuremath{\mathcal{G}}}_{\mu\nu}\right)^{\left(1\right)}\cdot k=-\left(\text{\ensuremath{\mathcal{G}}}_{\mu\nu}\right)^{(2)}\cdot[h,h]+K_{\mu\nu},
\end{equation}
where the final form of the $K_{\mu\nu}$ tensor is 
\begin{eqnarray}
	&  & K_{\mu\nu}=-\frac{1}{2}h_{\mu\nu}\left(R\right)^{(1)}\cdot h-\frac{h}{2}(\text{\ensuremath{\mathcal{G}}}{}_{\mu\nu})^{(1)}\cdot h-\frac{3}{4}\bar{\nabla}_{\nu}h^{\alpha\beta}\bar{\nabla}_{\mu}h_{\alpha\beta}+\frac{1}{2}\bar{\nabla}_{\alpha}h_{\mu\beta}\bar{\nabla}^{\alpha}h_{\nu}^{\beta}\nonumber \\
	&  & -\frac{1}{2}\bar{\nabla}_{\alpha}h_{\mu\beta}\bar{\nabla}^{\beta}h_{\nu}^{\alpha}+\frac{1}{2}\bar{\nabla}_{\alpha}\left(h_{\beta\nu}\bar{\nabla}_{\mu}h^{\alpha\beta}+h_{\beta\mu}\bar{\nabla}_{\nu}h^{\alpha\beta}+h^{\alpha\beta}\bar{\nabla}_{\beta}h_{\nu\mu}-\bar{\nabla}^{\alpha}\left(h_{\nu}^{\beta}h_{\beta\mu}\right)\right)\nonumber \\
	&  & -\frac{1}{4}h\bar{\nabla}_{\nu}\bar{\nabla}_{\mu}h+\frac{1}{2}\bar{\nabla}_{\nu}\left(h\bar{\nabla}_{\mu}h\right)-\frac{1}{4}\bar{\nabla}_{\alpha}(h_{\nu}^{\alpha}\bar{\nabla}_{\mu}h+h_{\mu}^{\alpha}\bar{\nabla}_{\nu}h-h_{\mu\nu}\bar{\nabla}^{\alpha}h)\nonumber \\
	&  & +\frac{c}{2}(2-D)\bar{\nabla}_{\nu}\bar{\nabla}_{\mu}h_{\alpha\beta}^{2}+\frac{d}{2}(2-D)\bar{\nabla}_{\nu}\bar{\nabla}_{\mu}h^{2}-\frac{2\Lambda}{D-2}\left(h_{\mu\beta}h_{\nu}^{\beta}+\bar{g}_{\mu\nu}(ch_{\alpha\beta}^{2}+dh^{2})\right)\nonumber \\
	&  & -\frac{1}{2}h^{\alpha\beta}\bar{\nabla}_{\nu}\bar{\nabla}_{\mu}h_{\alpha\beta}-\frac{1}{2}\bar{g}_{\mu\nu}\left(-\frac{5}{4}\bar{\nabla}^{\sigma}h^{\alpha\beta}\bar{\nabla}_{\sigma}h_{\alpha\beta}+\frac{1}{2}\bar{\nabla}_{\alpha}h_{\sigma\beta}\bar{\nabla}^{\beta}h^{\sigma\alpha}\right)\nonumber \\
	&  & -\frac{1}{2}\bar{g}_{\mu\nu}\left(+\frac{1}{2}h^{\alpha\beta}\bar{\nabla}_{\alpha}\bar{\nabla}_{\beta}h-h^{\alpha\beta}\bar{\square}h_{\alpha\beta}+\frac{1}{2}h\bar{\square}h+\frac{3}{4}\bar{\nabla}^{\sigma}h\bar{\nabla}_{\sigma}h\right)\nonumber \\
	&  & -\frac{1}{2}\bar{g}_{\mu\nu}\left(c(2-D)\bar{\square}h_{\alpha\beta}^{2}+d(2-D)\bar{\square}h^{2})-\bar{R}(ch_{\alpha\beta}^{2}+dh^{2})\right).
\end{eqnarray}
Using $\left(\text{\ensuremath{\mathcal{G}}}_{\mu\nu}\right)^{\left(1\right)}\cdot h=0$
and choosing the transverse traceless gauge for the sake of simplicity, which yields $\bar{\square}h_{\mu\nu}=\frac{4\Lambda}{(D-1)(D-2)}h_{\mu\nu}$, the $K$ tensor becomes 
\begin{eqnarray}
	&&K_{\mu\nu}=\frac{1}{2}h^{\alpha\beta}\bar{\nabla}_{\alpha}\bar{\nabla}_{\beta}h_{\nu\mu}-\frac{1}{2}\bar{\nabla}_{\alpha}h_{\mu\beta}\bar{\nabla}^{\alpha}h_{\nu}^{\beta}-\frac{1}{2}\bar{\nabla}_{\alpha}h_{\mu\beta}\bar{\nabla}^{\beta}h_{\nu}^{\alpha}\nonumber\\
	&&+\frac{1}{2}\bar{\nabla}_{\alpha}h_{\beta\nu}\bar{\nabla}_{\mu}h^{\alpha\beta}+\frac{1}{2}\bar{\nabla}_{\alpha}h_{\beta\mu}\bar{\nabla}_{\nu}h^{\alpha\beta}+\left(c(2-D)-\frac{1}{2}\right)h^{\alpha\beta}\bar{\nabla}_{\nu}\bar{\nabla}_{\mu}h_{\alpha\beta}\nonumber\\
	&&+\left(\frac{\Lambda}{(D-2)(D-2)}\left(cD^{2}+cD-6c+2\right)\right)\bar{g}_{\mu\nu}h_{\alpha\beta}^{2}\nonumber\\
	&&+\left(c(D-2)+\frac{5}{8}\right)\bar{g}_{\mu\nu}\bar{\nabla}^{\sigma}h^{\alpha\beta}\bar{\nabla}_{\sigma}h_{\alpha\beta}-\frac{1}{4}\bar{g}_{\mu\nu}\bar{\nabla}_{\alpha}h_{\sigma\beta}\bar{\nabla}^{\beta}h^{\sigma\alpha}\nonumber\\
	&&+\left(c(2-D)-\frac{3}{4}\right)\bar{\nabla}_{\nu}h^{\alpha\beta}\bar{\nabla}_{\mu}h_{\alpha\beta}-\frac{2\Lambda}{(D-1)(D-2)}h_{\mu\beta}h_{\nu}^{\beta}.
\end{eqnarray}
Using the following identities
\begin{equation}
	h^{\alpha\beta}\bar{\nabla}_{\alpha}\bar{\nabla}_{\beta}h_{\nu\mu}=\bar{\nabla}_{\alpha}\left(h^{\alpha\beta}\bar{\nabla}_{\beta}h_{\nu\mu}\right),
\end{equation}
\begin{equation}
	\bar{\nabla}_{\alpha}h_{\mu\beta}\bar{\nabla}^{\alpha}h_{\nu}^{\beta}=\bar{\nabla}_{\alpha}\left(h_{\mu\beta}\bar{\nabla}^{\alpha}h_{\nu}^{\beta}\right)-\frac{4\Lambda}{(D-1)(D-2)}h_{\mu\beta}h_{\nu}^{\beta},
\end{equation}
\begin{equation}
	\bar{\nabla}_{\alpha}h_{\mu\beta}\bar{\nabla}^{\beta}h_{\nu}^{\alpha}=\bar{\nabla}_{\alpha}\left(h_{\mu\beta}\bar{\nabla}^{\beta}h_{\nu}^{\alpha}\right)-\frac{2\Lambda D}{(D-1)(D-2)}h_{\mu\beta}h_{\nu}^{\beta},
\end{equation}
\begin{equation}
	\bar{\nabla}_{\alpha}h_{\beta\nu}\bar{\nabla}_{\mu}h^{\alpha\beta}=\bar{\nabla}_{\alpha}\left(h_{\beta\nu}\bar{\nabla}_{\mu}h^{\alpha\beta}\right)-\frac{2\Lambda D}{(D-1)(D-2)}h_{\mu\beta}h_{\nu}^{\beta}
\end{equation}
and
\begin{equation}
	\bar{\nabla}_{\alpha}h_{\beta\mu}\bar{\nabla}_{\nu}h^{\alpha\beta}=\bar{\nabla}_{\alpha}\left(h_{\beta\mu}\bar{\nabla}_{\nu}h^{\alpha\beta}\right)-\frac{2\Lambda D}{(D-1)(D-2)}h_{\mu\beta}h_{\nu}^{\beta},
\end{equation}
\begin{eqnarray}
	&&\left(c(2-D)-\frac{1}{2}\right)h^{\alpha\beta}\bar{\nabla}_{\nu}\bar{\nabla}_{\mu}h_{\alpha\beta}+\left(c(2-D)-\frac{3}{4}\right)\bar{\nabla}_{\nu}h^{\alpha\beta}\bar{\nabla}_{\mu}h_{\alpha\beta}\nonumber\\
	&&=\left(c(2-D)-\frac{1}{2}\right)\bar{\nabla}_{\nu}\left(h^{\alpha\beta}\bar{\nabla}_{\mu}h_{\alpha\beta}\right)-\frac{1}{4}\bar{\nabla}_{\nu}h^{\alpha\beta}\bar{\nabla}_{\mu}h_{\alpha\beta},
\end{eqnarray}
\begin{equation}
	\bar{g}_{\mu\nu}\bar{\nabla}^{\sigma}h^{\alpha\beta}\bar{\nabla}_{\sigma}h_{\alpha\beta}=\bar{\nabla}_{\sigma}\left(\bar{g}_{\mu\nu}h_{\alpha\beta}\bar{\nabla}^{\sigma}h^{\alpha\beta}\right)-\frac{4\Lambda}{(D-1)(D-2)}\bar{g}_{\mu\nu}h_{\alpha\beta}^{2},
\end{equation}
\begin{equation}
	\bar{g}_{\mu\nu}\bar{\nabla}_{\alpha}h_{\sigma\beta}\bar{\nabla}^{\beta}h^{\sigma\alpha}=\bar{\nabla}_{\alpha}\left(\bar{g}_{\mu\nu}h_{\sigma\beta}\bar{\nabla}^{\beta}h^{\sigma\alpha}\right)-\frac{2\Lambda D}{(D-1)(D-2)}\bar{g}_{\mu\nu}h_{\alpha\beta}^{2},
\end{equation}
we can express the $K$ tensor in a more compact form as
\begin{eqnarray}
	&&K_{\mu\nu}=\bar{\nabla}_{\alpha}H^{\alpha}\thinspace_{\mu\nu}+\frac{\Lambda}{(D-2)}\left(c(D-2)+\frac{1}{2}\right)\bar{g}_{\mu\nu}h_{\alpha\beta}^{2}-\frac{1}{4}\bar{\nabla}_{\nu}h^{\alpha\beta}\bar{\nabla}_{\mu}h_{\alpha\beta}\nonumber\\
	&&-\frac{\Lambda D}{(D-1)(D-2)}h_{\mu\beta}h_{\nu}^{\beta},
\end{eqnarray}
where
\begin{eqnarray}
	&&H^{\alpha}\thinspace_{\mu\nu}=\frac{1}{2}\left(h^{\alpha\beta}\bar{\nabla}_{\beta}h_{\nu\mu}+h_{\beta\nu}\bar{\nabla}_{\mu}h^{\alpha\beta}+h_{\beta\mu}\bar{\nabla}_{\nu}h^{\alpha\beta}-h_{\mu\beta}\bar{\nabla}^{\alpha}h_{\nu}^{\beta}-h_{\mu\beta}\bar{\nabla}^{\beta}h_{\nu}^{\alpha}\right)\nonumber\\
	&&-\frac{1}{4}\bar{g}_{\mu\nu}h_{\sigma\beta}\bar{\nabla}^{\beta}h^{\sigma\alpha}+\left(c(2-D)-\frac{1}{2}\right)\delta_{\nu}^{\alpha}h^{\sigma\beta}\bar{\nabla}_{\mu}h_{\sigma\beta}\nonumber\\&&+\left(c(D-2)+\frac{5}{8}\right)\bar{g}_{\mu\nu}h_{\sigma\beta}\bar{\nabla}^{\alpha}h^{\sigma\beta}.
\end{eqnarray}
In this gauge the coefficient $d$ is not fixed but can be set to zero.
$K_{\mu\nu}$ has a single parameter which one can choose to fix the
stability of the flat spacetime. In the Chapter 2, we use this expression to study the linearization stability of the Minkowski space by choosing $\Lambda =0$.

\newpage

\section{Appendix: ADM formalism of topologically massive gravity}
\label{chp:appendixc}

The topologically massive gravity is a higher order gravity theory that involves the third order derivative Cotton tensor, hence the ADM construction is somewhat cumbersome. In this appendix, we give a full account of this. What is also important is that as the action of the theory is only diffeomorphism invariant up to a boundary term, canonical ADM analysis should better be carried out at the field equation level. For the purpose of completeness we give the computation at the level of the action also. For
this purpose we need to find the ADM decomposition of Christoffel symbol,
the Ricci tensor, the scalar curvature and Cotton tensor. Let us compute the corresponding quantities step by step.

\subsection{ ADM split of the Christoffel symbol }

We denote the full $2+1$ dimensional metric with $g_{\mu\nu}$ and the ADM decomposition
of the metric is 
\begin{equation}
	ds^{2}=(n_{i}n^{i}-n^{2})dt^{2}+2n_{i}dtdx^{i}+\gamma_{ij}dx^{i}dx^{j},\label{eq:admmetric}
\end{equation}
where $n$ is lapse function and $n_{i}$ is shift vector both of which are functions of all coordinates. The spatial
indices can be raised and lowered with the $2$ dimensional spatial
metric $\gamma_{ij}$. We will denote the spacetime coordinates with the Greek
indices and the space coordinates with the Latin indices as $\mu,\nu,\rho,...=0,1,2$
and $i,j,k,...=1,2$ respectively. The components of the three dimensional metric tensor are then
$g_{00}=-(n^{2}-n_{i}n^{i})$, $g_{0i}=n_{i}$ and $g_{ij}=\gamma_{ij}$.
Similarly, components of the inverse metric are $g^{00}=-\frac{1}{n^{2}}$
, $g^{0i}=\frac{1}{n^{2}}n^{i}$ and $g^{ij}=\gamma^{ij}-\frac{1}{n^{2}}n^{i}n^{j}$. We 
define the extrinsic curvature tensor, $k_{ij}$, of the surface
\begin{equation}
	k_{ij}:=\frac{1}{2n}\left(\dot{\gamma}_{ij}-D_{i}n_{j}-D_{j}n_{i}\right),\label{eq:extrinsiccurvature}
\end{equation}
where $D_{i}$ is the covariant derivative which is compatible with
$\gamma$, namely $ D \gamma = 0 $ and an over dot denotes the time derivative.
$\Gamma$ denotes Christoffel symbol of the three dimensional space with the well known definition
\begin{equation}
	\Gamma_{\nu\rho}^{\mu}=\frac{1}{2}g^{\mu\sigma}\left(\partial_{\nu}g_{\rho\sigma}+\partial_{\rho}g_{\nu\sigma}-\partial_{\sigma}g_{\nu\rho}\right)
\end{equation}
and $\gamma$ denotes the Christoffel symbol of the two dimensional surface, which is compatible with the spatial metric $\gamma$ as
\begin{equation}
	\gamma_{ij}^{k}=\frac{1}{2}\gamma^{kp}\left(\partial_{i}\gamma_{jp}+\partial_{j}\gamma_{ip}-\partial_{p}\gamma_{ij}\right).
\end{equation}
Now we can find the relations between the Christoffel symbol components of the three and two dimensional spaces. We can express the three dimensional Christoffel connection which consist of only time components as
\begin{equation}
	\Gamma_{00}^{0}=\frac{1}{2}g^{0\sigma}\left(2\partial_{0}g_{\sigma0}-\partial_{\sigma}g_{00}\right),
\end{equation}
where $\sigma$ can be a space or time component as we expressed
above. Considering the possible cases we obtain 
\begin{equation}
	\Gamma_{00}^{0}=\frac{1}{2}g^{00}\partial_{0}g_{00}+\frac{1}{2}g^{0k}\left(2\partial_{0}g_{k0}-\partial_{k}g_{00}\right),
\end{equation}
and inserting the components of the metric and the inverse metric tensor we have
\begin{equation}
	\Gamma_{00}^{0}=\frac{1}{2n^{2}}(2n\dot{n}-n_{i}n_{j}\dot{\gamma}^{ij})+\frac{n^{k}}{2n^{2}}\left(2n\partial_{k}n-2\partial_{k}n_{i}n^{i}-n_{i}n_{j}\partial_{k}\gamma^{ij}\right).
\end{equation}
Expressing the partial derivatives in terms of covariant derivatives
and by using $\dot{\gamma}^{ij}=-\gamma^{ik}\gamma^{lj}\dot{\gamma}_{kl}=-2nk^{ij}-D^{i}n^{j}-D^{j}n^{i}$
we obtain the final result as 
\begin{equation}
	\Gamma_{00}^{0}=\frac{1}{n}\left(\dot{n}+n^{k}\left(\partial_{k}n+n^{i}k_{ik}\right)\right).
\end{equation}
Now let us compute the Christoffel symbol with only one lower space component.
It can be expressed as
\begin{equation}
	\Gamma_{0i}^{0}=\frac{1}{2}g^{00}\partial_{i}g_{00}+\frac{1}{2}g^{0k}\left(\partial_{0}g_{ki}+\partial_{i}g_{k0}-\partial_{k}g_{i0}\right),
\end{equation}
and substituting the corresponding metric tensor components in the last equation we have
\begin{equation}
	\Gamma_{0i}^{0}=\frac{1}{n}\left(\partial_{i}n+n^{k}k_{ik}\right).
\end{equation}
The Christoffel symbol which has only one lower time component can be expressed as
\begin{equation}
	\Gamma_{0j}^{i}=\frac{1}{2}g^{0i}\partial_{j}g_{00}+\frac{1}{2}g^{ik}\left(\partial_{0}g_{kj}+\partial_{j}g_{k0}-\partial_{k}g_{j0}\right),
\end{equation}
and in terms of covariant derivative it becomes
\begin{equation}
	\Gamma_{0j}^{i}=-\frac{1}{n}n^{i}\left(\partial_{j}n+k_{kj}n^{k}\right)+nk_{j}\thinspace^{i}+D_{j}n^{i}.
\end{equation}
With upper time and lower space components the Christoffel
connection is
\begin{equation}
	\Gamma_{ij}^{0}=\frac{1}{2}g^{0k}\left(\partial_{i}g_{j0}+\partial_{j}g_{i0}-\partial_{0}g_{ij}\right)+\frac{1}{2}g^{0k}\left(\partial_{i}g_{kj}+\partial_{j}g_{ki}-\partial_{k}g_{ij}\right)
\end{equation}
and it can be compactly expressed as
\begin{equation}
	\Gamma_{ij}^{0}=\frac{1}{n}k_{ij}.
\end{equation}
We can express the Christoffel symbol which consists of only space components as follows
\begin{equation}
	\Gamma_{ij}^{k}=\gamma_{ij}^{k}+\frac{1}{2}g^{k0}\left(\partial_{i}g_{0j}+\partial_{j}g_{0i}-\partial_{0}g_{ij}\right),
\end{equation}
or equivalently, this can be expresses as
\begin{equation}
	\Gamma_{ij}^{k}=\gamma_{ij}^{k}-\frac{n^{k}}{n}k_{ij}.
\end{equation}
The last component that we need to compute is the Christoffel symbol with upper space and lower time components, explicitly it is
\begin{equation}
	\Gamma_{00}^{i}=\frac{1}{2}g^{0i}\partial_{0}g_{00}+\frac{1}{2}g^{ik}\left(2\partial_{0}g_{0k}-\partial_{k}g_{00}\right),
\end{equation}
and substituting the metric tensor components it can also be expressed as 
\begin{eqnarray}
	\Gamma_{00}^{i}=\frac{n^{i}}{n^{2}}\left(-n\dot{n}+\dot{n}_{k}n^{k}-nn_{k}n_{r}k^{kr}-n_{k}n_{r}D^{k}n^{r}\right)\nonumber\\
	+\left(\gamma^{ik}-\frac{1}{n^{2}}n^{i}n^{k}\right)\left(\dot{n}_{k}+n\partial_{k}n-D_{k}n_{r}n^{r}\right).
\end{eqnarray}
After a lengthy calculation we obtain
\begin{equation}
	\Gamma_{00}^{i}=-\frac{1}{n}\left(n^{i}\dot{n}+n^{i}n_{k}n_{r}k^{kr}+n^{i}n^{k}\partial_{k}n\right)+\gamma^{ik}\left(\dot{n}_{k}+n\partial_{k}n-n^{r}D_{k}n_{r}\right),
\end{equation}
and by using $\gamma^{ik}\dot{n}_{k}=\dot{n}^{i}-n_{k}\dot{\gamma}^{ik}$
we can express the last equation as 
\begin{equation}
	\Gamma_{00}^{i}=-\frac{n^{i}}{n}\left(\dot{n}+n^{k}\left(\partial_{k}n+n^{l}k_{kl}\right)\right)+n\left(\partial^{i}n+2n^{k}k_{k}\thinspace^{i}\right)+\dot{n}^{i}+n^{k}D_{k}n^{i}.
\end{equation}
To compute the decomposition of the field equations, we need to compute additional tensor quantities such that Ricci tensor components, the scalar curvature and the Cotton tensor.

\subsection{ADM split of the Ricci tensor and the scalar curvature}

The three dimensional Ricci tensor which is compatible with the metric $g$ can be expressed as
\begin{equation}
	R_{\rho\sigma}=\partial_{\mu}\Gamma_{\rho\sigma}^{\mu}-\partial_{\rho}\Gamma_{\mu\sigma}^{\mu}+\Gamma_{\mu\nu}^{\mu}\Gamma_{\rho\sigma}^{\nu}-\Gamma_{\sigma\nu}^{\mu}\Gamma_{\mu\rho}^{\nu}.
\end{equation}
We can express the hypersurface projection of the three dimensional Ricci tensor as
\begin{eqnarray}
	R_{ij}=\partial_{0}\Gamma_{ij}^{0}+\partial_{k}\Gamma_{ij}^{k}-\partial_{i}\Gamma_{0j}^{0}-\partial_{i}\Gamma_{kj}^{k}+\left(\Gamma_{00}^{0}+\Gamma_{k0}^{k}\right)\Gamma_{ij}^{0}+\left(\Gamma_{0k}^{0}+\Gamma_{lk}^{l}\right)\Gamma_{ij}^{k}\nonumber\\
	-\Gamma_{0i}^{0}\Gamma_{0j}^{0}-\Gamma_{jk}^{0}\Gamma_{0i}^{k}-\Gamma_{0j}^{k}\Gamma_{ik}^{0}-\Gamma_{jm}^{k}\Gamma_{ik}^{m},\thinspace\thinspace\thinspace\thinspace\thinspace\thinspace\thinspace\thinspace\thinspace\thinspace\thinspace\thinspace\thinspace\thinspace\thinspace\thinspace\thinspace\thinspace\thinspace
\end{eqnarray}
and by inserting the expressions for the three dimensional Christoffel connection we obtain
\begin{eqnarray}
	R_{ij}=&&^{(2)}R_{ij}+kk_{ij}-2k_{ik}k_{j}^{k}\nonumber\\&&+\frac{1}{n}\left(\dot{k}_{ij}-n^{k}D_{k}k_{ij}-D_{i}\partial_{j}n-k_{kj}D_{i}n^{k}-k_{ki}D_{j}n^{k}\right),\label{eq:rij}
\end{eqnarray}
where $^{(2)}R_{ij}$ denotes the Ricci tensor of the hypersurface,
which is explicitly
\begin{equation}
	^{(2)}R_{ij}=\partial_{k}\gamma_{ij}^{k}-\partial_{i}\gamma_{kj}^{k}+\gamma_{kl}^{k}\gamma_{ij}^{l}-\gamma_{jl}^{k}\gamma_{ki}^{l}.
\end{equation}
The three dimensional Ricci tensor component which is orthogonal to the hypersurface is 
\begin{equation}
	R_{00}=\partial_{k}\Gamma_{00}^{k}-\partial_{0}\Gamma_{0k}^{k}+\Gamma_{k0}^{k}\Gamma_{00}^{0}-\Gamma_{k0}^{m}\Gamma_{0m}^{k}+\left(\Gamma_{mk}^{m}-\Gamma_{k0}^{0}\right)\Gamma_{00}^{k}
\end{equation}
and it can be written as
\begin{eqnarray}
	R_{00}=\frac{1}{n}n^{i}n^{j}\left(\dot{k}_{ij}-n^{k}D_{k}k_{ij}-D_{i}\partial_{j}n-2k_{kj}D_{i}n^{k}\right)-n^{2}k_{ij}^{2}\nonumber\\
	+n^{i}n^{j}\left(^{(2)}R_{ij}+kk_{ij}-2k_{ik}k_{j}^{k}\right)+n\left(D_{k}\partial^{k}n-\dot{k}-n^{k}D_{k}k+2n^{k}D_{m}k_{k}^{m}\right).\,\,\,
\end{eqnarray}
By using equation (\ref{eq:rij}) we can express the result in a compact form as
\begin{eqnarray}
	R_{00}=n^{i}n^{j}R_{ij}-n^{2}k_{ij}^{2}+nn^{k}\left(D_{m}k_{k}^{m}-D_{k}k\right)\nonumber\\+n\left(D_{k}\partial^{k}n-\dot{k}+n^{k}D_{m}k_{k}^{m}\right).\label{eq:r00}
\end{eqnarray}
The Ricci tensor with once time and once space projection is 
\begin{equation}
	R_{0i}=\partial_{k}\Gamma_{0i}^{k}-\partial_{0}\Gamma_{ki}^{k}+\Gamma_{k0}^{k}\Gamma_{0i}^{0}+\Gamma_{km}^{m}\Gamma_{0i}^{k}-\Gamma_{ik}^{0}\Gamma_{00}^{k}-\Gamma_{ik}^{m}\Gamma_{m0}^{k}
\end{equation}
and in terms of $2+1$ dimensional decomposition it can be written as
\begin{eqnarray}
	R_{0i}=\frac{1}{n}n^{j}\left(\dot{k}_{ij}-n^{k}D_{k}k_{ij}-D_{i}\partial_{j}n-k_{kj}D_{i}n^{k}-k_{ki}D_{j}n^{k}\right)\nonumber\\
	+n^{j}\left(^{(2)}R_{ij}+kk_{ij}-2k_{ik}k_{j}^{k}\right)+n\left(D_{i}k+D_{m}k_{i}^{m}\right),
\end{eqnarray}
which can be similarly expressed in terms of $R_{ij}$ by using the equation (\ref{eq:rij}) as
\begin{equation}
	R_{0i}=n^{j}R_{ij}+n\left(D_{m}k_{i}^{m}-D_{i}k\right).\label{eq:ri0}
\end{equation}
Since we know the decomposition of the three dimensional Ricci tensor components, we can compute the decomposition of the three dimensional scalar curvature $R$. Contracting the Ricci tensor with the inverse metric tensor yields the scalar curvature as
\begin{equation}
	R=R_{\mu\nu}g^{\mu\nu}=R_{00}g^{00}+2R_{0i}g^{0i}+R_{ij}g^{ij},
\end{equation}
and by using our results we have
\begin{equation}
	R={}^{(2)}R+k^{2}+k_{ij}^{2}+\frac{2}{n}\left(\dot{k}-D_{i}D^{i}n-n^{i}D_{i}k\right),\label{eq:r}
\end{equation}
where ${}^{(2)}R$ is scalar curvature of the hypersurface which is compatible with spatial metric $\gamma$.

\subsection{ADM split of the TMG action}

The ADM decomposition of the TMG action is known in the literature  \cite {deser_canonical}. Here, as a complementary exercise to our decomposition of the field equations, we will construct the decomposition of the action in the metric formulation. From the actual physical point of view, one should not have a physical difference between the two formulations. But because of the possible boundary terms, and due to the fact that the TMG action is only diffeomorphism invariant up to a boundary term, the canonical variables and the form of the actions look quite different. We have chosen to work withe field equations, since the action formulation involves tensor densities. But the following computation is still valuable and can be used for the Hamiltonian formulation of the theory. 

For simplicity let us analyse the TMG action in two parts. 

\subsection{The Einstein-Hilbert action}

The Einstein-Hilbert Lagrangian is 
\begin{equation}
	\text{\ensuremath{\mathscr{L}}}_{EH}=\sqrt{-g}\left(R-2\varLambda\right),
\end{equation}
where $\sqrt{-g}=n\sqrt{\gamma}$ and substituting the  three dimensional
curvature scalar, which was given in (\ref{eq:r}) in the action, we
arrive at
\begin{equation}
	\text{\ensuremath{\mathscr{L}}}_{EH}=n\sqrt{\gamma}\left(^{(2)}R+k^{2}+k_{ij}^{2}+\frac{2}{n}\left(\dot{k}-D_{i}D^{i}n-n^{i}D_{i}k\right)-2\varLambda\right).
\end{equation}
We can express the term which has explicit time dependence as
\begin{equation}
	\dot{k}\sqrt{\gamma}=\partial_{0}\left(k\sqrt{\gamma}\right)-k\partial_{0}\sqrt{\gamma},
\end{equation}
where
\begin{equation}
	\partial_{0}\sqrt{\gamma}=\frac{1}{2}\sqrt{\gamma}\gamma^{ij}\partial_{0}\gamma_{ij}=\sqrt{\gamma}\left(nk+D_{i}n^{i}\right)
\end{equation}
and then we obtain
\begin{equation}
	\dot{k}\sqrt{\gamma}=\partial_{0}\left(k\sqrt{\gamma}\right)-\sqrt{\gamma}\left(nk^{2}+kD_{i}n^{i}\right).
\end{equation}
Inserting the last expression in the action we have
\begin{eqnarray}
	\text{\ensuremath{\mathscr{L}}}_{EH}=n\sqrt{\gamma}\left(^{(2)}R-k^{2}+k_{ij}^{2}+\varLambda\right)+\partial_{0}\left(2\sqrt{\gamma}k\right)\\\nonumber-D_{i}\left(2\sqrt{\gamma}\left(D^{i}n+kn^{i}\right)\right),
\end{eqnarray}
where the last three terms are boundary terms. So up to boundary terms
ADM formalism of the Einstein-Hilbert action becomes
\begin{equation}
	\text{\ensuremath{\mathscr{L}}}_{EH}=\sqrt{-g}\left(R+\Lambda\right)=n\sqrt{\gamma}\left(^{(2)}R-k^{2}+k_{ij}^{2}-2\Lambda\right).\label{eq:ehaction}
\end{equation}

\subsection{The Chern-Simons action}

The Chern-Simons Lagrangian is
\begin{equation}
	\text{\ensuremath{\mathscr{L}}}_{CS}=\frac{1}{2\mu}\sqrt{-g}\epsilon^{\rho\nu\mu}\left(\Gamma_{\rho\gamma}^{\sigma}\partial_{\nu}\Gamma_{\mu\sigma}^{\gamma}+\frac{2}{3}\Gamma_{\rho\gamma}^{\sigma}\Gamma_{\nu\delta}^{\gamma}\Gamma_{\mu\sigma}^{\delta}\right).
\end{equation}
$2+1$ dimensional decomposition of the three index $\epsilon$ tensor
gives
\begin{eqnarray}
	&\text{\ensuremath{\mathscr{L}}}_{CS}=\frac{1}{2\mu}\sqrt{-g}\Biggl(\epsilon^{n0m}\left(\Gamma_{n\gamma}^{\sigma}\partial_{0}\Gamma_{m\sigma}^{\gamma}+\frac{2}{3}\Gamma_{n\gamma}^{\sigma}\Gamma_{0\delta}^{\gamma}\Gamma_{m\sigma}^{\delta}\right)&\\&+\epsilon^{0mn}\left(\Gamma_{0\gamma}^{\sigma}\partial_{m}\Gamma_{n\sigma}^{\gamma}+\frac{2}{3}\Gamma_{0\gamma}^{\sigma}\Gamma_{m\delta}^{\gamma}\Gamma_{n\sigma}^{\delta}\right)
	+\epsilon^{mn0}\left(\Gamma_{m\gamma}^{\sigma}\partial_{n}\Gamma_{0\sigma}^{\gamma}+\frac{2}{3}\Gamma_{m\gamma}^{\sigma}\Gamma_{n\delta}^{\gamma}\Gamma_{0\sigma}^{\delta}\right)\Biggr).&\nonumber
\end{eqnarray}
Here the terms with the coefficient $\frac{2}{3}$ are identical, by expressing these terms together we have
\begin{equation}
	\text{\ensuremath{\mathscr{L}}}_{CS}=\frac{1}{2\mu}\varepsilon^{mn}\left(\Gamma_{n\gamma}^{\sigma}\partial_{0}\Gamma_{m\sigma}^{\gamma}+\Gamma_{0\gamma}^{\sigma}\partial_{m}\Gamma_{n\sigma}^{\gamma}+\Gamma_{m\gamma}^{\sigma}\partial_{n}\Gamma_{0\sigma}^{\gamma}+2\Gamma_{n\gamma}^{\sigma}\Gamma_{0\delta}^{\gamma}\Gamma_{m\sigma}^{\delta}\right).
\end{equation}
Let us compute the ADM decomposition of the Chern-Simons action term by term. The first term can be decomposed as
\begin{equation}
	\varepsilon^{mn}\Gamma_{n\gamma}^{\sigma}\partial_{0}\Gamma_{m\sigma}^{\gamma}=\varepsilon^{mn}\left(\Gamma_{n0}^{0}\partial_{0}\Gamma_{m0}^{0}+\Gamma_{nk}^{0}\partial_{0}\Gamma_{m0}^{k}+\Gamma_{n0}^{k}\partial_{0}\Gamma_{mk}^{0}+\Gamma_{nl}^{k}\partial_{0}\Gamma_{mk}^{l}\right).
\end{equation}
In terms of the ADM decomposition the terms in the right hand side of the last equation can be recast as follows
\begin{eqnarray}
	&&\varepsilon^{mn}\Gamma_{n0}^{0}\partial_{0}\Gamma_{m0}^{0}\\&&=\frac{\varepsilon^{mn}}{n^{2}}\left(\left(\partial_{n}n+n^{k}k_{kn}\right)\partial_{m}\dot{n}+\partial_{n}n\partial_{0}\left(n^{l}k_{ml}\right)+n^{k}k_{kn}\partial_{0}\left(n^{l}k_{ml}\right)\right)\nonumber
\end{eqnarray}
and
\begin{eqnarray}
	&&\varepsilon^{mn}\Gamma_{nk}^{0}\partial_{0}\Gamma_{m0}^{k}=\varepsilon^{mn}\Biggl(\frac{1}{n^{3}}\biggl(\dot{n}n^{k}k_{kn}\partial_{m}n\biggr)\\
	&&-\frac{1}{n^{2}}\biggl(k_{kn}\partial_{0}\left(n^{k}\partial_{m}n\right)+n^{k}k_{kn}n^{l}\dot{k}_{lm}\biggr)+\frac{1}{n}k_{nk}\partial_{0}D_{m}n^{k}+k_{nk}\dot{k}_{m}\thinspace^{k}\Biggr)\nonumber 
\end{eqnarray}
and
\begin{eqnarray}
	&&\varepsilon^{mn}\Gamma_{n0}^{k}\partial_{0}\Gamma_{mk}^{0}=\varepsilon^{mn}\Biggl(-\frac{1}{n^{2}}\biggl(\dot{k}_{km}n^{k}\left(\partial_{m}n+n^{l}k_{ln}\right)+\dot{n}k_{km}D_{n}n^{k}\biggr)\nonumber\\&&+\frac{1}{n^{3}}\biggl(\dot{n}n^{k}k_{km}\partial_{n}n\biggr)
	+\frac{1}{n}\dot{k}_{mk}D_{n}n^{k}+\dot{k}_{mk}k_{n}\thinspace^{k}\Biggr).\thinspace
\end{eqnarray}
We also have the last piece as
\begin{eqnarray}
	&&\varepsilon^{mn}\Gamma_{nl}^{k}\partial_{0}\Gamma_{mk}^{l}=\varepsilon^{mn}\Biggl(\biggl(\frac{1}{n^{2}}\biggl(k_{nl}n^{k}\partial_{0}\left(n^{l}k_{mk}\right)+\dot{n}k_{km}n^{l}\gamma_{nl}^{k}\biggr)\nonumber\\
	&&-\frac{1}{n}\biggl(\gamma_{nl}^{k}\partial_{0}\left(n^{l}k_{mk}\right)+\dot{\gamma}_{mk}^{l}k_{nl}n^{k}\biggr)+\dot{\gamma}_{mk}^{l}\gamma_{nl}^{k}\Biggr).
\end{eqnarray}
Combining these we obtain the following identity
\begin{eqnarray}
	&&\varepsilon^{mn}\Gamma_{n\gamma}^{\sigma}\partial_{0}\Gamma_{m\sigma}^{\gamma}=\varepsilon^{mn}\Biggl(\frac{1}{n^{2}}\biggl(\partial_{n}n\partial_{m}\dot{n}+2\partial_{n}n\dot{n}^{k}k_{km}-\dot{n}k_{km}\partial_{n}n^{k}\biggr)\\
	&&+\frac{1}{n}\biggl(k_{nk}D_{m}\dot{n}^{k}+\dot{k}_{mk}\partial_{n}n^{k}-\gamma_{nl}^{k}\dot{n}^{l}k_{mk}\biggr)+\dot{k}_{mk}k_{n}\thinspace^{k}+k_{nk}\dot{k}_{m}\thinspace^{k}+\gamma_{nl}^{k}\dot{\gamma}_{mk}^{l}\Biggr).\nonumber
\end{eqnarray}
Similarly the second term can be decomposed as
\begin{equation}
	\varepsilon^{mn}\Gamma_{0\gamma}^{\sigma}\partial_{m}\Gamma_{n\sigma}^{\gamma}=\varepsilon^{mn}\left(\Gamma_{00}^{0}\partial_{m}\Gamma_{n0}^{0}+\Gamma_{0k}^{0}\partial_{m}\Gamma_{n0}^{k}+\Gamma_{00}^{k}\partial_{m}\Gamma_{nk}^{0}+\Gamma_{0l}^{k}\partial_{m}\Gamma_{nk}^{l}\right),
\end{equation}
where the right hand side of the equation can be expressed term by term as
\begin{eqnarray}
	\varepsilon^{mn}\Gamma_{00}^{0}\partial_{m}\Gamma_{n0}^{0}=\varepsilon^{mn}\Biggl(-\frac{1}{n^{3}}\biggl(\dot{n}n^{r}k_{rn}\partial_{m}n+n^{r}k_{rn}\partial_{m}nn^{k}\left(\partial_{k}n+n^{l}k_{lk}\right)\biggr)\nonumber\\
	+\frac{1}{n^{2}}\biggl(\dot{n}\partial_{m}\left(n^{r}k_{rn}\right)+n^{k}\left(\partial_{k}n+n^{l}k_{lk}\right)\partial_{m}\left(n^{r}k_{rn}\right)\biggr)\Biggr)\thinspace\thinspace\thinspace\thinspace\thinspace\thinspace\thinspace\thinspace\thinspace\thinspace\thinspace
\end{eqnarray}
and
\begin{eqnarray}
	&&\varepsilon^{mn}\Gamma_{0k}^{0}\partial_{m}\Gamma_{n0}^{k}=\varepsilon^{mn}\Biggl(\frac{1}{n^{3}}\biggl(n^{r}k_{rn}\partial_{m}nn^{k}\left(\partial_{k}n+n^{l}k_{lk}\right)\biggr)\\	&&+\frac{1}{n}\biggl(\left(\partial_{k}n+n^{l}k_{lk}\right)\left(\partial_{m}D_{n}n^{k}+\partial_{m}nk_{n}\thinspace^{k}\right)+\partial_{m}k_{n}\thinspace^{k}\left(\partial_{k}n+n^{l}k_{lk}\right)\biggr)
	\nonumber\\&&-\frac{1}{n^{2}}\biggl(n^{k}\left(\partial_{k}n+n^{l}k_{lk}\right)\partial_{m}\left(n^{r}k_{rn}\right)+\partial_{m}n^{k}\left(\partial_{k}n+n^{l}k_{lk}\right)\left(\partial_{n}n+k_{ln}n^{l}\right)\biggr)\Biggr)\nonumber
\end{eqnarray}
and
\begin{eqnarray}
	&&\varepsilon^{mn}\Gamma_{00}^{k}\partial_{m}\Gamma_{nk}^{0}=\varepsilon^{mn}\Biggl(\frac{1}{n^{3}}\biggl(n^{r}k_{kn}\partial_{m}nn^{k}\left(\partial_{r}n+n^{l}k_{lr}\right)+\dot{n}\partial_{m}nn^{k}k_{kn}\biggr)\nonumber\\
	&&-\frac{1}{n^{2}}\biggl(n^{k}n^{r}\partial_{m}k_{kn}\left(\partial_{r}n+n^{l}k_{lr}\right)+\partial_{m}nk_{kn}\bigl(\dot{n}^{k}+n^{l}D_{l}n^{k}\bigr)+\dot{n}n^{k}\partial_{m}k_{kn}\biggr)\nonumber\\
	&&+\frac{1}{n}\biggl(\partial_{m}k_{kn}\bigl(\dot{n}^{k}+n^{l}D_{l}n^{k}\bigr)-\partial_{m}nk_{n}\thinspace^{k}\left(\partial_{k}n+2n^{l}k_{lk}\right)\biggr)\nonumber\\&&+\partial_{m}k_{n}\thinspace^{k}\left(\partial_{k}n+2n^{l}k_{lk}\right)\Biggr)
\end{eqnarray}
also and
\begin{eqnarray}
	&&\varepsilon^{mn}\Gamma_{0l}^{k}\partial_{m}\Gamma_{nk}^{l}=\varepsilon^{mn}\Biggl(\frac{1}{n^{3}}\biggl(-n^{l}k_{kn}\partial_{m}nn^{k}\left(\partial_{l}n+n^{r}k_{lr}\right)\biggr)\nonumber\\
	&&+\frac{1}{n^{2}}\biggl(n^{k}\partial_{m}\left(k_{kn}n^{l}\right)\left(\partial_{l}n+n^{r}k_{lr}\right)+\partial_{m}nn^{l}k_{kn}D_{l}n^{k}\biggr)\nonumber\\
	&&-\frac{1}{n}\biggl(\partial_{m}\left(k_{kn}n^{l}\right)D_{l}n^{k}-\partial_{m}nk_{n}\thinspace^{k}n^{l}k_{lk}+n^{k}\partial_{m}\gamma_{nk}^{l}\left(\partial_{l}n+k_{rl}n^{r}\right)\biggr)\nonumber\\
	&&+\partial_{m}\gamma_{nk}^{l}\left(nk_{l}\thinspace^{k}+D_{l}n^{k}\right)-k_{l}\thinspace^{k}\partial_{m}\left(k_{kn}n^{l}\right)\Biggr).
\end{eqnarray}
Collecting the pieces one has the identity,
\begin{eqnarray}
	&&\varepsilon^{mn}\Gamma_{0\gamma}^{\sigma}\partial_{m}\Gamma_{n\sigma}^{\gamma}=\varepsilon^{mn}\Biggl(\partial_{m}k_{n}\thinspace^{k}\partial_{k}n+\partial_{m}k_{nk}\partial^{k}n+\partial_{m}k_{n}\thinspace^{k}n^{l}k_{kl}\nonumber\\
	&&+\partial_{m}k_{nk}n^{l}k_{l}\thinspace^{k}+D_{l}n^{k}\partial_{m}\gamma_{nk}^{l}-k_{l}\thinspace^{k}k_{nk}\partial_{m}n^{l}+nk_{l}\thinspace^{k}\partial_{m}\gamma_{n\sigma k}^{l}\nonumber\\
	&&+\frac{1}{n}\biggl(\gamma_{nl}^{k}\partial_{m}n^{l}\partial_{k}n+\gamma_{nr}^{k}\partial_{m}n^{r}n^{l}k_{lk}+\dot{n}^{k}\partial_{m}k_{nk}-D_{l}n^{k}k_{nk}\partial_{m}n^{l}\biggr)\nonumber\\
	&&+\frac{1}{n^{2}}\biggl(\dot{n}k_{kn}\partial_{m}n^{k}-\partial_{m}n^{k}\partial_{k}n\partial_{n}n-\partial_{m}n^{k}\partial_{n}nn^{l}k_{kl}-\dot{n}^{l}\partial_{m}nk_{ln}\biggr)\Biggr).\,\,\,\,\,\,\,\,\,\,\,\,
\end{eqnarray}

The third term can be decomposed as
\begin{equation}
	\varepsilon^{mn}\Gamma_{m\gamma}^{\sigma}\partial_{n}\Gamma_{0\sigma}^{\gamma}=\varepsilon^{mn}\left(\Gamma_{m0}^{0}\partial_{n}\Gamma_{00}^{0}+\Gamma_{mk}^{0}\partial_{n}\Gamma_{00}^{k}+\Gamma_{m0}^{k}\partial_{n}\Gamma_{0k}^{0}+\Gamma_{ml}^{k}\partial_{n}\Gamma_{0k}^{l}\right),
\end{equation}
where 
\begin{eqnarray}
	\varepsilon^{mn}\Gamma_{m0}^{0}\partial_{n}\Gamma_{00}^{0}=\varepsilon^{mn}\Biggl(-\frac{1}{n^{3}}\biggl(n^{l}k_{ml}n^{k}\partial_{n}n\left(\partial_{k}n+n^{l}k_{lk}\right)+n^{l}k_{ml}\partial_{n}n\dot{n}\biggr)\nonumber\\
	+\frac{1}{n^{2}}\biggl(n^{l}k_{ml}\partial_{n}\dot{n}+\partial_{m}n\partial_{n}\dot{n}+\left(\partial_{m}n+n^{l}k_{ml}\right)\partial_{n}\left(n^{k}\left(\partial_{k}n+n^{l}k_{lk}\right)\right)\biggr)\Biggr)\thinspace\thinspace\thinspace\thinspace\thinspace\thinspace\thinspace
\end{eqnarray}
and
\begin{eqnarray}
	&&\varepsilon^{mn}\Gamma_{mk}^{0}\partial_{n}\Gamma_{00}^{k}=\varepsilon^{mn}\Biggl(\frac{1}{n^{3}}\biggl(n^{k}k_{km}\partial_{n}n\left(\dot{n}+n^{l}\left(\partial_{l}n+n^{r}k_{rl}\right)\right)\biggr)\nonumber\\
	&&-\frac{1}{n^{2}}\biggl(k_{km}\partial_{n}\left(\dot{n}n^{k}\right)+k_{km}\partial_{n}\left(n^{k}n^{l}\left(\partial_{l}n+n^{r}k_{rl}\right)\right)\biggr)\nonumber\\
	&&+\frac{1}{n}\biggl(k_{km}\partial_{n}\dot{n}^{k}+k_{km}\partial_{n}n^{l}D_{l}n^{k}+k_{km}\partial_{n}n\left(\partial^{k}n+2n^{l}k_{l}\thinspace^{k}\right)\biggr)\nonumber\\
	&&+k_{km}n^{r}\partial_{n}D_{r}n^{k}+k_{km}\partial_{n}\left(\partial^{k}n+2n^{l}k_{l}\thinspace^{k}\right)\Biggr)
\end{eqnarray}
and
\begin{eqnarray}
	&&\varepsilon^{mn}\Gamma_{m0}^{k}\partial_{n}\Gamma_{0k}^{0}=\varepsilon^{mn}\Biggl(\frac{1}{n^{3}}\biggl(n^{k}\partial_{n}nk_{rm}n^{r}\left(\partial_{k}n+n^{l}k_{kl}\right)\biggr)\nonumber\\
	&&-\frac{1}{n^{2}}\biggl(\left(n^{k}\partial_{m}n+n^{k}k_{rm}n^{r}\right)\partial_{n}\left(\partial_{k}n+n^{l}k_{kl}\right)+\partial_{n}nD_{m}n^{k}\left(\partial_{k}n+n^{l}k_{kl}\right)\biggr)\nonumber\\
	&&+\frac{1}{n}\biggl(D_{m}n^{k}\left(\partial_{k}n+n^{l}k_{kl}\right)-k_{m}\thinspace^{k}\partial_{n}n\left(\partial_{k}n+n^{l}k_{kl}\right)\biggr)\nonumber\\&&+k_{m}\thinspace^{k}\partial_{n}\left(\partial_{k}n+n^{l}k_{kl}\right)\Biggr)\Biggr)
\end{eqnarray}
and
\begin{eqnarray}
	&&\varepsilon^{mn}\Gamma_{ml}^{k}\partial_{n}\Gamma_{0k}^{l}=\varepsilon^{mn}\Biggl(-\frac{1}{n^{3}}\biggl(n^{k}\partial_{n}nk_{lm}n^{l}\left(\partial_{k}n+n^{l}k_{kl}\right)\biggr)\nonumber\\&&+\frac{1}{n^{2}}\biggl(n^{l}\partial_{n}n\gamma_{ml}^{k}\left(\partial_{k}n+n^{l}k_{kl}\right)+n^{k}k_{lm}\partial_{n}\left(n^{l}\left(\partial_{k}n+n^{l}k_{kl}\right)\right)\nonumber\\&&-\frac{1}{n}\biggl(\gamma_{ml}^{k}\partial_{n}\left(n^{l}\left(\partial_{k}n+n^{l}k_{kl}\right)\right)+n^{k}k_{lm}\partial_{n}D_{k}n^{l}+n^{k}k_{lm}\partial_{n}nk_{k}\thinspace^{l}\biggr)\Biggr)\nonumber\\&&+\gamma_{ml}^{k}\partial_{n}\left(D_{k}n^{l}+nk_{k}\thinspace^{l}\right)-n^{k}\partial_{n}k_{k}\thinspace^{l}k_{lm}\biggr).
\end{eqnarray}
Then the third term becomes,
\begin{eqnarray}
	&&\varepsilon^{mn}\Gamma_{m\gamma}^{\sigma}\partial_{n}\Gamma_{0\sigma}^{\gamma}=\varepsilon^{mn}\Biggl(k_{m}\thinspace^{k}n^{l}\partial_{n}k_{kl}+\gamma_{mk}^{l}\partial_{n}D_{l}n^{k}+\gamma_{mk}^{l}\partial_{n}\left(nk_{l}\thinspace^{k}\right)\nonumber\\&&-k_{mk}n^{l}\partial_{n}k_{l}\thinspace^{k}+k_{m}\thinspace^{k}\partial_{n}\partial_{k}n+k_{mk}\partial_{n}\partial^{k}n+2k_{mk}n^{l}\partial_{n}k_{l}\thinspace^{k}+3k_{mk}\partial_{n}n^{l}k_{l}\thinspace^{k}\nonumber\\&&+\frac{1}{n}\biggl(\partial_{m}n^{l}\partial_{n}\left(\partial_{l}n+k_{lk}n^{k}\right)-\gamma_{mr}^{k}\partial_{n}n^{r}\left(\partial_{k}n+n^{l}k_{lk}\right)+D_{l}n^{k}k_{mk}\partial_{n}n^{l}+\partial_{n}\dot{n}^{k}k_{mk}\biggr)\nonumber\\&&\frac{1}{n^{2}}\biggl(-\dot{n}k_{km}\partial_{n}n^{k}+2\partial_{n}n^{k}\partial_{k}n\partial_{m}n+2\partial_{n}n^{k}\partial_{m}nn^{l}k_{kl}+\partial_{m}n\partial_{n}\dot{n}\biggr)\Biggr).
\end{eqnarray}
The last term also can be decomposed as
\begin{eqnarray}
	&&\varepsilon^{mn}\Gamma_{n\gamma}^{\sigma}\Gamma_{0\delta}^{\gamma}\Gamma_{m\sigma}^{\delta}=\varepsilon^{mn}\Biggl(\Gamma_{n0}^{0}\left(\Gamma_{m0}^{0}\Gamma_{00}^{0}+\Gamma_{0k}^{0}\Gamma_{m0}^{k}\right)+\Gamma_{nk}^{0}\left(\Gamma_{m0}^{0}\Gamma_{00}^{k}+\Gamma_{0l}^{k}\Gamma_{m0}^{l}\right)\nonumber\\
	&&+\Gamma_{n0}^{l}\left(\Gamma_{ml}^{0}\Gamma_{00}^{0}+\Gamma_{0k}^{0}\Gamma_{ml}^{k}\right)+\Gamma_{nk}^{l}\left(\Gamma_{ml}^{0}\Gamma_{00}^{k}+\Gamma_{0r}^{k}\Gamma_{ml}^{r}\right)\Biggr),\thinspace\thinspace\thinspace\thinspace\thinspace\thinspace\thinspace\thinspace\thinspace\thinspace\thinspace
\end{eqnarray}
where the decomposition of the right hand side of the equation can be expressed term by term as
\begin{eqnarray}
	&&\varepsilon^{mn}\Gamma_{n0}^{0}\left(\Gamma_{m0}^{0}\Gamma_{00}^{0}+\Gamma_{0k}^{0}\Gamma_{m0}^{k}\right)=\varepsilon^{mn}\Biggl(\frac{1}{n^{2}}\biggl(D_{m}n^{k}\left(\partial_{n}n+k_{ln}n^{l}\right)\left(\partial_{k}n+n^{r}k_{rk}\right)\biggr)\nonumber\\
	&&+\frac{1}{n}\biggl(k_{m}\thinspace^{k}\left(\partial_{n}n+k_{ln}n^{l}\right)\left(\partial_{k}n+n^{r}k_{rk}\right)\biggr)\Biggr)\thinspace\thinspace\thinspace\thinspace\thinspace\thinspace\thinspace\thinspace\thinspace\thinspace\thinspace
\end{eqnarray}
and
\begin{eqnarray}
	&&\varepsilon^{mn}\Gamma_{nk}^{0}\left(\Gamma_{m0}^{0}\Gamma_{00}^{k}+\Gamma_{0l}^{k}\Gamma_{m0}^{l}\right)\nonumber\\&&=\varepsilon^{mn}\Biggl(-\frac{1}{n^{3}}\biggl(\partial_{m}nk_{ln}n^{l}\dot{n}\biggr)+k_{nr}\left(k_{m}\thinspace^{k}D_{k}n^{r}+k_{k}\thinspace^{r}D_{m}n^{k}\right)\nonumber\\
	&&+\frac{1}{n^{2}}\biggl(\dot{n}^{k}k_{nk}\left(\partial_{m}n+k_{lm}n^{l}\right)-D_{m}n^{k}n^{l}k_{nl}\left(\partial_{k}n+n^{r}k_{rk}\right)\biggr)\nonumber\\
	&&+\frac{1}{n}\biggl(k_{n}\thinspace^{k}\left(\partial_{m}n+2k_{lm}n^{l}\right)\left(\partial_{k}n+n^{r}k_{rk}\right)+D_{m}n^{k}D_{k}n^{r}k_{nr}\biggr)\Biggr)\,\,\,\,\,\,\,\,\,
\end{eqnarray}
and
\begin{eqnarray}
	&&\varepsilon^{mn}\Gamma_{n0}^{l}\left(\Gamma_{ml}^{0}\Gamma_{00}^{0}+\Gamma_{0k}^{0}\Gamma_{ml}^{k}\right)=\varepsilon^{mn}\Biggl(-\frac{1}{n^{3}}\biggl(\partial_{n}nk_{lm}n^{l}\dot{n}\biggr)\nonumber\\
	&&+\frac{1}{n^{2}}\biggl(\dot{n}k_{mk}D_{n}n^{k}-\gamma_{mp}^{k}n^{p}\left(\partial_{n}n+k_{ln}n^{l}\right)\left(\partial_{k}n+n^{r}k_{rk}\right)\biggr)\nonumber\\
	&&+\frac{1}{n}\biggl(\gamma_{mp}^{k}D_{n}n^{p}\left(\partial_{k}n+n^{r}k_{rk}\right)\biggr)+\gamma_{mp}^{k}k_{k}\thinspace^{p}\left(\partial_{k}n+n^{r}k_{rk}\right)\Biggr)
\end{eqnarray}
and
\begin{eqnarray}
	&&\varepsilon^{mn}\Gamma_{nk}^{l}\left(\Gamma_{ml}^{0}\Gamma_{00}^{k}+\Gamma_{0r}^{k}\Gamma_{ml}^{r}\right)\nonumber\\&&=\varepsilon^{mn}\Biggl(\gamma_{nl}^{p}k_{pm}\left(\partial^{l}n+n^{r}k_{r}\thinspace^{l}\right)-\gamma_{mp}^{k}k_{nr}n^{p}k_{k}\thinspace^{r}+\gamma_{mp}^{k}\gamma_{nl}^{p}\left(\partial_{k}n^{l}+nk_{k}\thinspace^{l}\right)\nonumber\\&&+\frac{1}{n^{2}}\biggl(-\dot{n}\gamma_{np}^{k}n^{p}k_{km}-\dot{n}^{r}k_{rn}n^{k}k_{km}+\gamma_{mp}^{k}n^{p}k_{ln}n^{l}\left(\partial_{k}n+n^{r}k_{rk}\right)\biggr)\nonumber\\&&-\frac{1}{n}\biggl(k_{n}\thinspace^{k}n^{l}k_{lm}\left(\partial_{k}n+n^{r}k_{rk}\right)+\gamma_{mp}^{k}\gamma_{nl}^{p}n^{l}\left(\partial_{k}n+n^{r}k_{rk}\right)\nonumber\\&&-\gamma_{np}^{k}\dot{n}^{p}k_{mk}+\gamma_{mp}^{k}n^{p}k_{nl}D_{k}n^{l}\biggr)\Biggr).
\end{eqnarray}
Finally collecting all the pieces together $2+1$ dimensional decomposition of the last term is
\begin{eqnarray}
	&&\varepsilon^{mn}\Gamma_{n\gamma}^{\sigma}\Gamma_{0\delta}^{\gamma}\Gamma_{m\sigma}^{\delta}=\varepsilon^{mn}\Biggl(\frac{1}{n}\biggl(\gamma_{ml}^{k}\partial_{n}n^{l}\left(\partial_{k}n+n^{r}k_{rk}\right)+D_{l}n^{k}k_{nk}\partial_{m}n^{l}+\gamma_{nk}^{r}k_{mr}\dot{n}^{k}\biggr)\nonumber\\
	&&+\frac{1}{n^{2}}\biggl(\dot{n}k_{km}\partial_{n}n^{k}+\partial_{m}n^{k}\partial_{k}n\partial_{n}n+\partial_{m}n^{k}\partial_{n}nn^{l}k_{kl}+\partial_{m}n\dot{n}^{k}k_{nk}\biggr)\nonumber\\
	&&+k_{m}\thinspace^{k}k_{nl}D_{k}n^{l}+k_{k}\thinspace^{l}k_{nl}\partial_{m}n^{k}+\gamma_{mr}^{k}\gamma_{nl}^{r}\left(D_{k}n^{l}+nk_{k}\thinspace^{l}\right)+\gamma_{mr}^{k}k_{n}\thinspace^{r}\left(k_{kl}n^{l}+\partial_{k}n\right)\nonumber\\
	&&+\gamma_{nl}^{r}k_{mr}\left(\partial^{l}n+n^{k}k_{k}\thinspace^{l}\right)\Biggr).
\end{eqnarray}

Using these terms we obtain the ADM decomposition of the Chern-Simons action as follows
\begin{eqnarray}
	&&\text{\ensuremath{\mathscr{L}}}_{CS}=\frac{1}{2\mu}\varepsilon^{mn}\Biggl(\frac{1}{n^{2}}\biggl(\partial_{n}n\dot{n}^{k}k_{mk}-\partial_{m}n^{k}\partial_{k}n\partial_{n}n-\partial_{m}n^{k}\partial_{n}nn^{l}k_{kl}+\dot{n}k_{ln}\partial_{m}n^{l}\biggr)\nonumber\\
	&&+\frac{1}{n}\biggl(\partial_{m}n^{k}\partial_{n}\left(\partial_{k}n+k_{kl}n^{l}\right)+\dot{n}^{k}\partial_{m}k_{nk}+\dot{k}_{km}\partial_{n}n^{k}\biggr)\nonumber\\
	&&+\dot{k}_{mk}k\thinspace^{k}+k_{nk}\dot{k}_{m}\thinspace^{k}+\gamma_{nl}^{k}\dot{\gamma}_{mk}^{l}+2D_{m}k_{nk}\left(\partial^{k}n+n^{l}k_{l}\thinspace^{k}\right)\nonumber\\
	&&+2k_{mk}D_{n}\left(\partial^{k}n+n^{l}k_{l}\thinspace^{k}\right)+\partial_{m}\gamma_{nk}^{l}\left(D_{l}n^{k}+nk_{k}\thinspace^{l}\right)\\
	&&+2\gamma_{mr}^{k}\gamma_{nl}^{r}\left(D_{k}n^{l}+nk_{k}\thinspace^{l}\right)+\gamma_{mr}^{k}\partial_{n}\left(D_{k}n^{r}+nk_{k}\thinspace^{r}\right)+2k_{m}\thinspace^{k}k_{nl}D_{k}n^{l}\Biggr).\nonumber
	\label{eq:tmgaction1}
\end{eqnarray}
Here the terms in the first two lines can be written as surface terms and therefore we will not take them into account in the canonical analysis but they are needed for the purposes of conserved charge computation. The first two terms in the third line are identical which is obvious from
\begin{equation}
	\varepsilon^{mn}k_{nk}\dot{k}_{m}\thinspace^{k}=\partial_{0}\left(\varepsilon^{mn}k_{nk}k_{m}\thinspace^{k}\right)-\varepsilon^{mn}\dot{k}_{nk}k_{m}\thinspace^{k}=\varepsilon^{mn}\dot{k}_{mk}k_{n}\thinspace^{k},
\end{equation}
where the total time derivative term vanishes due to the antisymmetry. Now let us focus on the non tensorial terms and try to express them in a tensorial form. The terms with the two dimensional Christoffel symbol can be written as
\begin{eqnarray}
	&&\varepsilon^{mn}\biggl(\gamma_{nl}^{k}\dot{\gamma}_{mk}^{l}+\partial_{m}\gamma_{nk}^{l}\left(D_{l}n^{k}+nk_{k}\thinspace^{l}\right)+2\gamma_{mr}^{k}\gamma_{nl}^{r}\left(D_{k}n^{l}+nk_{l}\thinspace^{k}\right)\nonumber\\&&+\gamma_{mr}^{k}\partial_{n}\left(D_{k}n^{r}+nk_{k}\thinspace^{r}\right)\biggr)\nonumber\\&&=\varepsilon^{mn}\biggl(\gamma_{nl}^{k}\dot{\gamma}_{mk}^{l}+2\left(D_{l}n^{k}+nk_{l}\thinspace^{k}\right)\left(\partial_{m}\gamma_{nk}^{l}+\gamma_{mr}^{l}\gamma_{nk}^{r}\right)\biggr).
\end{eqnarray}
By using the definition of the two dimensional Riemann tensor, which is explicitly
\begin{equation}
	^{(2)}R^{l}\thinspace_{kmn}=\frac{1}{2}\left(\delta_{m}^{l}\gamma_{kn}-\delta_{n}^{l}\gamma_{km}\right){}^{(2)}R\label{eq:2driemann},
\end{equation}
we can write
\begin{equation}
	\varepsilon^{mn}\left(D_{l}n^{k}+nk_{l}\thinspace^{k}\right)\left(\partial_{m}\gamma_{nk}^{l}+\gamma_{mr}^{l}\gamma_{nk}^{r}\right)=\frac{1}{2}\varepsilon^{mn}\left(D_{l}n^{k}+nk_{l}\thinspace^{k}\right)^{(2)}R^{l}\thinspace_{kmn}
\end{equation}
and it reduces to
\begin{equation}
	\varepsilon^{mn}\left(D_{l}n^{k}+nk_{l}\thinspace^{k}\right)\left(\partial_{m}\gamma_{nk}^{l}+\gamma_{mr}^{l}\gamma_{nk}^{r}\right)=\frac{1}{2}\varepsilon^{mn}D_{m}n_{n}{}^{(2)}R.
\end{equation}
The remaining term that we need to compute is the term which involves
a time derivative of the two dimensional Christoffel symbol and
it can be expressed as
\begin{eqnarray}
	&&\varepsilon^{mn}\gamma_{nl}^{k}\dot{\gamma}_{mk}^{l}=\frac{1}{2}\varepsilon^{mn}\biggl(\gamma_{nl}^{k}2\dot{\gamma}^{lr}\gamma_{nl}^{k}\gamma_{mk}^{p}\gamma_{pr}-\dot{\gamma}_{kr}\partial_{m}\left(\gamma_{nl}^{k}\gamma^{lr}\right)\nonumber\\&&-\dot{\gamma}_{mr}\partial_{k}\left(\gamma_{nl}^{k}\gamma^{lr}\right)+\dot{\gamma}_{km}\partial_{r}\left(\gamma_{nl}^{k}\gamma^{lr}\right)\biggr).
\end{eqnarray}
Ignoring the surface terms and using the metric compatibility condition we have
\begin{eqnarray}
	&&\varepsilon^{mn}\gamma_{nl}^{k}\dot{\gamma}_{mk}^{l}=\frac{1}{2}\varepsilon^{mn}\Biggl(2\dot{\gamma}^{lr}\gamma_{nl}^{k}\gamma_{mk}^{p}\gamma_{pr}-\dot{\gamma}_{kr}\gamma^{lr}\partial_{m}\gamma_{nl}^{k}+\dot{\gamma}_{kr}\gamma^{pr}\gamma_{mp}^{l}\gamma_{nl}^{k}\nonumber\\
	&&-\dot{\gamma}_{mr}\gamma^{lr}\partial_{k}\gamma_{nl}^{k}+\dot{\gamma}_{mr}\gamma^{pr}\gamma_{pk}^{l}\gamma_{nl}^{k}+\dot{\gamma}_{mr}\gamma^{pl}\gamma_{pk}^{r}\gamma_{nl}^{k}\nonumber\\
	&&+\dot{\gamma}_{mk}\gamma^{pr}\partial_{p}\gamma_{nr}^{k}-\dot{\gamma}_{mk}\gamma^{pr}\gamma_{pr}^{l}\gamma_{nl}^{k}-\dot{\gamma}_{mk}\gamma^{pr}\gamma_{pl}^{l}\gamma_{nr}^{k}\Biggr).
\end{eqnarray}
Again by using the expression of the two dimensional Riemann tensor which was given in equation (\ref{eq:2driemann}) we have
\begin{equation}
	\varepsilon^{mn}\left(-\dot{\gamma}_{kr}\gamma^{lr}\partial_{m}\gamma_{nl}^{k}+\dot{\gamma}_{kr}\gamma^{pr}\gamma_{mp}^{l}\gamma_{nl}^{k}\right)=-\frac{1}{2}\varepsilon^{mn}\dot{\gamma}_{mn}{}^{(2)}R=0
\end{equation}
and  we can express some of the remaining terms as
\begin{equation}
	\varepsilon^{mn}\left(\dot{\gamma}_{mk}\gamma^{pr}\partial_{p}\gamma_{nr}^{k}+\dot{\gamma}_{mr}\gamma^{pl}\gamma_{pk}^{r}\gamma_{nl}^{k}-\dot{\gamma}_{mk}\gamma^{pr}\gamma_{pr}^{l}\gamma_{nl}^{k}\right)=\varepsilon^{mn}\dot{\gamma}_{mk}\gamma^{pr}\partial_{n}\gamma_{pr}^{k}.
\end{equation}
Finally we obtain
\begin{eqnarray}
	&&2\varepsilon^{mn}\gamma_{nl}^{k}\dot{\gamma}_{mk}^{l}=\varepsilon^{mn}\Bigl(2\dot{\gamma}^{lr}\gamma_{nl}^{k}\gamma_{mk}^{p}\gamma_{pr}\nonumber\\&&+\dot{\gamma}_{mk}\gamma^{pr}\left(\partial_{n}\gamma_{pr}^{k}-\gamma_{pl}^{l}\gamma_{nr}^{k}\right)-\dot{\gamma}_{mr}\gamma^{lr}\left(\partial_{k}\gamma_{nl}^{k}-\gamma_{np}^{k}\gamma_{kl}^{p}\right)\Bigr)\thinspace\thinspace
\end{eqnarray}
and we can write the term that involves a time derivative of the connection
in a symmetric form as,
\begin{equation}
	\varepsilon^{mn}\gamma_{nl}^{k}\dot{\gamma}_{mk}^{l}=:A^{ab}\dot{\gamma}_{ab},
\end{equation}
where we have defined a new two index non-tensor quantity which can be expressed as 
\begin{eqnarray}
	&&A^{ab}=\frac{1}{4}\biggl(-2\varepsilon^{mn}\gamma_{nl}^{k}\gamma_{mk}^{b}\gamma^{la}+\varepsilon^{an}\biggl(\gamma^{pr}\left(\partial_{n}\gamma_{pr}^{b}-\gamma_{pl}^{l}\gamma_{nr}^{b}\right)\nonumber\\&&-\gamma^{rb}\left(\partial_{k}\gamma_{nr}^{k}-\gamma_{np}^{k}\gamma_{kr}^{p}\right)\biggr)\biggr)+a\leftrightarrow b.
\end{eqnarray}
Consequently all two dimensional Christoffel symbol terms together can be written as
\begin{eqnarray}
	\frac{1}{2\mu}\varepsilon^{mn}\biggl(\gamma_{nl}^{k}\dot{\gamma}_{mk}^{l}+\partial_{m}\gamma_{nk}^{l}\left(D_{l}n^{k}+nk_{k}\thinspace^{l}\right)+2\gamma_{mr}^{k}\gamma_{nl}^{r}\left(D_{k}n^{l}+nk_{l}\thinspace^{k}\right)\nonumber\\+\gamma_{mr}^{k}\partial_{n}\left(D_{k}n^{r}+nk_{k}\thinspace^{r}\right)\biggr)=\frac{1}{2\mu}\left(A^{ab}\dot{\gamma}_{ab}+\varepsilon^{mn}D_{m}n_{n}{}^{(2)}R\right).
\end{eqnarray}
Substituting the result in the ADM decomposition of the Chern-Simons
action, which is given in (\ref{eq:tmgaction1}) we obtain
\begin{eqnarray}
	\text{\ensuremath{\mathscr{L}}}_{CS}=\frac{1}{2\mu}\varepsilon^{mn}\Biggl(2\dot{k}_{mk}k_{n}\thinspace^{k}+2D_{m}k_{nk}\left(\partial^{k}n+n^{l}k_{l}\thinspace^{k}\right)+2k_{mk}D_{n}\left(\partial^{k}n+n^{l}k_{l}\thinspace^{k}\right)\nonumber\\
	+D_{m}n_{n}{}^{(2)}R+2k_{m}\thinspace^{k}k_{nl}D_{k}n^{l}\Biggr)+\frac{1}{2\mu}A^{ab}\dot{\gamma}_{ab}\thinspace\thinspace\thinspace\thinspace\thinspace\thinspace\thinspace\thinspace\thinspace\thinspace\thinspace\thinspace\thinspace\thinspace\thinspace\thinspace\thinspace\thinspace\thinspace\thinspace\thinspace\thinspace\thinspace\thinspace\thinspace\thinspace\thinspace\thinspace\thinspace\thinspace\thinspace\thinspace
\end{eqnarray}
and by using $\dot{\gamma}_{ab}=2nk_{ab}+D_{a}n_{b}+D_{b}n_{a}$ we
have
\begin{eqnarray}
	&&\text{\ensuremath{\mathscr{L}}}_{CS}=\frac{1}{\mu}\varepsilon^{mn}\Biggl(D^{k}\left(D_{m}k_{nk}n\right)+D_{n}\left(k_{mk}\partial^{k}n\right)-D^{k}\left(nD_{n}k_{mk}\right)\nonumber\\
	&&+D_{n}\left(k_{mk}\left(n^{l}k_{l}\thinspace^{k}\right)\right)+D_{k}\left(k_{m}\thinspace^{k}k_{nl}n^{l}\right)+D_{m}\left(\frac{1}{2}n_{n}{}^{(2)}R\right)-\frac{1}{2}n_{n}\partial_{m}{}^{(2)}R\nonumber\\
	&&-D_{k}\left(k_{m}\thinspace^{k}k_{nl}\right)n^{l}+\dot{k}_{mk}k_{n}\thinspace^{k}+D_{m}k_{nk}n^{l}k_{l}\thinspace^{k}\nonumber\\
	&&-D_{n}k_{mk}n^{l}k_{l}\thinspace^{k}-nD^{k}D_{m}k_{nk}+nD^{k}D_{n}k_{mk}\Biggr)\nonumber\\
	&&+\frac{1}{\mu}\Biggl(A^{ab}\gamma_{ab}^{r}n_{r}-\partial_{a}A^{ab}n_{b}+A^{ab}nk_{ab}+\partial_{a}\left(A^{ab}n_{b}\right)\Biggr).
\end{eqnarray}

Finally up to boundary terms the ADM decomposition of the Chern-Simons Lagrangian is
\begin{eqnarray}
	&&\text{\ensuremath{\mathscr{L}}}_{CS}=\frac{1}{\mu}\Biggl(\varepsilon^{mn}\dot{k}_{mk}k_{n}\thinspace^{k}+n\left(A^{ab}k_{ab}+2\varepsilon^{mn}D^{k}D_{n}k_{mk}\right)\Biggr)\\
	&&+\frac{1}{\mu}n_{a}\Biggl(\frac{1}{2}\varepsilon^{an}\partial_{n}{}^{(2)}R-2\varepsilon^{mn}D_{n}k_{mk}k{}^{ak}-\varepsilon^{mn}D_{k}\left(k_{m}\thinspace^{k}k_{n}\thinspace^{a}\right)-\left(\partial_{b}A^{ba}-A^{cb}\gamma_{cb}^{a}\right)\Biggr).\nonumber\label{eq:tmgaction}
\end{eqnarray}

\subsection{The TMG action}

Combining all the above results the Lagrangian of the topologically massive gravity up to a boundary term in the ADM formulation reads
\begin{eqnarray}
	&&\text{\ensuremath{\mathscr{L}}}_{TMG}=n\sqrt{\gamma}\left(-2\Lambda+{}^{(2)}R-k^{2}+k_{ab}^{2}\right)+v^{ab}\left(\dot{\gamma}_{ab}-2nk_{ab}-2D_{a}n_{b}\right)\nonumber\\&&+\frac{1}{\mu}\sqrt{\gamma}\left(\epsilon^{mn}\dot{k}_{mk}k_{n}\thinspace^{k}+n\left(\gamma^{-\frac{1}{2}}A^{ab}k_{ab}+2\epsilon^{mn}D^{k}D_{n}k_{mk}\right)\right)\nonumber\\&&+\frac{1}{\mu}\sqrt{\gamma}n_{a}\biggl(\frac{1}{2}\epsilon^{an}\partial_{n}{}^{(2)}R-2\epsilon^{mn}D_{n}k_{mk}k{}^{ak}-\epsilon^{mn}D_{k}\left(k_{m}\thinspace^{k}k_{n}\thinspace^{a}\right)\nonumber\\&&-\gamma^{-\frac{1}{2}}\left(\partial_{b}A^{ba}-A^{cb}\gamma_{cb}^{a}\right)\biggr),\label{eq:tmgactionfinal}
\end{eqnarray}
where we have introduced the Lagrange multiplier $v^{ab}$. As usual we can  define the conjugate momenta of the two dimensional metric tensor as
\begin{equation}
	\pi^{ab}=\frac{\delta\text{\ensuremath{\mathscr{L}}}_{TMG}}{\delta\dot{\gamma}_{ab}}=v^{ab}
\end{equation}
and the conjugate momenta of the extrinsic curvature tensor as 
\begin{equation}
	\varPi^{ab}=\frac{\delta\text{\ensuremath{\mathscr{L}}}_{TMG}}{\delta\dot{k}_{ab}}=\frac{1}{\mu}\sqrt{\gamma}\epsilon^{an}k_{n}\thinspace^{b}.\label{eq:pii}
\end{equation}

\subsection{The Hamiltonian of TMG}

It is obvious from equation (\ref{eq:pii}) that $k$ and $\varPi$
are not independent, therefore the system is constrained. Or to stress this: we cannot solve the velocities $\dot{k}_{ab}$ in terms of their momenta 	$\varPi^{ab}$, so we have to include
this dependency as a constraint equation by introducing Lagrange multipliers
in the Hamiltonian. Therefore the Hamiltonian of TMG is
\begin{equation}
	\text{\ensuremath{\mathscr{H}}}_{TMG}=\pi^{ab}\dot{\gamma}_{ab}+\varPi^{ab}\dot{k}_{ab}-\text{\ensuremath{\mathscr{L}}}_{TMG}+f_{ab}\left(\varPi^{ab}-\frac{1}{\mu}\sqrt{\gamma}\epsilon^{an}k_{n}\thinspace^{b}\right),
\end{equation}
here just like $A^{ab}$ , $\pi^{ab}$ is not a tensor. Inserting
the TMG action, which was expressed in equation (\ref{eq:tmgactionfinal}),
Hamiltonian of the TMG becomes
\begin{eqnarray}
	&&\text{\ensuremath{\mathscr{H}}}_{TMG}=\partial_{a}\left(\pi^{ab}n_{b}\right)+f_{ab}\left(\varPi^{ab}-\frac{1}{\mu}\sqrt{\gamma}\epsilon^{an}k_{n}\thinspace^{b}\right)\nonumber\\&&+n\sqrt{\gamma}\biggl(2\Lambda-{}^{(2)}R+k^{2}-k_{ab}^{2}-\frac{2}{\mu}\epsilon^{mn}D^{k}D_{n}k_{mk}+\gamma^{-\frac{1}{2}}\left(2\pi^{ab}-\frac{1}{\mu}A^{ab}\right)k_{ab}\biggr)\nonumber\\&&+\sqrt{\gamma}n_{a}\biggl(-\frac{1}{2\mu}\epsilon^{an}\partial_{n}{}^{(2)}R+\frac{1}{\mu}\epsilon^{mn}\left(2D_{n}k_{mk}k{}^{ak}+D_{k}\left(k_{m}\thinspace^{k}k_{n}\thinspace^{a}\right)\right)\nonumber\\&&+\gamma^{-\frac{1}{2}}\partial_{b}\left(\frac{1}{\mu}A^{ba}-2\pi^{ab}\right)+\gamma^{-\frac{1}{2}}\left(\frac{1}{\mu}A^{bc}-2\pi^{cb}\right)\gamma_{cb}^{a}\biggr).
\end{eqnarray}
We can introduce the following tensor
\begin{equation}
	P^{ab}=\gamma^{-\frac{1}{2}}\left(2\pi^{ab}-\frac{1}{\mu}A^{ab}\right)
\end{equation}
to recast the Hamiltonian as
\begin{eqnarray}
	&&\text{\ensuremath{\mathscr{H}}}_{TMG}= n\sqrt{\gamma}\left(2\Lambda-{}^{(2)}R+k^{2}-k_{ab}^{2}-\frac{2}{\mu}\epsilon^{mn}D^{k}D_{n}k_{mk}+P^{ab}k_{ab}\right)\nonumber\\
	&&+\sqrt{\gamma}n_{a}\left(-\frac{1}{2\mu}\epsilon^{an}\partial_{n}{}^{(2)}R+\frac{2}{\mu}\epsilon^{mn}D_{n}k_{mk}k{}^{ak}+\frac{1}{\mu}\epsilon^{mn}D_{k}\left(k_{m}\thinspace^{k}k_{n}\thinspace^{a}\right)+D_{b}P^{ab}\right)\nonumber\\&&+f_{ab}\left(\varPi^{ab}-\frac{1}{\mu}\sqrt{\gamma}\epsilon^{an}k_{n}\thinspace^{b}\right),\label{eq:hamiltoniantmg}
\end{eqnarray}
where $f_{ab}$ is a Lagrange multiplier.

\subsection{The Constraint equations of TMG }

We can now obtain the constraint equations of TMG from the above Hamiltonian. Variations of $n_{a}$, $n$ and $f_{ab}$ yields three constraint equations.

The Hamiltonian constraint equation can be obtained from the variation with respect to $n$ as
\begin{equation}
	\left(\Phi_{1}\right)=\frac{\delta\text{\ensuremath{\mathscr{H}}}_{TMG}}{\delta n}=-\frac{2}{l^{2}}-{}^{(2)}R+k^{2}-k_{ab}^{2}-\frac{2}{\mu}\epsilon^{mn}D^{k}D_{n}k_{mk}+P^{ab}k_{ab},
\end{equation}
and the momentum constraint equations can be obtained from the variation with respect to $n_{a}$ as
\begin{eqnarray}
	&&\left(\Phi_{2}\right)^{a}=\frac{\delta\text{\ensuremath{\mathscr{H}}}_{TMG}}{\delta n_{a}}=\frac{2}{\mu}\epsilon^{mn}D_{n}k_{mk}k{}^{ak}+\frac{1}{\mu}\epsilon^{mn}D_{k}\left(k_{m}\thinspace^{k}k_{n}\thinspace^{a}\right)\nonumber\\&&-\frac{1}{2\mu}\epsilon^{an}\partial_{n}{}^{(2)}R-D_{b}P^{ba}
\end{eqnarray}
and also we have an additional constraint equation which comes from the variation of the Hamiltonian with respect to Lagrange multiplier $f_{ab}$ as
\begin{equation}
	\left(\Phi_{3}\right)^{ab}=\frac{\delta\text{\ensuremath{\mathscr{H}}}_{TMG}}{\delta f_{ab}}=\varPi^{ab}-\frac{1}{2\mu}\sqrt{\gamma}\left(\epsilon^{an}k_{n}\thinspace^{b}+\epsilon^{bn}k_{n}\thinspace^{a}\right).
\end{equation}

In contrast to the constraint equations directly obtained from the field equations, it seems there arise three constraint equations from the Hamiltonian formalism of TMG. But actually  the last constraint equation does not say anything new about the theory, it only repeats the relation between the extrinsic curvature tensor and its conjugate momenta, where the relation is obvious in equation (\ref{eq:pii}).

\subsection{ADM split of the Cotton tensor}

To carry out the full ADM analysis, we need to split the Cotton tensor on the hypersurface and normal to the hypersurface. In local coordinates the Cotton tensor is defined as
\begin{equation}
	C_{\mu\nu}=\frac{1}{2}\epsilon{}^{\rho\alpha\beta}\left(g_{\mu\rho}\nabla_{\alpha}S_{\beta\nu}+g_{\nu\rho}\nabla_{\alpha}S_{\beta\mu}\right),
\end{equation}
here $\epsilon{}^{\rho\alpha\beta}$ is the antisymmetric tensor in interchanging
of any two indices and it can be expressed in terms of the two dimensional antisymmetric tensor, $\epsilon^{mn}$, where $\epsilon{}^{0mn}=\epsilon{}^{n0m}=\epsilon{}^{mn0}=\frac{1}{n}\epsilon^{mn}$
and $\epsilon^{mn}=\gamma^{-\frac{1}{2}}\varepsilon^{mn}.$ Also $S_{\mu\nu}$
is the Schouten tensor which is defined in three dimensions as
\begin{equation}
	S_{\mu\nu}=R_{\mu\nu}-\frac{1}{4}g_{\mu\nu}R.
\end{equation}
Let us find the $2+1$ dimensional decomposition of the Cotton tensor step by step. Since there are summations over Greek indices we can decompose them as summations over time and space components. Decomposing $\rho$ gives
\begin{equation}
	C_{\mu\nu}=\frac{1}{2}\epsilon{}^{0\alpha\beta}\left(g_{\mu0}\nabla_{\alpha}S_{\beta\nu}+g_{\nu0}\nabla_{\alpha}S_{\beta\mu}\right)+\frac{1}{2}\epsilon{}^{k\alpha\beta}\left(g_{\mu k}\nabla_{\alpha}S_{\beta\nu}+g_{\nu k}\nabla_{\alpha}S_{\beta\mu}\right)
\end{equation}
and decomposing $\alpha$ and $\beta$ indices gives
\begin{eqnarray}
	C_{\mu\nu}=&&\frac{1}{2}\epsilon{}^{0mn}\left(g_{\mu0}\nabla_{m}S_{n\nu}+g_{\nu0}\nabla_{m}S_{n\mu}\right)\\
	&&+\frac{1}{2}\epsilon{}^{k0m}\left(g_{\mu k}\nabla_{0}S_{m\nu}+g_{\nu k}\nabla_{0}S_{m\mu}\right)+\frac{1}{2}\epsilon{}^{km0}\left(g_{\mu k}\nabla_{m}S_{0\nu}+g_{\nu k}\nabla_{m}S_{0\mu}\right)\nonumber.
\end{eqnarray}
Using the expression for three index $\epsilon$ tensor and renaming
the indices we can write
\begin{eqnarray}
	C_{\mu\nu}=\frac{1}{2n}\epsilon{}^{mn}\left(g_{\mu0}\nabla_{m}S_{n\nu}+g_{\nu0}\nabla_{m}S_{n\mu}+g_{\mu n}\nabla_{0}S_{m\nu}\right.\nonumber\\\left.+g_{\nu n}\nabla_{0}S_{m\mu}-g_{\mu n}\nabla_{m}S_{0\nu}-g_{\nu n}\nabla_{m}S_{0\mu}\right)
\end{eqnarray}
and in terms of the partial derivatives it is
\begin{eqnarray}
	C_{\mu\nu}=&&\frac{1}{2n}\epsilon{}^{mn}\Biggl(g_{\mu0}\left(\partial_{m}S_{n\nu}-\varGamma_{m\nu}^{0}S_{n0}-\varGamma_{m\nu}^{k}S_{nk}\right)\\&&+g_{\nu0}\left(\partial_{m}S_{n\mu}-\varGamma_{m\mu}^{0}S_{n0}-\varGamma_{m\mu}^{k}S_{nk}\right)\nonumber\\
	&&+g_{\mu n}\left(\partial_{0}S_{m\nu}-\varGamma_{0\nu}^{0}S_{m0}-\varGamma_{0\nu}^{k}S_{mk}-\partial_{m}S_{0\nu}+\varGamma_{m\nu}^{0}S_{00}+\varGamma_{m\nu}^{k}S_{0k}\right)\nonumber\\
	&&+g_{\nu n}\left(\partial_{0}S_{m\mu}-\varGamma_{0\mu}^{0}S_{m0}-\varGamma_{0\mu}^{k}S_{mk}-\partial_{m}S_{0\mu}+\varGamma_{m\mu}^{0}S_{00}+\varGamma_{m\mu}^{k}S_{0k}\right)\Biggr)\nonumber.
\end{eqnarray}
By using the general expression above we can obtain the Cotton tensor
components. Hypersurface projection of the Cotton tensor is 
\begin{eqnarray}
	C_{ij}=&&\frac{1}{2n}\epsilon{}^{mn}\Bigl(g_{i0}\left(\partial_{m}S_{nj}-\varGamma_{mj}^{0}S_{n0}-\varGamma_{mj}^{k}S_{nk}\right)\\
	&&+g_{in}\left(\partial_{0}S_{mj}-\varGamma_{0j}^{0}S_{m0}-\varGamma_{0j}^{k}S_{mk}-\partial_{m}S_{0j}+\varGamma_{mj}^{0}S_{00}+\varGamma_{mj}^{k}S_{0k}\right)\Bigr)+i\leftrightarrow j\nonumber
\end{eqnarray}
and projection once to the normal and once to the surface yields
\begin{eqnarray}
	C_{0i}=&&\frac{1}{2n}\epsilon{}^{mn}\Biggl(g_{00}\left(\partial_{m}S_{ni}-\varGamma_{mi}^{0}S_{n0}-\varGamma_{mi}^{k}S_{nk}\right)\\&&+g_{i0}\left(\partial_{m}S_{n0}-\varGamma_{m0}^{0}S_{n0}-\varGamma_{m0}^{k}S_{nk}\right)\nonumber\\
	&&+g_{0n}\left(\partial_{0}S_{mi}-\varGamma_{0i}^{0}S_{m0}-\varGamma_{0i}^{k}S_{mk}-\partial_{m}S_{0i}+\varGamma_{mi}^{0}S_{00}+\varGamma_{mi}^{k}S_{0k}\right)\nonumber\\
	&&+g_{in}\left(\partial_{0}S_{m0}-\varGamma_{00}^{0}S_{m0}-\varGamma_{00}^{k}S_{mk}-\partial_{m}S_{00}+\varGamma_{m0}^{0}S_{00}+\varGamma_{m0}^{k}S_{0k}\right)\Biggr).\nonumber
\end{eqnarray}
Projecting twice to the surface yields
\begin{eqnarray}
	C_{00}=&&\frac{1}{n}\epsilon{}^{mn}\Biggl(g_{00}\left(\partial_{m}S_{n0}-\varGamma_{m0}^{0}S_{n0}-\varGamma_{m0}^{k}S_{nk}\right)\\
	&&+g_{0n}\left(\partial_{0}S_{m0}-\varGamma_{00}^{0}S_{m0}-\varGamma_{00}^{k}S_{mk}-\partial_{m}S_{00}+\varGamma_{m0}^{0}S_{00}+\varGamma_{m0}^{k}S_{0k}\right)\Biggr)\nonumber.
\end{eqnarray}
Before finding the explicit expressions for the components of the
Cotton tensor we need to express the components of the Schouten tensor.
From the definition of the Schouten tensor, one has
\begin{equation}
	S_{ij}=R_{ij}-\frac{1}{4}\gamma_{ij}R
\end{equation}
and by using the equations (\ref{eq:ri0}, \ref{eq:r}) projection
once to the surface and once to normal yields
\begin{equation}
	S_{i0}=n^{j}S_{ij}+n\left(D_{m}k_{i}^{m}-D_{i}k\right)
\end{equation}
and from equations (\ref{eq:r00}, \ref{eq:r}) projection twice normal to the surface gives
\begin{eqnarray}
	S_{00}=n^{i}n^{j}S_{ij}-n^{2}k_{ij}^{2}+nn^{k}\left(D_{m}k_{k}^{m}-D_{k}k\right)\nonumber\\+n\left(D_{k}\partial^{k}n-\dot{k}+n^{k}D_{m}k_{k}^{m}\right)+\frac{1}{4}n^{2}R.
\end{eqnarray}
Before finding the Cotton tensor components, let us compute some frequently appearing expressions. These are
\begin{equation}
	\partial_{m}S_{nj}-\varGamma_{mj}^{0}S_{n0}-\varGamma_{mj}^{k}S_{nk}=D_{m}S_{nj}-k_{mj}\left(D_{r}k_{n}^{r}-D_{n}k\right),
\end{equation}
\begin{eqnarray}
	\partial_{m}S_{n0}-\varGamma_{m0}^{0}S_{n0}-\varGamma_{m0}^{k}S_{nk}=n^{r}\left(D_{m}S_{nr}-k_{mr}\left(D_{s}k_{n}^{s}-D_{n}k\right)\right)\nonumber\\+n\left(D_{m}D_{r}k_{n}^{r}-k_{m}^{k}S_{kn}\right),
\end{eqnarray}
\begin{eqnarray}
	\partial_{0}S_{mj}-\varGamma_{0j}^{0}S_{m0}-\varGamma_{0j}^{k}S_{mk}=\dot{S}_{mj}-nk_{j}^{k}S_{mk}-S_{mk}D_{j}n^{k}\nonumber\\-\left(\partial_{j}n+n^{r}k_{rj}\right)\left(D_{s}k_{m}^{s}-D_{m}k\right)
\end{eqnarray}
and
\begin{eqnarray}
	-\partial_{m}S_{0j}+\varGamma_{mj}^{0}S_{00}+\varGamma_{mj}^{k}S_{0k}=D_{m}\left(-n^{r}S_{rj}-n\left(D_{r}k_{j}^{r}-D_{j}k\right)\right)\nonumber\\
	+k_{mj}\left(D_{k}\partial^{k}n-\dot{k}+n^{k}D_{s}k_{k}^{s}+n\left(\frac{1}{4}R-k_{rs}^{2}\right)\right),
\end{eqnarray}
\begin{eqnarray}
	\partial_{0}S_{m0}-\varGamma_{00}^{0}S_{m0}-\varGamma_{00}^{k}S_{mk}=n\left(\partial_{0}D_{r}k_{m}^{r}-\partial_{m}\dot{k}-S_{m}^{k}\left(\partial_{k}n+n^{r}k_{rk}\right)\right)\nonumber\\n^{r}\left(\dot{S}_{mr}-nk_{j}^{k}S_{mk}-S_{mk}D_{r}n^{k}-\left(\partial_{r}n+n^{s}k_{sr}\right)\left(D_{k}k_{m}^{k}-D_{m}k\right)\right)
\end{eqnarray}
and
\begin{eqnarray}
	-\partial_{m}S_{00}+\varGamma_{m0}^{0}S_{00}+\varGamma_{m0}^{k}S_{0k}=n^{r}D_{m}\left(-n^{s}S_{rs}-n\left(D_{s}k_{r}^{s}-\partial_{r}k\right)\right)\nonumber\\
	+n^{r}k_{mr}\left(D_{k}\partial^{k}n-\dot{k}+n^{k}D_{s}k_{k}^{s}+n\left(\frac{1}{4}R-k_{rs}^{2}\right)\right)+n\partial_{m}n\left(k_{rs}^{2}-\frac{1}{4}R\right)\nonumber\\
	-n^{k}\partial_{m}n\left(D_{s}k_{k}^{s}-D_{k}k\right)-n\partial_{m}\left(D_{k}\partial^{k}n-\dot{k}+n^{k}D_{s}k_{k}^{s}\right)-\frac{1}{4}n^{2}\partial_{m}R\nonumber\\
	+2n^{2}k_{rs}D_{m}k^{rs}+nk_{m}^{k}S_{kr}n^{r}+n^{2}k_{m}^{k}\left(D_{s}k_{k}^{s}-D_{k}k\right).\thinspace\thinspace\thinspace\thinspace\thinspace\thinspace\thinspace\thinspace\thinspace\thinspace\thinspace
\end{eqnarray}
Making use of these identities, we can finally obtain the components of the Cotton tensor as follows: full hypersurface projection of the Cotton tensor becomes
\begin{eqnarray}
	&&C_{ij}=\frac{1}{2n}\epsilon{}^{mn}\Biggl(n_{i}\left(D_{m}S_{nj}-k_{mj}\left(D_{r}k_{n}^{r}-\partial_{n}k\right)\right)\nonumber\\
	&&+\gamma_{in}\dot{S}_{mj}-nk_{j}^{k}S_{mk}-S_{mk}D_{j}n^{k}-\left(\partial_{j}n+n^{r}k_{rj}\right)\left(D_{s}k_{m}^{s}-\partial_{m}k\right)\nonumber\\
	&&+D_{m}\left(-n^{r}S_{rj}-n\left(D_{r}k_{j}^{r}-D_{j}k\right)\right)\nonumber\\
	&&+k_{mj}\left(D_{k}\partial^{k}n-\dot{k}+n^{k}D_{s}k_{k}^{s}+\frac{n}{4}R-nk_{rs}^{2}\right)\Biggr)+i\leftrightarrow j\label{eq:cij}
\end{eqnarray}
and projection once to the surface once normal to the surface becomes
\begin{eqnarray}
	C_{i0}&=&n^{j}C_{ij}+\frac{1}{2}\epsilon{}^{mn}\Biggl(-n\left(D_{m}S_{ni}-k_{mi}\left(D_{r}k_{n}^{r}-\partial_{n}k\right)\right)\\ &&+n_{i}\left(D_{m}D_{r}k_{n}^{r}-k_{m}^{k}S_{kn}\right)+\gamma_{in}\partial_{0}D_{r}k_{m}^{r}-S_{m}^{k}\left(\partial_{k}n+n^{r}k_{rk}\right)\nonumber\\&&-D_{m}D_{k}\partial^{k}n-D_{m}\left(n^{k}D_{s}k_{k}^{s}\right)+k_{m}^{k}S_{kr}n^{r}
	+\partial_{m}n\left(k_{rs}^{2}-\frac{1}{4}R\right)\nonumber\\&&+n\left(2k_{rs}D_{m}k^{rs}-\frac{1}{4}\partial_{m}R+k_{m}^{k}\left(D_{r}k_{k}^{r}-\partial_{k}k\right)\right)\Biggr).\nonumber\thinspace\thinspace\thinspace\thinspace\thinspace\thinspace\thinspace\thinspace\thinspace\thinspace\thinspace\thinspace\thinspace\thinspace\thinspace\thinspace
\end{eqnarray}
Projecting twice to the normal of the surface yields
\begin{eqnarray}
	C_{00}&=&n^{i}n^{j}C_{ij}+\epsilon{}^{mn}\Biggl(-nn^{r}\left(D_{m}S_{nr}-k_{mr}\left(D_{s}k_{n}^{s}-D_{n}k\right)\right)\\&&+\left(n_{i}n^{i}-n^{2}\right)\left(D_{m}D_{r}k_{n}^{r}-k_{m}^{k}S_{kn}\right)+n_{n} \partial_{0}D_{r}k_{m}^{r}-S_{m}^{k}\left(\partial_{k}n+n^{r}k_{rk}\right)\nonumber\\
	&&-D_{m}D_{k}\partial^{k}n-D_{m}\left(n^{k}D_{s}k_{k}^{s}\right)+k_{m}^{k}S_{kr}n^{r}\nonumber\\
	&&+\partial_{m}n\left(k_{rs}^{2}-\frac{1}{4}R\right)+n\left(2k_{rs}D_{m}k^{rs}-\frac{1}{4}\partial_{m}R+k_{m}^{k}\left(D_{r}k_{k}^{r}-\partial_{k}k\right)\right)\Biggr)\nonumber.
\end{eqnarray}
For notational simplicity, let us define the following tensors
\begin{equation}
	A_{mni} :=D_{m}S_{ni}-k_{mi}\left(D_{r}k_{n}^{r}-\partial_{n}k\right),
\end{equation}
\begin{equation}
	B_{mn}:=D_{m}D_{r}k_{n}^{r}-k_{m}^{k}S_{kn},
\end{equation}
\begin{equation}
	E_{m}:=2k_{rs}D_{m}k^{rs}-\frac{1}{4}\partial_{m}R+k_{m}^{k}\left(D_{r}k_{k}^{r}-\partial_{k}k\right),
\end{equation}
\begin{eqnarray}
	C_{m}:=\partial_{0}D_{r}km-S_{m}^{k}\left(\partial_{k}n+n^{r}k_{rk}\right)-D_{m}D_{k}\partial^{k}n-D_{m}\left(n^{k}D_{s}k_{k}^{s}\right)\nonumber\\+k_{m}^{k}S_{kr}n^{r}+\partial_{m}n\left(k_{rs}^{2}-\frac{1}{4}R\right)
\end{eqnarray}
and reexpress $C_{i0}$ and $C_{00}$ in terms of $C_{ij}$ and the  above tensors as follows
\begin{equation}
	C_{i0}=n^{j}C_{ij}+\frac{1}{2}\epsilon{}^{mn}\left(-nA_{mni}+n_{i}B_{mn}+\gamma_{in}\left(C_{m}+nE_{m}\right)\right)\label{eq:ci0}
\end{equation}
and
\begin{eqnarray}
	&&C_{00}=n^{i}n^{j}C_{ij}\\&&+\epsilon{}^{mn}\left(-nn^{r}A_{mnr}+\left(n_{r}n^{r}-n^{2}\right)B_{mn}+n_{n}\left(C_{m}+nE_{m}\right)\right).\nonumber\label{eq:c00}
\end{eqnarray}

Obtaining the last two equations has been our main goal in this part, let us now move on to the field equations.

\subsection{ADM split of the TMG field equations}

Matter-coupled cosmological TMG field equations are
\begin{equation}
	\mathscr{E}_{\mu\nu}=G_{\mu\nu}+\Lambda g_{\mu\nu}+\frac{1}{\mu}C_{\mu\nu}=\kappa\tau_{\mu\nu},
\end{equation}
where $G_{\mu\nu}$ is the three dimensional Einstein tensor with
the definition
\begin{equation}
	G_{\mu\nu}=R_{\mu\nu}-\frac{1}{2}g_{\mu\nu}R.
\end{equation}
By using the Cotton and Ricci tensor components and the scalar
curvature, we obtain the ADM decomposition of the TMG field equation components. Full hypersurface projection of the field equation is
\begin{equation}
	\text{\ensuremath{\mathscr{E}}}_{ij}=S_{ij}-\frac{1}{4}\gamma_{ij}R+\Lambda\gamma_{ij}+\frac{1}{\mu}C_{ij}=\kappa\tau_{ij},
\end{equation}
while projection once to the surface and once normal to the surface gives
\begin{eqnarray}
	\mathscr{E}_{0i}&=&n^{j}\mathscr{E}_{ij}+n\left(D_{r}k_{i}^{r}-\partial_{i}k\right)\nonumber\\&&+\frac{1}{2\mu}\epsilon{}^{mn}\left(n_{i}B_{mn}-nA_{mni}+\gamma_{in}\left(C_{m}+nE_{m}\right)\right)=\kappa\tau_{0i}
\end{eqnarray}
and projection twice to the normal yields
\begin{eqnarray}
	\mathscr{E}_{00}=2n^{i}\mathscr{E}_{0i}-n^{i}n^{j}\mathscr{E}_{ij}+n\left(D_{k}\partial^{k}n-\dot{k}+n^{k}D_{k}k+n\left(\frac{1}{2}R-k_{rs}^{2}\right)\right)\nonumber\\-\Lambda n^{2}-\frac{1}{\mu}\epsilon{}^{mn}n^{2}B_{mn}=\kappa\tau_{00}.
\end{eqnarray}
The field equation components can be used to find the constraint equations
by setting $\text{\ensuremath{\mathscr{E}}}_{ij}$ to $\kappa\tau_{ij}$
and $\text{\ensuremath{\mathscr{E}}}_{0i}$ to $\kappa\tau_{0i}$.
Then we obtain the momentum constraint equation of TMG as
\begin{eqnarray}
	\Phi_{i}:=n\left(D_{r}k_{i}^{r}-\partial_{i}k\right)+\frac{1}{2\mu}\epsilon{}^{mn}\left(-nA_{mni}+n_{i}B_{mn}+\gamma_{in}\left(C_{m}+nE_{m}\right)\right)\nonumber\\=\kappa\left(\tau_{0i}-n^{j}\tau_{ij}\right)
\end{eqnarray}
and  the Hamiltonian constraint equation as
\begin{eqnarray}
	\Phi:=D_{k}\partial^{k}n-\dot{k}+n^{k}D_{k}k+n\left(\frac{1}{2}R-k_{rs}^{2}\right)-\Lambda n-\frac{1}{\mu}\epsilon{}^{mn}nB_{mn}\nonumber\\=\kappa\left(\tau_{00}-2n^{i}\tau_{0i}+n^{i}n^{j}\tau_{ij}\right).
\end{eqnarray}
Substituting the corresponding tensors, the explicit form of the momentum constraint equation is
\begin{multline}
	\Phi_{i}=n\left(D_{r}k_{i}^{r}-\partial_{i}k\right)+\frac{1}{2\mu}\epsilon{}^{mn}\Biggl(-n\left(D_{m}S_{ni}-k_{mi}\left(D_{r}k_{n}^{r}-\partial_{n}k\right)\right)\\+n_{i}\left(D_{m}D_{r}k_{n}^{r}-k_{m}^{k}S_{kn}\right)+\gamma_{in}\left\{ n\left(2k_{rs}D_{m}k^{rs}-\frac{1}{4}\partial_{m}R+k_{m}^{k}\left(D_{r}k_{k}^{r}-\partial_{k}k\right)\right)\right.\\+\partial_{0}D_{r}k_{m}^{r}-S_{m}^{k}\left(\partial_{k}n+n^{r}k_{rk}\right)-D_{m}D_{k}\partial^{k}n-D_{m}\left(n^{k}D_{s}k_{k}^{s}\right)+k_{m}^{k}S_{kr}n^{r}\\\left.\left.+\partial_{m}n\left(k_{rs}^{2}-\frac{1}{4}R\right)\right\} \right)=\kappa\left(\tau_{0i}-n^{j}\tau_{ij}\right)\label{eq:momentumconstraint}
\end{multline}
and  the explicit form of the Hamiltonian constraint equation is
\begin{eqnarray}
	\Phi=D_{k}\partial^{k}n-\dot{k}+n^{k}D_{k}k+n\left(\frac{1}{2}R-k_{rs}^{2}\right)-\frac{1}{\mu}\epsilon{}^{mn}n\left(D_{m}D_{r}k_{n}^{r}-k_{m}^{k}S_{kn}\right)\nonumber\\
	-\Lambda n=\kappa\left(\tau_{00}-2n^{i}\tau_{0i}+n^{i}n^{j}\tau_{ij}\right).\thinspace\thinspace\thinspace\thinspace\thinspace\thinspace\thinspace\thinspace\thinspace\thinspace\thinspace\thinspace\thinspace\thinspace\thinspace\thinspace\thinspace\thinspace\thinspace\thinspace
\end{eqnarray}
By using the ADM formulation of the scalar curvature, $R=2\left(3\Lambda-\kappa\tau\right)$
for the TMG case, the Hamiltonian constraint equation reduces to
\begin{eqnarray}
	\Phi=\frac{1}{2}\left(^{(2)}R+k^{2}-k_{ij}^{2}\right)-\Lambda-\frac{1}{\mu}\epsilon{}^{mn}\left(D_{m}D_{r}k_{n}^{r}-k_{m}^{k}S_{kn}\right)\nonumber\\=\frac{\kappa}{n^{2}}\left(\tau_{00}-2n^{i}\tau_{0i}+n^{i}n^{j}\tau_{ij}\right).\label{eq:hamiltonianconstraint}
\end{eqnarray}

\subsection{Constraint equations and field equations in Gaussian normal coordinates}

Up to this point we have worked with generic lapse and shift functions. Dues to the complicated expressions in this generic form, let us restrict the discussion to the Gaussian normal coordinates. Of course this is always possible in a small neighborhood of the spacetime manifold. Coordinate singularities might develop which would make these coordinates unsuitable, but this does not change the relevant discussion here. In the Gaussian normal coordinates, one has 
\begin{equation}
	ds^{2}=-dt^{2}+\gamma_{ij}dx^{i}dx^{j},
\end{equation}
where we took $n=1$, $n_{i}=0$ and with this setting  the Hamiltonian constraint equation which is given with equation
(\ref{eq:hamiltonianconstraint}) reduces to
\begin{equation}
	\Phi=\frac{1}{2}\left(^{(2)}R+k^{2}-k_{ij}^{2}\right)-\Lambda-\frac{1}{\mu}\epsilon{}^{mn}\left(D_{m}D_{r}k_{n}^{r}-k_{m}^{k}S_{kn}\right)=\kappa\tau_{00}
\end{equation}
and the momentum constraint equations given in equation (\ref{eq:momentumconstraint}) reduce to
\begin{eqnarray}
	\Phi_{i}&&=D_{r}k_{i}^{r}-\partial_{i}k+\frac{1}{2\mu}\epsilon{}^{mn}\Biggl(-D_{m}S_{ni}+k_{mi}\left(D_{r}k_{n}^{r}-\partial_{n}k\right)\\
	&&+\gamma_{in}\left(2k_{rs}D_{m}k^{rs}-\frac{1}{4}\partial_{m}R+k_{m}^{k}\left(D_{r}k_{k}^{r}-\partial_{k}k\right)+\partial_{0}D_{r}k_{m}^{r}\right)\Biggr)=\kappa\tau_{0i}\nonumber.
\end{eqnarray}
The extrinsic curvature tensor of the surface in these coordinates simply becomes
$k_{ij}=\frac{1}{2}\dot{\gamma}_{ij}$ and from equation (\ref{eq:rij}),
three dimensional Ricci tensor can be expressed as
\begin{equation}
	R_{ij}=^{(2)}R_{ij}+kk_{ij}-2k_{ik}k_{j}^{k}+\dot{k}_{ij}.
\end{equation}
From equation (\ref{eq:r}) scalar curvature reduces to
\begin{equation}
	R=^{(2)}R+k^{2}+k_{ij}^{2}+2\dot{k}.
\end{equation}
We can compute the Schouten tensor by using these expressions and find
the expressions of the constraints in the Gaussian normal
coordinates. As a final result for the Hamiltonian constraint equation
we obtain the following expression 
\begin{eqnarray}
	-\frac{1}{2\mu}\epsilon^{mn}\left(D_{m}D^{k}\dot{\gamma}_{kn}-\frac{1}{2}\dot{\gamma}_{sm}\gamma^{ks}\left(^{(2)}R_{kn}-\dot{\gamma}_{kp}\dot{\gamma}_{sn}\gamma^{ps}-\ddot{\gamma}_{kn}\right)\right)\nonumber\\+\frac{1}{8}\dot{\gamma}_{ij}\left(\dot{\gamma}_{ab}\gamma^{ab}\gamma^{ij}+\dot{\gamma}^{ij}\right)=\tau_{00}+\Lambda-\frac{1}{2}{}^{(2)}R\label{eq:gnc_hc}
\end{eqnarray}
and for the momentum constraint equations we have
\begin{eqnarray}
	-\frac{1}{8\mu}\epsilon^{mn}\left(\dot{\gamma}_{ab}\gamma^{ab}D_{m}\dot{\gamma}_{in}-2\gamma^{ks}D_{m}\left(\dot{\gamma}_{kn}\dot{\gamma}_{si}\right)+2D_{m}\ddot{\gamma}_{in}-\dot{\gamma}_{mi}D^{k}\dot{\gamma}_{kn}\right)\nonumber\\
	+\frac{1}{8\mu}\epsilon^{m}\thinspace_{i}\left(2\dot{\gamma}^{kp}D_{k}\dot{\gamma}_{pm}-\dot{\gamma}^{ks}D_{m}\dot{\gamma}_{ks}+2D^{k}\ddot{\gamma}_{km}-\dot{\gamma}_{mk}\gamma^{kl}D^{p}\dot{\gamma}_{pl}\right)\nonumber\\+\frac{1}{2}\left(D^{k}\dot{\gamma}_{ki}-\gamma^{ab}D_{i}\dot{\gamma}_{ab}\right)=\tau_{0i}+\frac{1}{2\mu}\epsilon^{mn}D_{m}{}^{(2)}R_{ni}.
\end{eqnarray}
Furthermore  if we take a conformally flat $2D$ metric on $\Sigma$, which is always possible, we have $\gamma_{ij}=e^{\varphi}\delta_{ij}$,
where $\varphi=\varphi(t,x_{i})$ and the metric of the Gausian normal coordinates becomes
\begin{equation}
	ds^{2}=-dt^{2}+e^{\varphi}\delta_{ij}dx^{i}dx^{j}.\label{eq:metricc}
\end{equation}
The Ricci tensor of the two dimensional surface is then
\begin{equation}
	^{(2)}R_{ij}=-\frac{1}{2}\gamma_{ij}e^{-\varphi}\nabla^{2}\varphi,
\end{equation}
where $\nabla^{2}=\partial_{x}^{2}+\partial_{y}^{2}=\partial_{k}\partial_{k}$
here. The scalar curvature of the hypersurface $\Sigma$ can be expressed as
\begin{equation}
	^{(2)}R=-e^{-\varphi}\partial_{k}\partial_{k}\varphi=-\frac{1}{2}e^{-\varphi}\left(2D_{k}\partial_{k}\varphi+\partial_{k}\varphi\partial_{k}\varphi\right).
\end{equation}
The Hamiltonian and the momentum constraints can be expressed as the following
equations respectively
\begin{equation}
	\frac{1}{4}\dot{\varphi}^{2}=\kappa\tau_{00}+\Lambda-\frac{1}{2}{}^{(2)}R,\label{eq:hcignc}
\end{equation}
\begin{equation}
	-\frac{1}{2}\partial_{i}\dot{\varphi}=\kappa\tau_{0i}+\frac{1}{2\mu}\epsilon^{mn}D_{m}{}^{(2)}R_{ni}.\label{eq:mcignc}
\end{equation}
The relation between the Ricci tensor and the scalar curvature of the hypersurface is
\begin{equation}
	^{(2)}R_{ij}=\frac{1}{2}\gamma_{ij}{}^{(2)}R
\end{equation}
and by using this relation we can express the momentum constraint equations as follows
\begin{equation}
	\partial_{i}\dot{\varphi}=-2\kappa\tau_{0i}-\frac{1}{2\mu}\epsilon^{m}\thinspace_{i}D_{m}{}^{(2)}R.\label{eq:mcgnc}
\end{equation}
By using the equation (\ref{eq:hcignc}) we can write
\begin{equation}
	^{(2)}R=2\left(\kappa\tau_{00}+\Lambda\right)-\frac{1}{2}\dot{\varphi}^{2}
\end{equation}
and substituting the last expression in equation (\ref{eq:mcgnc}) we obtain
\begin{equation}
	\partial_{i}\dot{\varphi}=-2\kappa\tau_{0i}-\frac{1}{\mu}\epsilon^{m}\thinspace_{i}\kappa\tau_{00}+\frac{1}{2\mu}\epsilon^{m}\thinspace_{i}\dot{\varphi}\partial_{m}\dot{\varphi}.\label{eq:final1}
\end{equation}
For simplicity let us define the current tensor $J$ as 
\begin{equation}
	J_{i}:=2\kappa\tau_{0i}+\frac{1}{\mu}\epsilon^{m}\thinspace_{i}\kappa\tau_{00}\label{eq:ji}
\end{equation}
and reexpress the constraint equation as
\begin{equation}
	\partial_{i}\dot{\varphi}=-J_{i}+\frac{1}{2\mu}\epsilon^{m}\thinspace_{i}\dot{\varphi}\partial_{m}\dot{\varphi}.
\end{equation}
Contracting the last equation with $\epsilon^{ki}$ gives
\begin{equation}
	\partial_{i}\dot{\varphi}=\frac{2\mu}{\dot{\varphi}}\epsilon_{i}\thinspace^{m}J_{m}+\frac{2\mu}{\dot{\varphi}}\epsilon_{i}\thinspace^{m}\dot{\varphi}\partial_{m}\dot{\varphi},
\end{equation}
which has the same left hand side with the equation (\ref{eq:final1}). This yields the equality
\begin{equation}
	-J_{i}+\frac{1}{2\mu}\epsilon^{m}\thinspace_{i}\dot{\varphi}\partial_{m}\dot{\varphi}=\frac{2\mu}{\dot{\varphi}}\epsilon_{i}\thinspace^{m}J_{m}+\frac{2\mu}{\dot{\varphi}}\epsilon_{i}\thinspace^{m}\dot{\varphi}\partial_{m}\dot{\varphi}.
\end{equation}
Let us reexpress our result by using the definition of the J tensor
which was given in equation (\ref{eq:ji}) as
\begin{eqnarray}
	\frac{2\mu}{\dot{\varphi}}\epsilon^{m}\thinspace_{i}\partial_{m}\dot{\varphi}\left(1+\frac{\dot{\varphi}^{2}}{4\mu^{2}}\right)=\frac{2}{\dot{\varphi}}\left(\partial_{i}+\frac{\dot{\varphi}}{2\mu}\epsilon^{m}\thinspace_{i}\partial_{m}\right)\kappa\tau_{00}\nonumber\\+\frac{4\mu}{\dot{\varphi}}\left(\epsilon_{i}\thinspace^{m}+\frac{\dot{\varphi}}{2\mu}\delta^{m}\thinspace_{i}\right)\kappa\tau_{0m}.\label{eq:final2}
\end{eqnarray}
Note that in the vacuum case, the right hand side of the last equation
becomes zero and the only solution is $\varphi_{0}=c\,t$ with a constant
$c$ which is the de Sitter solution and the constant can be obtained from
the trace equation as $c=\frac{2}{l}$.

\subsection{Perturbations around the de Sitter space}

We can compute the perturbed constraint equations around the background solution given by $\varphi_{0}$. The perturbation is defined as $\varphi=\varphi_{0}+\delta\varphi$.
Since $\varphi_{0}=c\,t$ any arbitrary spatial derivative of the $\varphi_{0}$
will be zero ($\partial_{i}\dot{\varphi}_{0}=0$). Nonvanishing terms
of the perturbation of the final result which is given in equation
(\ref{eq:final2}) is then
\begin{eqnarray}
	\frac{2\mu}{\dot{\varphi_{0}}}\epsilon^{m}\thinspace_{i}\partial_{m}\dot{\delta\varphi}\left(1+\frac{\dot{\varphi_{0}}^{2}}{4\mu^{2}}\right)=\frac{2}{\dot{\varphi}}\left(\partial_{i}+\frac{\dot{\varphi}}{2\mu}\epsilon^{m}\thinspace_{i}\partial_{m}\right)\kappa\delta\tau_{00}\nonumber\\+\frac{4\mu}{\dot{\varphi}}\left(\epsilon_{i}\thinspace^{m}+\frac{\dot{\varphi}}{2\mu}\delta^{m}\thinspace_{i}\right)\kappa\delta\tau_{0m},
\end{eqnarray}
where we can express the result in terms of the solution constant
$c$ as
\begin{eqnarray}
	\frac{2\mu}{c}\epsilon^{m}\thinspace_{i}\partial_{m}\dot{\delta\varphi}\left(1+\frac{c^{2}}{4\mu^{2}}\right)=\frac{2}{c}\left(\partial_{i}+\frac{c}{2\mu}\epsilon^{m}\thinspace_{i}\partial_{m}\right)\kappa\delta\tau_{00}\nonumber\\+\frac{4\mu}{c}\left(\epsilon_{i}\thinspace^{m}+\frac{c}{2\mu}\delta^{m}\thinspace_{i}\right)\kappa\delta\tau_{0m}.\label{eq:final3}
\end{eqnarray}

\subsection{From de Sitter to anti de Sitter}

Wick rotating the parameters as $x_{i}\rightarrow ix_{i}$ , $t\rightarrow it$,
$c\rightarrow ic$, one obtains the AdS space and the Gaussian normal coordinates metric (\ref{eq:metricc}) becomes
\begin{equation}
	ds^{2}=dt^{2}-e^{-ct}\left(dx^{2}+dx^{2}\right).
\end{equation}
At the chiral point $\mu^{2}l^{2}=1$, we have $c^{2}=-\frac{4}{l^{2}}$
and left hand side of the perturbed equation (\ref{eq:final3}) is
zero but since the right hand side of the equation is not zero, there
exists a linearization instability at this point.

\subsection{TMG field equations in Gaussian normal coordinates}

We can also analyse the ADM decomposition of the TMG field equations
in Gaussian normal coordinates. The ADM decomposition of the Cotton tensor,
which is given in equation (\ref{eq:cij}), reduces to
\begin{eqnarray}
	C_{ij}=&&\frac{1}{2}\epsilon{}^{mn}\Biggl(\gamma_{in}\Biggl(\dot{S}_{mj}-k_{j}^{k}S_{mk}-D_{m}\left(D_{r}k_{j}^{r}-D_{j}k\right)\\&&-k_{mj}\left(\dot{k}-\frac{1}{4}R+k_{rs}^{2}\right)\Biggr)\Biggr)+i\leftrightarrow j\nonumber
\end{eqnarray}
and in terms of the metric components it can be expressed as
\begin{equation}
	C_{ij}=\frac{1}{4}\epsilon{}^{mn}\left(\gamma_{in}D_{m}\partial_{j}\dot{\varphi}+\gamma_{jn}D_{m}\partial_{i}\dot{\varphi}\right).
\end{equation}
Now we can express the TMG field equations as follows
\begin{equation}
	\mathscr{E}_{ij}=-\frac{1}{4}\gamma_{ij}\left(\dot{\varphi}^{2}+2\ddot{\varphi}\right)+\Lambda\gamma_{ij}+\frac{1}{4\mu}\left(\epsilon{}^{m}\thinspace_{i}D_{m}\partial_{j}\dot{\varphi}+\epsilon{}^{m}\thinspace_{j}D_{m}\partial_{i}\dot{\varphi}\right)
\end{equation}
and writing the covariant derivatives as partial derivatives we
have 
\begin{eqnarray}
	\mathscr{E}_{ij}=-\frac{1}{4}\gamma_{ij}\left(\dot{\varphi}^{2}+2\ddot{\varphi}\right)+\Lambda\gamma_{ij}+\frac{1}{8\mu}\Biggl(2\epsilon{}^{m}\thinspace_{i}\partial_{m}\partial_{j}\dot{\varphi}+2\epsilon{}^{m}\thinspace_{j}\partial_{m}\partial_{i}\dot{\varphi}\nonumber\\
	-\epsilon{}^{m}\thinspace_{i}\left(\partial_{m}\varphi\partial_{j}\dot{\varphi}+\partial_{j}\varphi\partial_{m}\dot{\varphi}\right)-\epsilon{}^{m}\thinspace_{j}\left(\partial_{m}\varphi\partial_{i}\dot{\varphi}+\partial_{i}\varphi\partial_{m}\dot{\varphi}\right)\Biggr).
\end{eqnarray}

%%%%%%%%%%%%%%%%%%%%%%%%%%%%%%%%%%%%%%%

\newpage

\section{ Acknowledgments} 

I would like to express my special appreciation and thanks to my advisor Prof. Dr. Bayram Tekin, for his excellent supervision, patient guidance, encouragement, friendly attitude and for being accessible with a continuous support. Without his guidance, it could not be easy for me to improve my talents and perspective to physics problems. I cannot imagine a better and professional advisor, I have learned so many things from him and I am grateful for this.

I would also like to thank my committee members, Prof. Dr. Atalay Karasu, Assoc. Prof. Dr. {Ali Ula\c{s}  Özgür Ki\c{s}isel}, Assoc. Prof. Dr. {Tahsin \c{C}a\u{g}r{\i} \c{S}i\c{s}man} and Assoc. Prof. Dr. {\c{C}etin  \c{S}entürk} for serving as my committe members even at times of hardship. 

I would also like to thank the following people who I have interacted throughout my graduate school years: Nadire Nayir, Esin Kenar, Fatma Acar, Dilek K{\i}z{\i}lören, \"{U}mmügül Erözbek Güngör, Merve Demirta\c{s}, Suat Dengiz, \.{I}brahim Güllü, Gök\c{c}en Deniz Özen, \c{S}ahin Kürek\c{c}i and Ercan K{\i}l{\i}\c{c}arslan.

A special thanks to my family members for always supporting me. Words cannot express how grateful I am to my mother Selime Alta\c{s}, my father Özcan Alta\c{s}, my sister Ay\c{c}a Can, my brother \c{S}ükrü Erdim Alta\c{s} and other family members Or\c{c}un Emrah Can, Esra Alta\c{s} and Alp Ahmet Can.\\

I would like to thank my husband Ali Kirac{\i} for his continuous support and help in the process of writing this thesis and always being there for me, making my life easier and beautiful. I'm so lucky to have such a wonderful family.

\newpage

\addcontentsline{toc}{section}{References}

\end{document}